\documentclass{article}
\usepackage{epsfig}
\usepackage{psfrag}
\usepackage{upgreek}
\usepackage{subfigure}
\usepackage{amsmath,amssymb,amscd,amsthm}
\title{
Discrete integrable systems and Poisson algebras from cluster maps}
\author{Allan Fordy\thanks{School of Mathematics,
University
of Leeds, Leeds LS2 9JT. ~~E-mail: a.p.fordy@leeds.ac.uk}
$\,$ and Andrew Hone\thanks{School of Mathematics,
Statistics and Actuarial Science, University of
Kent, Canterbury CT2 7NF, U.K. ~~E-mail: A.N.W.Hone@kent.ac.uk}
}

\newcommand{\br}{\begin{rem}}
\newcommand{\er}{\end{rem}}
\newcommand{\bex}{\begin{exa}}
\newcommand{\eex}{\end{exa}}
\newcommand{\bd}{\begin{Def}}
\newcommand{\ed}{\end{Def}}
\newcommand{\bt}{\begin{theorem}}
\newcommand{\et}{\end{theorem}}
\newcommand{\bl}{\begin{lemma}}
\newcommand{\el}{\end{lemma}}
\newcommand{\be}{\begin{equation}}
\newcommand{\ee}{\end{equation}}
\newcommand{\bea}{\begin{eqnarray}}
\newcommand{\eea}{\end{eqnarray}}
\newcommand{\pa}{\partial}
\newcommand{\nn}{\nonumber}
\newcommand{\adots}{\mathinner{\mkern2mu\raise1pt\hbox{.}\mkern2mu
\raise4pt\hbox{.}\mkern2mu\raise7pt\hbox{.}\mkern1mu}}

\newcommand{\beq}{\begin{equation}}
\newcommand{\eeq}{\end{equation}}
\newcommand{\bear}{\begin{array}}
\newcommand{\eear}{\end{array}}
\newcommand\la{{\lambda}}

\newcommand\al{{\alpha}}

\newcommand\om{{\omega}}

\newcommand\dd{\mathrm{d}}
\newtheorem{thm}{Theorem}[section]

\newtheorem{propn}[thm]{Proposition}

\newtheorem{rem}[thm]{Remark}

\newtheorem{exa}[thm]{Example}

\newtheorem{conje}[thm]{Conjecture}

\newtheorem{lem}[thm]{Lemma}

\newtheorem{cor}[thm]{Corollary}

\newenvironment{prf}{\trivlist \item [\hskip
\labelsep {\bf Proof:}]\ignorespaces}{\qed \endtrivlist}

\theoremstyle{remark}

\newcommand{\RA}{{\mathrm A}}
\newcommand{\RB}{{\mathrm B}}
\newcommand{\RC}{{\mathrm C}}
\newcommand{\RD}{{\mathrm D}}

\newcommand{\CH}{{\cal H}}

\newcommand{\Q}{{\mathbb Q}}
\newcommand{\Z}{{\mathbb Z}}
\newcommand{\C}{{\mathbb C}}

\begin{document}

\maketitle

\begin{abstract}
We consider nonlinear recurrences generated from cluster mutations applied
to quivers that have the property of being cluster mutation-periodic
with period 1. Such quivers were completely classified by
Fordy and Marsh, who characterised them in terms of the
skew-symmetric matrix that defines the quiver. The associated nonlinear
recurrences are equivalent to birational maps, and we explain how these maps
can be endowed with an invariant Poisson bracket and/or presymplectic structure.

Upon applying the algebraic entropy test, we are led to a series of conjectures
which imply that the entropy of the cluster maps can be determined from their
tropical analogues, which leads to a sharp classification result. Only four special
families of these maps should have zero entropy.
These families are examined in detail, with many explicit examples given,
and we show how they lead to discrete dynamics that is integrable
in the Liouville-Arnold sense.
\end{abstract}


\section{Introduction}

Cluster algebras were first developed
by Fomin and Zelevinsky more than a decade ago \cite{fz1}.
Their structure arises in diverse parts of mathematics and theoretical physics,
including Lie theory, quantum algebras, Teichm\"uller theory,
discrete integrable systems and T- and Y-systems.
One of the original motivations for cluster algebras came from a series of observations made by Michael Somos and
others (see \cite{gale}), concerning nonlinear recurrence
relations of the form
\beq
\label{recf}
x_{n+N} \, x_n = F(x_{n+1}, \ldots ,  x_{n+N-1}),
\eeq
where $F$ is a polynomial in $N-1$ variables. The original observation of Somos was that certain
choices of $F$ lead to integer sequences when all $N$ initial values are chosen to be 1. This was
explained by the further observation that for such special $F$ the recurrence (\ref{recf}) exhibits
the \textit{Laurent phenomenon}, meaning that all iterates are Laurent polynomials
in the initial data with integer coefficients.
One of the most well-known examples is the Somos-4 recurrence given by
\beq\label{somos4}
x_{n+4}\, x_n=x_{n+3}x_{n+1}+x_{n+2}^2,
\eeq
which generates the sequence $1,1,1,1,2,3,7,23,59,314,1529,8209,83313,\ldots$\footnote{
{\tt http://oeis.org/A006720}}
starting from four initial 1s, while if the initial data $x_1,x_2,x_3,x_4$ are viewed
as variables then the iterates $x_n$ belong to the Laurent polynomial ring
$\Z[x_1^{\pm 1}, x_2^{\pm 1}, x_3^{\pm 1}, x_4^{\pm 1}]$.

In this paper we consider recurrences of the general form
\beq\label{arec}
x_{n+N}\, x_n = \prod_{a_j\geq 0}x_{n+j}^{a_j}+ \prod_{a_j\leq 0}x_{n+j}^{-a_j},
\eeq
where the indices in each product lie in the range $1\leq j\leq N-1$, with the
exponents $(a_1,...,a_{N-1})$ forming an integer $(N-1)$-tuple which is
palindromic, so that $a_j = a_{N-j}$.  
The main purpose of this paper is to identify which
recurrences of the form (\ref{arec}) can be regarded as
finite-dimensional discrete integrable systems,
in the sense of the
standard Liouville-Arnold definition of integrability
for maps \cite{maeda, veselov}.
The latter  requires  that a map should preserve a Poisson bracket, as well as having
sufficiently many first integrals that commute with respect to this bracket.

A quiver is a graph consisting of a number of nodes together
with arrows between the nodes.
In \cite{fordy_marsh}, Fordy and Marsh showed how recurrences of the form
(\ref{arec}) are produced from sequences of mutations in cluster
algebras defined by quivers with a special periodicity property with
respect to mutations. They define a quiver $Q$ with $N$ nodes to be
\textit{cluster mutation-periodic with period} $m$ if
$\mu_m\cdots \mu_{2}\cdot \mu_1 Q = \rho^m Q$,
where $\mu_j$ denotes quiver mutation at node $j$ and $\rho$ denotes
a cyclic permutation of the nodes. Associated with the quiver
mutation there is a corresponding cluster mutation
acting on a cluster ${\bf x}=(x_1, \ldots ,x_N)$ in a
(coefficient-free) cluster algebra. In the period 1 case, $m=1$,
the action of a suitable ordered sequence of cluster mutations
on cluster variables is precisely equivalent to iteration
of a recurrence of the form (\ref{arec}). A complete
classification of period 1 quivers is given in \cite{fordy_marsh}:
any such quiver produces a recurrence, and conversely any
recurrence of the form (\ref{arec}) determines a
cluster mutation-periodic quiver with period 1.

\subsection{Outline of the paper}

In the next section, we briefly review how the recurrences (\ref{arec}), or the
equivalent birational maps in dimension $N$, come about from cluster mutations.
The main object is the $N\times N$ skew-symmetric integer matrix $B$ corresponding to the quiver,
which not only defines the exponents appearing in (\ref{arec}), but also
produces a presymplectic form which is invariant under the map; this is the two-form
introduced in \cite{gsvduke}.  When det$\, B\neq 0$, the form is symplectic, so the
map automatically has a nondegenerate Poisson bracket. The main result of section 2
is Theorem \ref{torusred}, which states that (even if det$\,B= 0$)   it is always possible to reduce (\ref{arec}) to a
symplectic map, possibly on a space of lower dimension. This provides us with the appropriate setting
in which to consider Liouville integrability in the rest of the paper.

In section \ref{intmaps} we consider the recurrences (\ref{arec}) in the light of
the algebraic entropy test \cite{bellon_viallet}.
We give details of a series of conjectures which show that
the algebraic entropy can be determined explicitly from the
tropical version of  (\ref{arec}), expressed in terms of the max-plus algebra.
From the point of view of the rest of the paper, the main result is a corollary of these
conjectures (Theorem \ref{zeroe}), which classifies the cases with zero
entropy into four families, labelled (i)-(iv).
All of the maps in family (i) have periodic orbits, and it is a trivial task to
show that they are Liouville integrable maps.
The majority of the rest of the paper is devoted to families (ii), (iii) and (iv).

Section \ref{primit} is concerned with the family (ii), which arises from
cluster mutations applied to the primitive quivers, denoted $P_{N}^{(q)}$,
which are defined for each positive integer $N$ and $q=1,\ldots , \lfloor{N/2}\rfloor$.
These were introduced in
\cite{fordy_marsh}, where they were shown to be the building blocks of all
mutation-periodic quivers with period 1. They are also equivalent to
affine $A$-type Dynkin quivers (or copies of such),  and it was 
shown by Fordy and Marsh that the cluster variables in this case satisfy linear
recurrence relations with constant coefficients. (This was
subsequently shown for the general case of cluster algebras associated with affine Dynkin quivers, in
\cite{assem} and \cite{keller_scher}.) Here we give a new proof of these  linear recurrences, which
relies on additional linear relations with periodic coefficients, and their associated
monodromy matrices.  These periodic quantities are the key to the Liouville integrability of the maps
in family (ii).  A large number of new examples of integrable maps arise in this construction and
are explicitly presented.

Section \ref{pert} deals with the family (iii), each member of which arises from a quiver
which is a deformation of a  primitive $P_{N}^{(q)}$,
for a particular $q$ and $N$. The general properties of this family are very close
to those of the primitives. In particular, the cluster variables satisfy linear
recurrences with constant coefficients, and there are additional
linear relations with periodic coefficients. Once again,
associated monodromy arguments, and Poisson subalgebras
defined by the periodic quantities, are the key to
the Liouville integrability of members of this family. Again, many new examples of integrable maps arise and
are explicitly presented.

The family (iv) consists of Somos-type recurrences with three terms, typified by (\ref{somos4}).
In section \ref{somosmaps} we outline some different approaches to understanding
the Liouville integrability of the maps in this family, such as making reductions of the
Hirota-Miwa equation and its Lax pair and by deriving higher bilinear relations with
constant coefficients.

Some of our results were announced previously in \cite{sigma}.

\section{Symplectic maps from cluster recurrences} \label{torusaction}
\setcounter{equation}{0}


Given a recurrence, a major problem is to find an appropriate
symplectic or Poisson structure which is invariant under the action of the corresponding
finite-dimensional map.  Remarkably, in the case of the  cluster recurrences (\ref{arec})
this problem can be solved algorithmically.

\subsection{Recurrences from periodic quivers}

A quiver $Q$ with $N$ nodes and no 1-cycles or 2-cycles
admits quiver mutation. The mutation $\mu_k$ at node $k$ produces
a new quiver $\tilde{Q}=\mu_kQ$ which is obtained as follows:
(i) reverse all arrows in/out of node $k$; (ii) if there are $p$ arrows from
node $j$ to node $k$, and $q$ arrows from node $k$ to node $\ell$,
then add $pq$ arrows from node $j$ to node $\ell$; (iii) remove
any 2-cycles created in step (ii).

An $N\times N$ skew-symmetric matrix $B$ with integer matrix elements
$b_{j\ell}$ defines a quiver $Q$ with $N$ nodes, without 1-cycles or 2-cycles.
Matrix mutation applied at the vertex $k$ is also denoted $\mu_k$,
and starting from $B$ it produces a new matrix
$\tilde{B}=\mu_kB=(\tilde{b}_{j\ell})$ defined by
\beq\label{matmut}
\tilde{b}_{j\ell} =
\left\{
\begin{array}{ll}
-b_{j\ell}\qquad & \mathrm{if} \quad
j=k\quad  \mathrm{or}\quad \ell=k, \\
b_{j\ell} + \frac{1}{2}(|b_{jk}|b_{k\ell} + b_{jk}|b_{k\ell}|) & \mathrm{otherwise.} \end{array}
\right.
\eeq
This matrix mutation is equivalent
to the action of $\mu_k$ on $Q$, via quiver mutation, producing the new
quiver $\tilde{Q}=\mu_k Q$.
In addition to the matrix mutation, the cluster variables
${\bf x}=(x_1,x_2,\ldots,x_N)$ are transformed
by $\mu_k$ to a new cluster
$\tilde{{\bf x}}=(\tilde{x}_1,\tilde{x}_2,\ldots,\tilde{x}_N)$ in such a way that all variables
except $x_k$ are left unaltered, so that $\tilde{x}_j= x_j, \; j\neq k$,
and the exchange relation corresponding to the mutation $\mu_k$ is conveniently written as
\beq \label{exchange}
\tilde{x}_k x_k = \prod_{j=1}^N x_j^{[b_{k,j}]_+} + \prod_{j=1}^N x_j^{[-b_{k,j}]_+},
\eeq
where
$
[b]_+=\max (b,0).
$
Note the identity
$
\frac{1}{2}(a|b|-|a|b) = a[-b]_+-b[-a]_+.
$

In what follows, we require that the $N\times N$ skew-symmetric matrix $B$ defines an $N$-node quiver $Q$ that is
cluster mutation-periodic with period 1. All such matrices were classified
in \cite{fordy_marsh}.  Cluster mutation-periodicity, in the case that the period is 1, means
that after applying a single step of mutation at one of the nodes, $\mu_1$ say, the quiver $\tilde{Q}$
is the same as the quiver $\rho Q$ obtained
from $Q$ by applying the cyclic permutation  $\rho$, 
given by $\rho: \, (1,2,3,\ldots,N)\mapsto (N,1,2,\ldots,N-1)$, such that the number of arrows
from $j$ to $k$ in $Q$ is the same as the number of arrows from $\rho^{-1}(j)$ to $\rho^{-1}(k)$ in $\rho Q$.
This periodicity requirement corresponds to explicit
conditions on the matrix elements of $B$, namely that
\beq\label{mutper}
\tilde{b}_{j+1,k+1}=b_{jk}
\eeq
for all $j,k$, where the indices are read modulo $N$.
Since $\rho^N = \mathrm{id}$, from
the periodicity it is clear that $Q$ is preserved
by the composition of $N$ mutations that cycle around its nodes,
i.e. $\mu_N\cdots\mu_2\cdot \mu_1Q =Q$.

Starting from $B$, one constructs the $N$th order recurrence relation
\beq\label{crec}
x_{n+N}x_n = \prod_{j=1}^{N-1} x_{n+j}^{[b_{1,j+1}]_+} + \prod_{j=1}^{N-1} x_{n+j}^{[-b_{1,j+1}]_+}, \qquad n=1,2,3,\ldots.
\eeq
If $B$ satisfies the conditions (\ref{mutper}), then
iterating the recurrence (\ref{crec}) starting from the initial data
$(x_1,\ldots,x_N)$ is
precisely equivalent to applying cluster mutation $\mu_1$ at the vertex 1, followed by the subsequent mutations
$\mu_2, \mu_3,\ldots, \mu_j, \mu_{j+1},\ldots$, 
and so on. The recurrence relation (\ref{crec}) is clearly
reversible, in the sense that (as long as neither is zero) it can be solved both for $x_{n+N}$, to iterate forwards, and
for $x_n$, to iterate backwards. This means that the index $n$ in (\ref{crec}) is allowed to take all values $n\in\Z$, and also
that iteration of the recurrence is equivalent to iteration of the birational map   $\varphi$
from $\C^N$ to itself,
defined by
\beq \label{bir}
\varphi: \qquad
\left(\begin{array}{c}
x_1 \\
x_2 \\
\vdots  \\
x_{N-1} \\
x_{N}
\end{array} \right)
\longmapsto \left(\begin{array}{c}
x_2 \\
x_3 \\
\vdots  \\
x_{N} \\
x_{N+1}
\end{array} \right),
\qquad \mathrm{where} \qquad
x_{N+1}=\frac{
\prod_{j=1}^{N-1} x_{j+1}^{[b_{1,j+1}]_+} + \prod_{j=1}^{N-1} x_{j+1}^{[-b_{1,j+1}]_+},
 }{x_1}.
\eeq
One can decompose the map (\ref{bir}) as 
$\varphi =\rho^{-1}\cdot \mu_1$, where in
terms of the cluster ${\bf x}=(x_j)$ the map $\mu_1$ sends
$(x_1, x_2, \ldots , x_N)$ to $(\tilde{x}_1, x_2,    \ldots , x_N)$, with
$\tilde{x}_1$ defined according to the exchange relation (\ref{exchange})
with $k=1$, and $\rho^{-1}$ sends $(x_1, x_2, \ldots , x_N)$ to $( x_2,    \ldots , x_N, x_1)$. Due
to the periodicity requirement on $B$, we have $\rho^{-1}\cdot \mu_1 \, (B) = B$, so the action
of $\varphi$ on this matrix is trivial.

\subsection{The Gekhtman-Shapiro-Vainshtein bracket}\label{gsv}

In \cite{gsv} it was
shown that very general cluster algebras admit a linear space of Poisson brackets of log-canonical
type, compatible with the cluster maps generated by mutations, and having the form
\beq\label{logcan}
\{x_j,x_k\}=c_{jk}\, x_{j}x_k
\eeq
for some skew-symmetric constant coefficient matrix $C=(c_{jk})$.
Compatibility of the Poisson structure means that
the cluster transformations $\mu_i$ given by (\ref{exchange}) correspond
to a {\em change of coordinates}, $\tilde{\bf{ x}}=\mu_i ({\bf x}) $, with their bracket
also being log-canonical,
\beq\label{logcant}
\{ \tilde{x}_j,\tilde{x}_k\}=\tilde{c}_{jk}\,\tilde{x}_{j}\tilde{x}_k,
\eeq
for another skew-symmetric constant matrix $\tilde{C}=(\tilde{c}_{jk})$.

Our viewpoint is to regard the cluster transformation as a {\em birational map}
${\bf x}\mapsto \tilde{\bf{ x}}=\varphi ({\bf x})$ in $\C^N$, and require a Poisson
structure that is {\it invariant} with respect to $\varphi$ (not just {\em covariant}).
Therefore, in (\ref{logcant}) we require $\tilde{C}=C$.
However, there may not be a non-trivial log-canonical Poisson bracket that is covariant or invariant under
cluster transformations.

\bex \label{affineA2}
{\em
Corresponding to (\ref{bir}) with $N=3$, the matrix $B$ and birational map on $\C^3$ are given by
\beq \label{affA2}
B=\left(\bear{ccc}
0 & -1 & -1 \\
1 & 0 & -1 \\
1 & 1 & 0 \eear\right),\qquad
\left(\begin{array}{c}
x_1 \\
x_2 \\
x_{3}
\end{array} \right)
\longmapsto \left(\begin{array}{c}
x_2 \\
x_3 \\
(x_{2}x_3+1)/x_1
\end{array} \right).
\eeq
Suppose that there is an invariant Poisson bracket of the form  (\ref{logcan}).
The condition $\varphi^*x_j = x_{j+1}$, implies that $c_{j+\ell,k+\ell}=c_{jk}$
for all indices $j,k,\ell$ in the appropriate range, so $C$ is a Toeplitz matrix:
$$
C= \left(\bear{ccc}
0 & \al & \beta  \\
-\al  & 0 & \al \\
-\beta & -\al & 0 \eear\right).
$$
Upon taking the bracket of both sides of the relation
$x_4x_1=x_2x_3+1$ with $x_2$, one finds that $(\al-\beta )x_1x_2x_4 = -\al x_2^2x_3$,
so $(\al-\beta )(x_2^2x_3+ x_2)=-\al x_2^2x_3$,
which gives $\al = \beta = 0$, so the bracket is trivial.
}
\eex
\begin{rem}[Non-invariance of symplectic leaves] \label{sleaf}
{\em
Even in the case where the map $\varphi$ does admit an invariant log-canonical
Poisson bracket, it may be degenerate, and in that case $\varphi$
need not preserve the symplectic leaves of the bracket. For instance, the Somos-4 recurrence
(\ref{somos4})  has the invariant Poisson bracket \cite{honelaur}
\beq\label{s4br}
\{\, x_j,x_k\,\} = (k-j)\,x_jx_k,
\eeq
which has rank two, but the two independent Casimirs $x_1x_3/x_2^2$, $x_2x_4/x_3^2$
are not fixed by the action of $\varphi$  (see Example
\ref{s4e} below), so the symplectic leaves are not preserved. The analogous observation
for Somos-5 appears in \cite{sigma}.
}
\end{rem}
In general, we shall see that it is more useful
to start with a two-form in the variables $x_j$,  rather than
a Poisson bivector field corresponding to a bracket.

\subsection{Symplectic forms for cluster maps}

Given a skew-symmetric matrix $B$, one can define the log-canonical two-form
\beq \label{omega}
\om =\sum_{j<k} \frac{b_{jk}}{x_jx_k}\dd x_j\wedge \dd x_k,
\eeq
which is just the constant skew-form
$\om = \sum_{j<k} b_{jk}\, \dd z_j\wedge \dd z_k$, written in the logarithmic coordinates $z_j=\log x_j$,
so it is evidently closed, but may be degenerate.  In  \cite{gsvduke} (see also \cite{fockgon}) it was shown
that for a cluster algebra defined by a skew-symmetric integer matrix $B$,
this two-form is compatible with cluster transformations, in the sense that
under a mutation map $\mu_i : \, {\bf x} \mapsto \tilde{ {\bf x}}$, it transforms as
$\mu_i^* \omega =\sum_{j<k}\tilde{b}_{jk} \dd\log\tilde{x}_j\wedge \dd\log\tilde{x}_k$.

In the case that the matrix $B$ is nondegenerate, $\om$ turns out to be a symplectic form for the map
$\varphi$, but in general it is a presymplectic form. For the purposes of our discussion it is
convenient to present some formulae from  \cite{fordy_marsh}, which give the following result.

\begin{lem} \label{symplectic}
Let $B$ be a skew-symmetric integer matrix.
The following conditions are equivalent.

\noindent {\em (i)} The matrix $B$ defines a cluster mutation-periodic quiver with period 1.

\noindent {\em (ii)} The matrix elements $B$ satisfy the relations
\beq\label{reln1}
b_{j,N}=b_{1,j+1}, \qquad j=1,\ldots , N-1,
\eeq
and
\beq\label{reln2}
b_{j+1,k+1}=b_{j,k}+b_{1,j+1} [-b_{1,k+1}]_+  - b_{1,k+1} [-b_{1,j+1}]_+  ,
\qquad 1\leq j,k \leq N-1.
\eeq

\noindent {\em (iii)} The two-form $\om$ is preserved by the map $\varphi$, i.e. $\varphi^* \om =\om$.
\end{lem}

\begin{prf} The proof of (i)$\iff$(ii) follows straight from the definition of periodicity (see the proof of
Theorem 6.1 in \cite{fordy_marsh}).
The implication (ii)$\implies$(iii) in Lemma \ref{symplectic} is a consequence of Theorem 2.1 in \cite{gsvduke},
after applying the permutation $\rho^{-1}$ to the coordinates.  The reverse implication follows from a direct
calculation, which is omitted.
\end{prf}

The formulae (\ref{reln1}) and (\ref{reln2}) entail that for
a period 1 cluster mutation-periodic quiver, the matrix $B$ is completely determined by
the elements in its first row, so that each recurrence of the form (\ref{arec}) with
palindromic exponents corresponds to a matrix $B$.
Theorem 6.1 in \cite{fordy_marsh} is equivalent to the following formula for $b_{jk}$:
\beq \label{aform}
b_{jk}=-b_{kj}=
a_{k-j}
+\sum_{\ell = 1}^{\max (j-1,N-k)}
a_\ell[-a_{\ell+k-j}]_+
-a_{\ell+k-j} [-a_\ell]_+ ,
\qquad 1\leq j< k\leq N,
\eeq
where the $a_j=b_{1,j+1}$ for $j=1,\ldots ,N-1$ form a
palindromic integer $(N-1)$-tuple, such that
$a_j=a_{N-j}$. Apart from being skew-symmetric,  $B$ is also symmetric about the skew diagonal,
i.e. $b_{jk}=b_{N-k+1,N-j+1}$.

Henceforth when we refer to a  recurrence of the form
(\ref{crec}), and the corresponding
matrix $B=(b_{jk})$, we assume that the conditions
(\ref{reln1}) and (\ref{reln2}) hold.
\bex [Corollary 2.2 in \cite{honelaur}] {\em
For integer values of $c\geq 0$, the skew-symmetric matrix 
\beq \label{s4gen}
B=\left(\bear{cccc} 0 & -1 & c & -1 \\
                    1 & 0  & -(c+1) & c \\
                    -c & (c+1) & 0 & -1 \\
                     1 & -c & 1 & 0
\eear\right),
\eeq
satisfies the conditions (\ref{reln1}) and (\ref{reln2}),
which means that it defines a period 1 cluster mutation-periodic quiver $Q$,
in the sense of \cite{fordy_marsh}.
Thus, by Lemma \ref{symplectic}, for each $c$ the map $\varphi$ corresponding to the recurrence
\beq \label{7betarec}
x_{n+4}\, x_{n} = x_{n+3}\, x_{n+1} +   x_{n+2}^c
\eeq
preserves the two-form
\beq \label{7sympm1is1}
\omega
= 
  -\left( \frac{\dd x_1 \wedge \dd x_2}{x_1x_2}+\frac{\dd x_1 \wedge \dd x_4}{x_1x_4}+
\frac{\dd x_3 \wedge \dd x_4}{x_3x_4}\right)
+c \left(\frac{\dd x_1 \wedge \dd x_3}{x_1x_3}+ \frac{\dd x_2 \wedge \dd x_4}{x_2x_4}\right)
-(c+1) \frac{\dd x_2 \wedge \dd x_3}{x_2x_3}.
\eeq
}
\eex

\begin{rem}[Invariant volume form]
\label{volume}{\em
For all maps $\varphi$ of the form (\ref{bir}),
or of the more general form (\ref{recf}), the volume $N$-form
\beq\label{vol}
\Omega = \frac{\dd x_1\wedge \ldots \wedge \dd x_N}{\prod_{j=1}^Nx_j},
\eeq
is invariant up to a sign, depending on the parity of $N$,
i.e. $\varphi^* \Omega = (-1)^N\Omega$. In the case that
$\om$ is nondegenerate, which can only happen for
even $N=2K$, up to overall scale this volume form is
the Poincar\'e invariant
$\om^K=\om\wedge\ldots \wedge\om$ ($K$ terms).
}
\end{rem}

If the matrix $B$ is nondegenerate,
then (up to rescaling by an overall constant) the associated log-canonical Poisson bracket
(\ref{logcan}) is given by the dual bivector field
\beq\label{biv}
\mathcal{J} = \sum_{j<k} c_{jk}\, x_jx_k\, \frac{\partial}{\partial x_j}\wedge \frac{\partial}{\partial x_k}
\qquad \mathrm{with} \qquad C=(c_{jk})=B^{-1}.
\eeq
In the case that $\det \,B =0$, it is necessary to consider
a projection to a lower-dimensional space with a symplectic form, as follows.

\begin{thm} \label{torusred}
The map $\varphi$ is symplectic whenever $B$ is nondegenerate.
For $\mathrm{rank} \, B = 2K \leq N$,
there is a rational map $\pi$ and a symplectic
birational map
$\hat\varphi$ such that the 
diagram
\beq \label{cd}
\begin{CD}
\C^N @>\varphi >> \C^N\\
@VV\pi V @VV\pi V\\
\C^{2K} @>\hat{\varphi}>> \C^{2K}
\end{CD}
\eeq
is commutative, with a log-canonical
symplectic form $\hat\om$ on $\C^{2K}$ that satisfies
$\pi^*\hat\om = \om$.
\end{thm}
The proof of this theorem will occupy most of the rest of this subsection.

To begin with we consider the null distribution of $\om$, which (away from
the hyperplanes $x_j=0$) is generated by $N-2K$ independent commuting vector fields, each of which is of the form
\beq\label{null}
\frac{\partial }{\partial t} = \sum_{j=1}^N u_jx_j \, \frac{\partial}{\partial x_j}
\qquad \mathrm{for} \qquad {\bf u}=(u_j)\in\mathrm{ker} \, B.
\eeq
Since this is an integrable distribution, Frobenius' theorem gives local coordinates
$t_1, \ldots , t_{N-2K} , y_1,\ldots ,y_{2K}$ such that the integral manifolds of the null distribution
are given by $y_j =$constant,  $j=1,\ldots ,2K$.
The coordinates $y_j$ must be invariants for these commuting vector fields, and can be chosen as linear functions of
the logarithmic coordinates $z_j = \log x_j$, but for our purposes it is more convenient
to take functions of the form
$$
y= {\bf x}^{\bf v} :=\prod_j x_j^{v_j} \qquad \iff \qquad ({\bf u}, {\bf v})=0 \quad \forall \, {\bf u}\in \mathrm{ker}\, B ,
$$
where $(\, , )$ denotes the standard scalar product. This yields a log-canonical symplectic form in terms of $y_j$, which is the generalised Weil-Petersson form in \cite{gsvduke}.

Alternatively, one can consider integral curves of the vector fields (\ref{null}), each of which
is the orbit of the point ${\bf x}$ under the action of the algebraic torus $\C^*$, which acts by scaling the coordinates $x_j$. We denote this
one-parameter group action by
\beq\label{scale}
{\bf x}\rightarrow \la^{\bf u}\cdot {\bf x}=(\la^{u_j}x_j), \qquad \la\in\C^*.
\eeq
Combining $N-2K$ independent vector fields of the form (\ref{null}), which (without loss of generality) can be defined by choosing
$N-2K$ independent vectors ${\bf u}=(u_j)\in\mathrm{ker}\, B$ with components $u_j\in\Z$,
we see that the integral manifold through a generic point ${\bf x}$ is the same as its orbit
under the scaling action of the algebraic torus $(\C^*)^{N-2K}$. The coordinates $y_j={\bf x}^{{\bf v}_j}$ are the invariants under
these scaling transformations, and, by choosing  the ${\bf v}_j$ to be integer vectors, define a rational map $\pi$.
\begin{lem} \label{symp}
Suppose that the integer vectors ${\bf v}_1, \ldots , {\bf v}_{2K}$ form a basis for
$\mathrm{im}\, B$, and define the rational map
$$
\bear{lrcl}
\pi : \quad & \C^N & \longrightarrow & \C^{2K} \\
            & {\bf x} & \longmapsto  & {\bf y}, \qquad y_j ={\bf x}^{{\bf v}_j}, \quad j=1,\ldots, 2K.
\eear
$$
Then the $y_j$ are a complete set of Laurent monomial invariants for all of the
scaling transformations (\ref{scale})
defined by integer vectors ${\bf u}\in \mathrm{ker}\, B$,
and there is a log-canonical symplectic form
\beq\label{yform}
\hat{\om} = \sum_{j<k} \frac{\hat{b}_{jk}}{y_jy_k}\dd y_j\wedge \dd y_k, \qquad
with \qquad \pi^*\hat\om = \om .
\eeq
\end{lem}
\begin{prf}
Since the integer matrix $B$ is skew-symmetric, the vector space
$\Q^N$ has an orthogonal direct sum decomposition
$\Q^N =  \mathrm{im}\, B \oplus \mathrm{ker}\, B$. The scaling action on Laurent monomials
gives $\la^ {{\bf u}}\cdot {\bf x}^{{\bf v}} = \prod_j \la^{u_jv_j} x_j^{v_j}
=\la^{({\bf u},{\bf v})}  {\bf x}^{{\bf v}}$, hence ${\bf x}^{{\bf v}}$ is
invariant under the overall action of $(\C^*)^{N-2K}$ if and only if $({\bf u},{\bf v})=0$
for all ${\bf u}\in \mathrm{ker}\, B$, so ${\bf v}\in\mathrm{im}\, B$, and a basis
of $\mathrm{im}\, B$ gives $2K$ independent monomial invariants.
Now extend the basis of $\mathrm{im}\, B$ to a basis $\{{\bf v}_j\}_{j=1}^{N}$ for
$\Q^N$,  so that ${\bf v}_j\in\mathrm{ker}\, B$ for
$2K+1\leq j\leq N$, and let $M$ be the matrix whose rows consist of the $N$ basis vectors.
Then one finds the block matrix
$$
B^\natural = M^{-T}BM^{-1}
=\left(\bear{cc} \hat{B} & \mathbf{0} \\  \mathbf{0} &  \mathbf{0} \eear \right)
$$
(with $M^{-T}=(M^{-1})^T$, and $T$ denoting transpose)
where $ \hat{B} = (\hat{b}_{jk}) $ is a nondegenerate skew-symmetric $2K\times 2K$ matrix.
Defining the two-form (\ref{yform}) in terms of $ \hat{B}$ gives
$ \pi^*\hat\om  =\om$, as required.
\end{prf}

The map $\pi$ and the specific form of $\hat\om$ depend on the choice of integer basis for  $\mathrm{im}\, B$.
Here we consider rational maps between fixed affine spaces, even
if these are not defined everywhere.  In the case where
$B$ is nondegenerate, the diagram (\ref{cd}) is trivial
for $\pi =\mathrm{id}$, but there are other non-trivial choices of $\pi$,
corresponding to different integer bases for $\mathrm{im}\, B$.

The following classical theorem (which is a special case of Theorem IV.1
in \cite{newman}) provides a canonical choice of $\pi$, and via Lemma \ref{symp} gives Darboux
coordinates for the presymplectic form $\om$. The proof in \cite{newman} is constructive.

\begin{thm}\label{normalform}
If $B$ is a skew-symmetric matrix of rank $2K$
in $\mathrm{Mat}_N(\Z )$ then there are integers $h_1,\ldots ,h_K$
and a unit matrix $M\in\mathrm{Mat}_N(\Z )$
such that $B=M^TB^\natural M$, where
$$
B^\natural = h_1 \, \left(\bear{cc} 0 & 1 \\ -1 & 0 \eear\right) \oplus
h_2 \, \left(\bear{cc} 0 & 1 \\ -1 & 0 \eear\right) \oplus
\ldots  \oplus
h_K \, \left(\bear{cc} 0 & 1 \\ -1 & 0 \eear\right) \oplus \mathbf{0},
$$
with
$h_j|h_{j+1}$ for $j=1,\ldots ,K-1$.
\end{thm}

In the above result, the first $2K$ rows of the unit matrix $M$ provide a $\Z$-basis for the
$\Z$-module   $\mathrm{im}\, B_\Z=\mathrm{im}\, B\cap \Z^N$
(a sublattice of $\Z^N$),
and the remaining
$N-2K$ rows provide a  $\Z$-basis for $\mathrm{ker}\, B_\Z$. For our purposes, it is only
the choice of integer vectors spanning $\mathrm{im}\, B$ that matters for the definition of
the map $\pi$, and if the integers $h_j$ are not all 1 then it is not essential
to have a $\Z$-basis for $\mathrm{im}\, B_\Z$.

The second part of the proof of Theorem \ref{torusred} involves showing
that, when $B$ is degenerate, it is possible to choose
integer vectors ${\bf v}_j$ spanning $\mathrm{im}\, B$
in a way 
that is compatible with the map $\varphi$.
In other words,   $\varphi$ should reduce to
a symplectic map  $\hat\varphi$ in the coordinates $y_j$,
and  we require that the latter map should also be birational.
The next result provides two sets of sufficient conditions
on the vectors ${\bf v}_j$ which ensure birationality of the
map  $\hat\varphi$.
\begin{lem} \label{ablem}
Let $\{{\bf v}_j\}_{j=1}^{2K}$  
be integer vectors that span  $\mathrm{im}\, B$, and suppose that
the columns of $B$ belong to
$<{\bf v}_1,\ldots, {\bf v}_{2K}>_\Z$. If either

{\em (a)} $\mathrm{im}\, B_\Z=<{\bf v}_1,\ldots, {\bf v}_{2K}>_\Z$, or

{\em (b)} each ${\bf v}_j$ belongs to the $\Z$-span of the columns of $B$,

\noindent
then for $y_j({\bf x}) = {\bf x}^{{\bf v}_j}$, there exist rational functions
$f_j({\bf y}), \, f^\dagger_j({\bf y}) \in \Q ({\bf y})=\Q (y_1,\ldots ,y_{2K})$
such that $\varphi^* y_j({\bf x}) =\pi^* f_j({\bf y})$,
$(\varphi^{-1})^* y_j({\bf x}) =\pi^* f^\dagger_j({\bf y})$
for $j=1,\ldots , 2K$, and the map $\hat\phi: \, {\bf y} \mapsto (f_j({\bf y}))$
is birational  and symplectic.
\end{lem}

\begin{prf}
Let $y({\bf x})={\bf x}^{\bf v}$ be one of the coordinate functions $y_j$ in the
image of the map $\pi$, with $\eta=\log y$, and denote the columns of $B$ by ${\bf w}_1,\ldots , {\bf w}_N$.
Then
$$
\varphi^* \eta ({\bf x})=\sum_{k=1}^N v_k \varphi^* z_k
=\sum_{k=1}^{N-1}v_k z_{k+1} + v_N\Big(  -z_1+\sum_{j=2}^N [-b_{1j}]_+z_j +
\log (1+ \exp F)\Big),
$$
with $F$ given by $F:=\sum_{k=1}^{N-1} b_{1,k+1}z_{k+1}$.
By the initial assumption on the ${\bf v}_j$,
we can write ${\bf w}_1=\sum_{j=1}^{2K} c_j {\bf v}_j$ for $c_j\in\Z$,
which implies that $\exp F=\pi^*\exp \sum_j c_j \eta_j =\pi^*\prod_j y_j^{c_j}$,
the pullback of a rational function of ${\bf y}$. Then we have
\beq\label{etaf}
\varphi^* \eta ({\bf x})= ({\bf m}, {\bf z}) +  \log (1+ \exp F)^{v_N},
\eeq
where ${\bf m}$ is an integer vector with components $m_1=-v_N$,
$m_j=v_{j-1}+[-b_{1j}]_+v_N$ for $2\leq j\leq N$.

By definition the ${\bf w}_j$ span $\mathrm{im}\, B$, so we can
write ${\bf v}=\sum_{k=1}^Nd_k{\bf w}_k$ for some $d_k\in\Q$, which
in components gives $v_j=\sum_{k=1}^Nd_kb_{jk}$ for $j=1,\ldots,N$.
By using (\ref{reln1}) and (\ref{reln2}) one obtains 
$$
m_j = \sum_{k=1}^Nd_k(b_{j-1,k}+[-b_{1j}]_+b_{Nk})
=\sum_{k=1}^{N-1}d_k(b_{j,k+1}+[-b_{1,k+1}]_+b_{j1})-d_Nb_{j1}
$$
for $j\geq 2$, while $m_1=-\sum_{k=1}^Nd_k b_{Nk}= \sum_{k=1}^{N-1}d_k b_{1,k+1}$,
implying that
$$
{\bf m} =\Big(\sum_{k=1}^{N-1}d_k[-b_{1,k+1}]_+ - d_N\Big){\bf w}_1 +\sum_{j=2}^Nd_{j-1}{\bf w}_j.
$$
Hence we see that
${\bf m}\in\mathrm{im}\, B_\Z$. In case (a), ${\bf m}=\sum_{j=1}^{2K} \tilde{c}_j {\bf v}_j$
for $\tilde{c}_j\in\Z$, so substituting in (\ref{etaf}) and exponentiating yields
$\varphi^* y({\bf x}) = \pi^* f$ with
$$
f({\bf y}) = \prod_j y_j^{\tilde{c}_j} \, \Big(1+\prod_j y_j^{c_j}\Big)^{v_N}\in \Q ({\bf y}),
$$
as required.
In case (b), on the other hand, each $d_k\in\Z$ and the ${\bf w}_j$ are in
$<{\bf v}_1,\ldots, {\bf v}_{2K}>_\Z$, so again one can expand ${\bf m}$
in this basis with coefficients  $\tilde{c}_j\in\Z$, and the same formula holds for $f$.
For the inverse map $\varphi^{-1}:\, (x_1,\ldots , x_N)\mapsto (x_0,x_1,\ldots, x_{N-1})$,
with
$$
x_0=\frac{1}{x_N}\left(\prod_{j=1}^{N-1}x_j^{[b_{1,j+1}]_+}+\prod_{j=1}^{N-1}x_j^{[-b_{1,j+1}]_+}\right),
$$
the fact that $(\varphi^{-1})^* y({\bf x}) = \pi^* f^\dagger$ with $f^\dagger ({\bf y})\in \Q ({\bf y})$
follows from an almost identical calculation.

The rational map $\hat\varphi$ is defined by the
$2K$ functions $f_j({\bf y})$, with a rational inverse $\hat\varphi^{-1}$ given by the functions
$f^\dagger_j({\bf y})$, and by construction $\hat\varphi \cdot \pi = \pi \cdot \varphi$.
Combining part (iii) of Lemma \ref{symplectic} together with the result of Lemma \ref{symp}
gives $\varphi^*\om -\om = \varphi^*\pi^*\hat\om-\pi^*\hat\om=\pi^*(\hat\varphi^*\hat\om-\hat\om)=0$,
so  $\hat\varphi^*\hat\om=\hat\om$ as required. Thus the proof of the lemma and the proof
of Theorem \ref{torusred} are complete.
\end{prf}
\begin{rem}{\em
Theorem \ref{normalform} provides a basis of $\mathrm{im}\, B$
which satisfies condition (a) above.
To satisfy condition (b), one can choose any $2K$ independent columns (or rows) of $B$,
corresponding to the $\tau$-coordinates of \cite{gsv, gsvduke}.
}
\end{rem}
To illustrate Theorem \ref{torusred}
we now present a couple of examples.
\bex [Somos-4]{\em
\label{s4e}
The Somos-4 recurrence (\ref{somos4})
is the special case $c=2$ of (\ref{7betarec}), in which case the skew-symmetric matrix $B$ of
(\ref{s4gen}) is degenerate, and $\mathrm{ker}\, B$ is spanned by the integer column vectors
$
\mathbf{u}_1= (1,1,1,1)^T ,
\;\;
\mathbf{u}_2= (1,2,3,4)^T .
$
By (\ref{scale}), each of these vectors generates a scaling action
on the phase space
$\C^4$, with weights given by their components, so that the torus
$(\C^*)^2$ acts via
$(x_1,x_2,x_3,x_4)\rightarrow (\la_1\la_2\, x_1 , \la_1\la_2^2\, x_2,
\la_1\la_2^3\, x_3 , \la_1\la_2^4\, x_4)$,
for  $(\la_1,\la_2)\in(\C^*)^2$.
Then $\mathrm{im}\, B=(\mathrm{ker}\, B)^\perp=<{\bf v}_1, {\bf v}_2>$,
with
$
{\bf v}_1=(1,-2,1,0)^T ,
\;\;
{\bf v}_2=(0,1,-2,1)^T,
$
whose components provide the exponents for the monomial invariants
\beq\label{s4yj}
y_1 = \frac{x_1x_3}{x_2^2}, \qquad y_2 = \frac{x_2x_4}{x_3^2}.
\eeq
(These monomials also provide two independent Casimirs for the degenerate
Poisson bracket (\ref{s4br}) mentioned above. The coefficients
of (\ref{s4br}) are obtained from the bivector ${\bf u}_1\wedge {\bf u}_2$.)
This defines the rational map $\pi : \C^4\rightarrow \C^2$ from the $x_j$ to the $y_j$.
Upon computing the pullback of $\varphi$ on these monomials, one has 
the map
\beq \label{s4map}
\hat{\varphi}: \qquad
\left(\begin{array}{c}
y_1 \\
y_2
\end{array} \right)
\longmapsto \left(\begin{array}{c}
y_2 \\
(y_2+1)/(y_1y_2^2)
\end{array} \right),
\eeq
which is of QRT type \cite{qrt1}, and preserves the symplectic form
\beq\label{canon}
\hat\om = 
\frac{1}{y_1y_2}\, \dd y_2\wedge \dd y_1
\eeq
where $\om = \pi^*\hat\om$, with $\om$  given by
the formula (\ref{7sympm1is1}) for $c=2$.
}
\eex
Note that, in the preceding example, the chosen basis is such that
every integer vector in im$\, B$ can be written as a $\Z$-linear
combination of the vectors ${\bf v}_1$, ${\bf v}_2$, so the condition
(a) in Lemma \ref{ablem} holds. At the same time, from the matrix
(\ref{s4gen}) with $c=2$ we see that the first
column of $B$ is ${\bf v}_2$, and the last column is $-{\bf v}_1$,
so condition (b) holds as well. The next example shows that
(a) and (b) can give inequivalent results.
\bex \label{viallet}
{\em
In \cite{honelaur}, a singularity confinement
pattern was used to obtain  the sixth-order recurrence
\beq
\label{sixthorder}
x_{n+6}\, x_n = (x_{n+5}\, x_{n+1})^2+ x_{n+4}^2x_{n+3}^4x_{n+2}^2,
\eeq
which is associated with mutations of a skew-symmetric matrix
of rank 2, namely
$$
B=\left(\bear{cccccc}
0 & -2 & 2 & 4 & 2 & -2 \\
2 & 0 & -6 & -6 & 0 & 2 \\
-2 & 6 & 0 & -6 & -6 & 4 \\
-4 & 6 & 6 & 0 & -6 & 2 \\
-2 & 0 & 6 & 6 & 0 & -2 \\
2 & -2 & -4 & -2 & 2 & 0
\eear\right).
$$
A basis of $\mathrm{im}\, B$
corresponding to case (a) of Lemma \ref{ablem} is given
in terms of the first and last columns of $B$ by
${\bf v}_1=-\frac{1}{2}{\bf w}_6$,  ${\bf v}_2=\frac{1}{2}{\bf w}_1$,
which gives the ${\bf y}$-coordinates
$y_j = x_jx_{j+4}/(x_{j+1}x_{j+2}^2x_{j+3})$, $j=1,2$,
and (up to rescaling) produces the same symplectic form as in
(\ref{canon}). The corresponding symplectic map is
\beq\label{vialletmap}
\hat{\varphi}: \qquad
\left(\begin{array}{c}
y_1 \\
y_2
\end{array} \right)
\longmapsto \left(\begin{array}{c}
y_2 \\
(y_2^2+1)/(y_1y_2)
\end{array} \right),
\eeq
whose singularity pattern under successive blowups was considered by Viallet
\cite{viallet}. The map (\ref{vialletmap})
has positive algebraic entropy, indicating
nonintegrability. (See Example \ref{trsix} in the next section.)

However, one can take a different basis, corresponding
to case (b) of Lemma \ref{ablem}, given by
${\bf v}_1'=-{\bf w}_6$,  ${\bf v}_2'={\bf w}_1$,
which is not a $\Z$-basis for $\mathrm{im}\, B_\Z$.
From this basis, one has the coordinates $(y_1',y_2')=(y_1^2,y_2^2)$,
and the map becomes
$$
\varphi' : \qquad
\left(\begin{array}{c}
y_1' \\
y_2'
\end{array} \right)
\longmapsto \left(\begin{array}{c}
y_2' \\
(y_2'+1)^2/(y_1'y_2')
\end{array} \right).
$$
The map from $(y_1,y_2)$ to   $(y_1',y_2')$ is  ramified,
as generically there are four pairs of values $(\pm y_1,\pm y_2)$
for each pair $(y_1',y_2')$, and
so the two maps $\hat\varphi$ and $\varphi '$ are not conjugate
to each other.
}
\eex

\section{Algebraic entropy and tropical recurrences} \label{intmaps}

\setcounter{equation}{0}

The deep connection between the integrability of maps and
various weak growth properties of the iterates has been appreciated for some time
(see \cite{veselov} and references).
In the case of rational maps, Bellon and Viallet \cite{bellon_viallet}
considered the growth of degrees of iterates, and used this to define a notion of
algebraic entropy. Each component of a rational map $\varphi$ in affine space
is a rational function of the coordinates, and the degree of the map, $d=\mathrm{deg}\,\varphi$,  is the maximum
of the degrees of the components.  By iterating the map $n$ times one gets a sequence of rational functions
whose degrees grow generically like $d^n$. At the $n$th step one can set $d_n=\mathrm{deg}\,\varphi^n$,
and then the algebraic entropy $\mathcal{E}$ of the map is defined to be
$
\mathcal{E}=\mathrm{lim}_{n\to\infty}\frac{1}{n}\log d_n.
$
Generically, for a map of degree $d$, the entropy is $\log d>0$, but for special maps there can be
cancellations in the rational functions that appear under iteration, which means that the entropy is smaller than expected.

It is conjectured that Liouville-Arnold integrability corresponds to zero algebraic entropy.
In an algebro-geometric setting, there are plausible
arguments which indicate that zero entropy should be a necessary condition for
integrability in the Liouville-Arnold sense \cite{bellon}. In the latter
setting, each iteration of the map
corresponds to a translation on an Abelian variety (the level set of the first integrals),
and the degree is a logarithmic height function, which necessarily grows like
$d_n\sim \mathrm{C}n^2$.

Often the algebraic entropy of a map can only be guessed
experimentally, by calculating the degree sequence $(d_n)$
up to some large $n$ and doing a numerical fit to a
linear recurrence. This is increasingly impractical as the dimension increases, and provides no proof that
the linear relation, with its corresponding entropy value, is correct. In dimension two, exact
results are possible via intersection theory \cite{takenawa, viallet}.

In the rest of this section we seek to isolate those recurrences with $\mathcal{E}=0$,
by finding a condition on the exponents which should be necessary and sufficient for
$\mathcal{E}>0$.
The main conjecture is the following.

\begin{conje} \label{entropyconj}
For a birational map given by  (\ref{bir}), corresponding to a recurrence
of the form   (\ref{crec}), the algebraic entropy is
$\mathcal{E}=\log |\la_{max}|$,
where, of all the roots of the two
polynomials
\beq\label{ppm}
P_{\pm}(\la )=\la^N + 1 - \sum_{j=1}^{N-1}[ \pm b_{1,j+1}]_+\la^j ,
\eeq
$\la_{max}$ is the one of largest magnitude.
The entropy is positive if and only if
\beq \label{maxe}
\max \left(\,  \sum_{j=1}^{N-1} [b_{1,j+1}]_+ \, , \, \sum_{j=1}^{N-1} [-b_{1,j+1}]_+ \, \right) \geq 3,
\eeq
\end{conje}

We now give very strong evidence for the above assertion, showing how it rests on a sequence
of other conjectures.
For the recurrences (\ref{crec}), the key to calculating their entropy is the Laurent phenomenon,
which leads to an exact recursion for the degrees of the denominators. The Laurent
property for the associated cluster algebra \cite{fz} implies that the  iterates
have the factorized form
$$x_n = \frac{\mathrm{N}_n({\bf x})} {\mathrm{M}_n({\bf x})},
\qquad \mathrm{with} \quad \mathrm{N}_n \in\Z [ {\bf x} ] = \Z [x_1,\ldots ,x_N], \quad
\mathrm{M}_n =\prod_{j=1}^N x_j^{d^{(j)}_n},
$$
where the  polynomials
$\mathrm{N}_n$ are not divisible by $x_j$ for $1\leq j\leq N$,
and
$\mathrm{M}_n$ are Laurent monomials.
A lower bound for the entropy
is provided by the growth
of degrees of denominators, and if the exponents
$d_n^{(j)}$ are all positive (for large enough $n$)
then  the monomial $\mathrm{M}_n$ is the denominator of
the Laurent polynomial $x_n$, and $\mathrm{N}_n$
is the numerator.

In addition to being Laurent polynomials, the form of the exchange relations (\ref{exchange})
in a cluster algebra means that all the cluster variables are given by subtraction-free rational
expressions in terms of any initial cluster ${\bf x}=(x_1,\ldots ,x_N)$. This implies that
the dynamics of the $\mathrm{M}_n$ can be decoupled from the $\mathrm{N}_n$.
\begin{propn} \label{troprec}
For all $n$, the exponent $d^{(j)}_n$ of each variable in the Laurent monomial $\mathrm{M}_n$
satisfies the recurrence
\beq \label{tropical}
d_{n+N}+d_n = \max \left( \, \sum_{j=1}^{N-1} [b_{1,j+1}]_+\, d_{n+j}\, , \, \sum_{j=1}^{N-1}  [-b_{1,j+1}]_+ \,d_{n+j}\, \right) ,
\eeq
with the initial conditions $d_1=-1$, $d_2=\ldots =d_N=0$ (up to shifting the index).
\end{propn}
\begin{prf}
Upon substituting $x_n = \mathrm{N}_n/\mathrm{M}_n$ into (\ref{crec}) and comparing  monomial factors on both sides one has
\beq \label{denoms}
\mathrm{M}_{n+N}\mathrm{M}_n = \mathrm{lcm}\left( \, \prod_{j=1}^{N-1}
\mathrm{M}_{n+j}^{[b_{1,j+1}]_+}\, , \,
\prod_{j=1}^{N-1} \mathrm{M}_{n+j}^{[-b_{1,j+1}]_+}\, \right) ,
\eeq
where $\mathrm{lcm}$ denotes the lowest common multiple.
To be more precise, for any sequence of Laurent polynomials $(x_n)$ with positive or negative coefficients,
the formula (\ref{denoms}) certainly holds provided that the two products on the right are not of equal degree
in any of the variables $x_j$, for $1\leq j\leq N$. If it happens that
$D:=\sum_{\ell =1}^{N-1} [b_{1,\ell+1}]_+\, d_{n+\ell}^{(j)}
=\sum_{\ell=1}^{N-1} [-b_{1,\ell+1}]_+\, d_{n+\ell}^{(j)}$ for some $j$,
then the coefficient of $x_j^{-D}$ in each of the two terms
on the right hand side of (\ref{crec}) is a non-zero subtraction-free rational
expression in the other variables $x_k$ with $1\leq k\leq N$, $k\neq j$,
and the sum of these two coefficients cannot vanish.
Hence no cancellations can occur between the two products of
numerators $\mathrm{N}_n$ on the right, and the formula (\ref{denoms}) always holds.
Taking the degree of each variable on the left and right hand sides of (\ref{denoms}) gives
the same recurrence (\ref{tropical}) in each case. From the initial data $x_1,\ldots,x_N$
for (\ref{crec}) it is clear that $d_{1}^{(1)}=-1$ and $d_{n}^{(1)}=0$ for $n=2, \ldots , N$,
and the degrees $d_{n}^{(j)}$ for $2\leq j \leq N$ have the same initial data but shifted along
by an appropriate number of steps.
\end{prf}
\begin{rem} \label{zelevinsky}{\em The recurrence (\ref{tropical})  is the tropical (or
ultradiscrete \cite{nobe}) analogue of the original
nonlinear recurrence (\ref{crec}), in terms of the max-plus algebra.
It is a special case of the recursion for the denominator vectors in a general cluster
algebra, which is stated without justification as equation (7.7) in \cite{fziv}.
}\end{rem}
\begin{conje}\label{numers}
Suppose that the sequence $d_n$ is not periodic and satisfies (\ref{tropical}) with the initial conditions
as in Proposition \ref{troprec}. Then

{\em (a)} $d_n>0$ for all $n>N$, and

{\em (b)} the total degree of the numerators satisfies
$\mathrm{deg}\,\mathrm{N}_n\sim \tilde{\mathrm{C}}\, d_n$ as $n\to\infty$,
for some constant $\tilde{\mathrm{C}}>0$.
\end{conje}
\begin{rem} {\em
Part (a) above follows from the first part of
Conjecture 7.4 in \cite{fziv}. Part (b) implies that
the growth of the denominators completely determines the algebraic
entropy, since the numerators grow at the same rate; it
should be a consequence of Proposition 6.1 in \cite{fziv} (graded homogeneity of cluster
variables).
}
\end{rem}
\bex[Tropical Somos-4] \label{ts4}
{\em
The tropical version of the Somos-4 recurrence  is
\beq\label{tsomos4}
d_{n+4}+d_{n}=\max (d_{n+3}+d_{n+1},2d_{n+2}).
\eeq
With initial conditions $d_1=-1$ and $d_2=d_3=d_4=0$ this  generates a sequence
that begins
$$
-1,0,0,0,1,1,2,3,3,5,6,7,9,10,12,14,15,18,20,22,25,27,30,33,\ldots ,
$$
which are the degrees
(in each of the variables $x_1,x_2,x_3,x_4$)  of the denominators of the Laurent polynomials
generated by (\ref{somos4}). The preceding sequence has quadratic growth,
$d_n\sim \mathrm{C}n^2$ as $n\to\infty$ (consistent with the growth
of logarithmic height on an elliptic curve),
so that the algebraic entropy is zero,
and this can be proved by considering the combination
\beq \label{ts4y}
Y_n=d_{n+2}-2d_{n+1}+d_{n},
\eeq
whose coefficients are the exponents in (\ref{s4yj}).
The sequence of quantities $Y_n$ is generated
by the tropical analogue of the map (\ref{s4map}),
that is
\beq \label{ts4map}
\left(\begin{array}{c}
Y_1 \\
Y_2
\end{array} \right)
\longmapsto \left(\begin{array}{c}
Y_2 \\
\left[ Y_2 \right]_+
-2Y_2-Y_1
\end{array} \right),
\eeq
and all of the orbits of the latter  are
periodic with period 8 (which is a special case of Nobe's
results on tropical QRT maps \cite{nobe}).
Thus, if $\mathcal{S}$ denotes the shift operator such that
$\mathcal{S}f_n = f_{n+1}$ for any function of the index $n$, then applying the operator $\mathcal{S}^8-1$
to both sides of (\ref{ts4y}) implies that the sequence of degrees $d_n$
satisfies a linear relation of order 10, namely
$
(\mathcal{S}^8-1)(d_{n+2}+d_{n}-2d_{n+1})=0.
$
All of the roots $\la$ of the characteristic polynomial %
corresponding to this linear relation have modulus $|\la |=1$,
and $\la =1$ is a triple root, which accounts for
the growth rate of $d_n$; the fact that $d_n=O(n^2)$ also follows directly
from (\ref{ts4y}), using $Y_n=O(1)$.
}
\eex
\bex\label{trsix}
{\em
The tropical version of the recurrence (\ref{sixthorder})
in Example \ref{viallet}  is
\beq
\label{tsixth}
d_{n+6}+ d_n = \max (\, 2d_{n+5}+ 2d_{n+1}\, , \,  2d_{n+4}+4d_{n+3} +  2d_{n+2}\, ).
\eeq
With the initial conditions $d_1=-1$ and $d_2=\ldots =d_5=0$, 
this generates a
degree sequence beginning
$$
-1,0,0,0,0,0,1,2,4,8,18,38,79, 164,342,712,1482,3084,6417,13356, \ldots ,
$$
that grows exponentially, such that
the entropy is
$\mathcal{E}=\log\lambda_{max}$, where
$\la_{max}\approx 2.08$ is the root of largest magnitude of the polynomial
$\la^4-\la^3-2\la^2-\la+1$. To see this, note that 
the map (\ref{vialletmap}), in recurrence form, is
$y_{n+2}\, y_n = y_{n+1} + y_{n+1}^{-1}$, so that its tropical analogue is
\beq
\label{tnonint}
Y_{n+2} +  Y_n = |Y_{n+1}|
.
\eeq
From the tropical version of Lemma \ref{ablem},
an appropriate choice of basis for $\mathrm{im}\, B$
gives the reduction from (\ref{tsixth}) to (\ref{tnonint}),
by setting
$Y_n= d_{n+4} -d_{n+3}-2d_{n+2}-d_{n+1}+d_n$.
It can be shown directly that all the orbits of (\ref{tnonint})
are periodic with period 9 \cite{period9},
and hence in this case the degrees $d_n$ satisfy a linear recurrence of order
13, that is
$
(\mathcal{S}^9-1)Y_n =
(\mathcal{S}^9-1)(\mathcal{S}^4-\mathcal{S}^3-2\mathcal{S}^2-\mathcal{S}+1)d_{n}=0.
$
From the periodicity of $Y_n$ it is clear that
$d_{n+4}-d_{n+3}-2d_{n+2}-d_{n+1}+d_n=O(1)$,
which implies that $d_n\sim \mathrm{C} \la_{max}^n$
for some $\mathrm{C}>0$.
}
\eex
%
%
%
Observe that in the two examples above, the
tropical recurrences (\ref{tsomos4}) and (\ref{tsixth}) both exhibit
periodic behaviour,
in the sense that the maximum on the right hand side is achieved
periodically by the first or the second entry.
If one writes ``$+$'' in the case where
$\sum_{j=2}^N [b_{1,j}]_+ > \sum_{j=2}^N [-b_{1,j}]_+$, and ``$-$'' otherwise,
then
for the tropical Somos-4 recurrence (\ref{tsomos4}) one finds that the
integer sequence in Example \ref{ts4} repeats the block pattern
``$+-+--+--$'' of length 8, while for (\ref{tsixth}) the repeating
symbolic sequence is the block
``$++--+++--$'' of length 9. The block length is in accord
with the period of the associated periodic maps
for the variables $Y_n$ in each case; for other choices of initial data
there are still repeating blocks of
the appropriate length, but the patterns can be different.
Based on a large amount of other numerical evidence, we are led to formulate
the following.
\begin{conje} \label{periodicity}
For each recurrence (\ref{tropical}) 
there exists some $k\geq 1$ such that for all
real solution sequences $(d_n)$,
the associated symbolic sequence corresponding to 
{\em max} is
eventually periodic with minimal period $k$.
\end{conje}
\begin{cor}\label{lintrop}
For large enough $n$,
every real solution of (\ref{tropical}) satisfies a linear
recurrence relation of order $kN$ with constant coefficients.
\end{cor}
The above corollary results from the fact that, given
Conjecture \ref{periodicity}, for large enough $n$
one can regard  (\ref{tropical})
as being
equivalent to the iteration of a linear recurrence relation whose
coefficients vary with period $k$,
and then the following result can be applied.
\begin{lem}\label{perlin}
Suppose that a sequence $(s_n)$ satisfies a linear recurrence of order
$\ell$ whose
coefficients are periodic of period $k$, say
$$
s_{n+\ell} = \sum_{r=0}^{\ell-1} c_{n}^{(r)}\, s_{n+r}, \qquad c_{n+k}^{(r)}=c_{n}^{(r)}.
$$
Then the terms of the sequence also satisfy a
linear recurrence of order $k\ell$ with constant coefficients, of the form
\beq\label{colin}
s_{n+k\ell} = \sum_{r=0}^{\ell-1} \tilde{c}_r\, s_{n+rk}.
\eeq
\end{lem}
\begin{prf}
This result should be well known
in the literature on linear recurrences, but for completeness
we sketch the proof.
It suffices to consider the $(\ell+1)\times (\ell+1)$ matrix
\beq\label{phi}
\tilde{\Phi}_n = \left(\bear{cccc} s_n & s_{n+1} & \ldots & s_{n+\ell } \\
                  s_{n+k} & s_{n+k+1} & \ldots & s_{n+k+\ell} \\
                  \vdots &   \vdots & \ddots  & \vdots \\
                  s_{n+k\ell} & s_{k\ell+1} & \ldots & s_{n+(k+1)\ell} \eear\right)
\eeq
which has vanishing determinant, by virtue of the fact that the vector
$(c_n^{(0)},\ldots, c_n^{(\ell-1)},-1)^T$
is in the right kernel. The recurrence with constant coefficients is obtained from 
the left kernel (the kernel of $\tilde{\Phi}_n^T$).
\end{prf}
In general, it is not easy to determine the period of the symbolic sequence, and
Conjecture \ref{periodicity} seems challenging.
The real orbits of the recurrences (\ref{tropical}) can be quite complicated.

A further refinement of Conjecture \ref{periodicity} is possible.
Let the matrices
$$
{\bf M}_\pm = \left(\bear{cc} {\mathbf 0}^T & -1 \\
                                           {\mathbf 1}_{N-1} & {\bf a}_\pm   \eear\right)
$$
be defined in terms of the palindromic vectors ${\bf a}_\pm$ of size $N-1$, which
correspond to the exponents $[\pm a_j]_+=[\pm b_{1,j+1}]_+$
(where $ {\mathbf 0}^T$ is a zero row vector of size $N-1$, and ${\mathbf 1}_{N-1}$
is the corresponding identity matrix), so that their characteristic polynomials
are $P_\pm$ as in (\ref{ppm}), respectively. Also, let
$$
{\mathbf \Pi} = \prod_{j=1,\ldots ,k}^{\rightarrow}{\bf M}_{\epsilon_j}
= {\bf M}_{\epsilon_1} \ldots {\bf M}_{\epsilon_k}
$$
be the path-ordered product corresponding to the symbolic
sequence of length $k$
for a particular orbit of (\ref{tropical}),
defined
by an appropriate choice of $\epsilon_j=\pm$, and let $\rho ({\bf M}_\pm)$
and $\rho ({\mathbf \Pi})$ denote the corresponding spectral radii.
\begin{conje}\label{spec} Let $k$ be the period
of Conjecture \ref{periodicity}. There are two possibilities:

{\em (1)} If $|\la_{max}|=\rho ({\bf M}_\pm)>\rho ({\bf M}_\mp)\geq 1$ then $k=1$,
with repeated symbol $\epsilon_1=\pm$, respectively.

{\em (2)} If $|\la_{max}|=\rho ({\bf M}_+)=\rho ({\bf M}_-)$ then
$\rho ({\mathbf \Pi})=|\la_{max}|^k$ with $k\geq 1$.
\end{conje}
To see how Conjecture  \ref{entropyconj}
now follows from all the rest, consider the matrix $\tilde{\Phi}_n$
in the proof of Lemma \ref{perlin}, for the case $s_n=d_n$ and $\ell=N$.
If $\Phi_n$ is the $N\times N$ submatrix such that
det$\, \Phi_n$ is the upper left connected minor of size $N$ in (\ref{phi}),
then a single iteration of (\ref{tropical}) gives
$\Phi_{n+1}=\Phi_n {\bf M}_{\epsilon_1}$ for some choice of
$\epsilon_1=\pm$. After $k$ iterations it follows from
Conjecture \ref{periodicity} that
$\Phi_{n+k}=\Phi_n {\bf \Pi}$, and if $n$ is large enough
then  $\Phi_{n+rk}=\Phi_n {\bf \Pi}^r$ for all $r$,
for some block of symbols $\epsilon_j$ of length $k$
that is fixed up to a cyclic permutation (which depends
on the choice of $n \bmod k$). By the Cayley-Hamilton
theorem, if
$\tilde{P}(\kappa ) = \kappa^N-\sum_{r=0}^{N-1}\tilde{c}_r\kappa^r$
is the characteristic polynomial of ${\bf \Pi}$
(which is independent of cyclic permutations of the
${\bf M}_{\epsilon_j}$),
then
$\tilde{P}(\mathcal{S}^k)\Phi_n = \Phi_n \tilde{P}({\bf \Pi})=0$,
which shows that $\Phi_n$, and hence $d_n$, satisfies the
recurrence (\ref{colin}) with $\ell =N$, for all $n$ large enough.
Thus the characteristic roots $\la$ for the growth of
$d_n$ satisfy $\tilde{P}(\la^k)=0$,
which implies that $d_n \sim \mathrm{C}|\la_{max}|^n$
for $|\la_{max}|=\rho({\bf \Pi} )^{1/k}>1$, or $d_n$ has polynomial
growth for $|\la_{max}|=1$. If $k=1$ holds,
then in either case 
(1) or case (2) of Conjecture \ref{spec},
${\bf \Pi}={\bf M}_+$ or ${\bf M}_-$, so $|\la_{max}|$
is given by one of the spectral radii of the matrices ${\bf M}_\pm$,
whichever is the larger,
and $\mathcal{E}=\log|\la_{max}|$. The condition in
case (2) is required to ensure that first part of the statement of
Conjecture \ref{entropyconj} holds
for $k>1$ as well.

For the second statement in Conjecture \ref{entropyconj}, note that
$P_\pm$ in (\ref{ppm}) are reciprocal   polynomials (their
coefficients are palindromic), so that $P_+(\la^{-1}) =\la^{-N}P_{+}(\la )$,
and similarly for $P_-$.
Let 
$$
S_{\pm} =                        
\sum_{j=1}^N [\pm b_{1,j+1}]_+,
$$
respectively. By the symmetry
of the matrices $B\to -B$ (or equivalently, the freedom to replace the
quiver $Q$ by its opposite), it can be assumed
without loss of generality that $S_+$ is the larger of the two,
so $S_{+}\geq S_{-}$,
and take $S_+\geq 3$ so that condition (\ref{maxe}) holds.
Now $P_+(0)=1$,
and $P_+(1)= 2-S_+\leq -1$, so
$P_+$ has a real root between $0$ and $1$. The reciprocal property implies
that the reciprocal of a root is also a root,
hence
$P_+$ has a real root larger than $1$, implying that $|\la_{max}|>1$
and the entropy is positive.

The cases for which (\ref{maxe}) does not hold are easily enumerated,
and it can be checked directly that $|\la_{max}|=1$ in these cases.
The main conclusion of this section is the following.
\begin{thm}\label{zeroe}
Suppose that Conjecture \ref{entropyconj} holds. Then up to the symmetry $S_+\leftrightarrow S_-$, there are only four
distinct choices of the pair $(S_+,S_-)$ for which the algebraic
entropy is zero, corresponding to four different families of
recurrences:

\noindent {\em (i)} $(S_+,S_-)=(1,0)$: For even $N=2m$ only, the recurrence is
\beq\label{per5}
x_{n+2m}\, x_n = x_{n+m} +1.
\eeq

\noindent {\em (ii)} $(S_+,S_-)=(2,0)$: For each $N\geq 2$ and
$1\leq q\leq\lfloor{N/2}\rfloor$,
the recurrence is
\beq \label{primrec}
x_{n+N}\, x_n = x_{n+N-q}\, x_{n+q} + 1.
\eeq

\noindent {\em (iii)} $(S_+,S_-)=(2,1)$: For even $N=2m$ only,
and  $1\leq q\leq m-1$,
the recurrence is
\beq \label{comprec}
x_{n+2m}\, x_n = x_{n+2m-q}\, x_{n+q} + x_{n+m}.
\eeq

\noindent {\em (iv)} $(S_+,S_-)=(2,2)$:
For each $N\geq 2$ and $1\leq p <q\leq
\lfloor{N/2}\rfloor$, the recurrence is
\beq \label{somosN}
x_{n+N}\, x_n = x_{n+N-p}\, x_{n+p} + x_{n+N-q}\, x_{n+q}.
\eeq
\end{thm}
The simplest is case (i), where the recurrence (\ref{per5})
decouples into $m$ independent copies of the Lyness map: all the orbits
are periodic, and the overall period of the sequence of $x_n$ is $5m$.
For each $n$ the function $F_n=x_n+x_{n+m}+x_{n+2m}+x_{n+3m}+x_{n+4m}$ is invariant, and can
be written as a function of $x_n$ and $x_{n+m}$ only, using (\ref{per5}). Moreover, the functions
$F_1,F_2,\ldots,F_m$ are independent and Poisson commute with respect to the bracket
corresponding to the symplectic form $\omega$, so trivially this system is also integrable
in the Liouville-Arnold sense.

The integrability of the families (ii),(iii) and (iv) above is discussed in subsequent sections.

\section{Linearisable recurrences from primitives}\label{primit}
\setcounter{equation}{0}

The primitives introduced in \cite{fordy_marsh} are the simplest examples of cluster mutation-periodic quivers with period 1,
and are the building blocks of all such quivers. The nonlinear recurrences that arise from the primitives have the form (\ref{primrec}),
corresponding to case (ii) in Theorem \ref{zeroe}, and can be rewritten as
\beq
x_{n+N}\, x_n = x_{n+p}\, x_{n+q} + 1, \qquad p+q=N,
\label{jkrec}
\eeq
so that for each $q=1,\ldots,\lfloor{N/2}\rfloor$ there is a different recurrence corresponding to the primitive $P_N^{(q)}$.  
When $p$ and $q$ are coprime, the associated quivers are mutation-equivalent to
the affine Dynkin quivers $\tilde{A}_{q,p}$,
corresponding to the $A_{N-1}^{(1)}$ Dynkin diagram with $q$ edges oriented clockwise and $p$ oriented
anticlockwise, while if
$\gcd (p,q)=k>1$ so that $p=k\hat{p}$, $q=k\hat{q}$, then the quiver is just the disjoint union of $k$ copies
of  $\tilde{A}_{\hat{q},\hat{p}}$. In the latter case, it is clear that the recurrence (\ref{jkrec}) is also
equivalent to $k$ copies of the recurrence of order $N/k$ corresponding to the coprime integers $\hat{p},\hat{q}$.
Hence it is sufficient to consider only the case where $p,q$ are coprime, which we will assume from now on.

The cluster algebras generated by affine $A$-type Dynkin quivers arise
from surfaces,  and are of finite mutation type \cite{finitemutation},
meaning that only a finite number of distinct
quivers is obtained under sequences of mutations (\ref{matmut}).
However, by the classification result in \cite{fz1}, these
cluster algebras are not themselves finite: the recurrence
(\ref{jkrec}) generates infinitely many cluster variables starting
from the initial cluster ${\bf x}=(x_1,\ldots ,x_N)$.

It was conjectured recently that the cluster variables in cluster algebras obtained
from affine Dynkin quivers satisfy linear recurrence relations, and this was proved
for all but the exceptional types \cite{assem}.
A different proof using cluster categories, valid for all affine Dynkin types,
was subsequently found by Keller and Scherotzke   \cite{keller_scher}.
In the case of $\tilde{A}_{q,p}$ (with coprime $p,q$)
a proof of the corresponding linear recurrence relations was already
given in \cite{fordy_marsh}.  Here we present a more direct
derivation of the linear recurrences arising from these affine $\tilde{A}_{q,p}$ quivers,
before using the Poisson structures from section 2 to explain the
Liouville integrability of the maps defined by (\ref{jkrec}).

\subsection{Linear relations with periodic coefficients}

The key to the properties of the nonlinear recurrence (\ref{jkrec}) is
the fact it can be written in the form
\beq \label{frieze}
\det \,\Psi_n = 1,\quad\mbox{where}\quad
           \Psi_n = \left( \bear{cc} x_n & x_{n+q} \\
                                 x_{n+p} & x_{n+N}
\eear\right).
\eeq
The identity (\ref{frieze}) is the frieze relation (see e.g. \cite{assem}); it implies that the iterates
of (\ref{jkrec}) form an infinite frieze.

Upon forming the matrix
\beq\label{3by3}
 \tilde{\Psi}_n = \left( \bear{ccc} x_n & x_{n+q} & x_{n+2q} \\
                                 x_{n+p} & x_{n+N}  & x_{n+N+q} \\
                                 x_{n+2p} & x_{n+N+p}  & x_{n+2N}
\eear\right),
\eeq
one can use the Dodgson condensation method \cite{1866-1} to expand the $3\times 3$ determinant
in terms of its $2\times 2$ minors, as
$$
\det\, \tilde{\Psi}_{n}=\frac{1}{x_{n+N}}\left( \det\Psi_n \, \det \Psi_{n+N}-\det\Psi_{n+q} \, \det \Psi_{n+p}\right)=0,
$$
by (\ref{frieze}). By considering the right and left kernels of $\tilde{\Psi}_{n}$, we are led to the following result.
\begin{lem} \label{JK}
The iterates of the recurrence (\ref{jkrec}) satisfy the linear relations
\beq\label{Jrec}
x_{n+2q}-J_n\, x_{n+q} + x_n = 0,
\eeq
\beq\label{Krec}
x_{n+2p}-K_n\, x_{n+p} + x_n = 0,
\eeq
whose coefficients are periodic functions of period $p,q$ respectively, that is
$$
J_{n+p}=J_n, \qquad K_{n+q}=K_n, \qquad  for\,\,  all \, \, n.
$$
\end{lem}

\begin{prf} A vector in the (right) kernel of $\tilde{\Psi}_{n}$, can be written as
${\bf k}_n = (\tilde{J}_n, -J_n, 1)^T$, where the third entry is normalised to 1 without
loss of generality. From the first two rows of the equation
$\tilde{\Psi}_{n}{\bf k}_n = 0$
we have the linear system
$$
\Psi_n \, \left(\bear{c} -\tilde{J}_n \\ J_n \eear\right)=\left(\bear{c} x_{n+2q} \\ x_{n+N+q} \eear\right).
$$
From Cramer's rule it follows that
$$
\tilde{J}_n=-(\det\Psi_n)^{-1}\,\left| \bear{cc} x_{n+2q} & x_{n+q} \\
                                 x_{n+N+q} & x_{n+N}
\eear\right| =(\det\Psi_n)^{-1}\det\Psi_{n+q}=1,
$$
and
$$
J_n = \left| \bear{cc} x_n & x_{n+2q}  \\
                                 x_{n+p} & x_{n+N+q}
\eear\right|,
$$
where we made use of (\ref{frieze}).
Hence the recurrence (\ref{Jrec}) is given by the first row of $\tilde{\Psi}_{n}{\bf k}_n = 0$. The second and
third rows of the latter equation provide the $2\times 2$ linear system
$$
\Psi_{n+p} \, \left(\bear{c} -\tilde{J}_n \\ J_n \eear\right)=\left(\bear{c} x_{n+N+q} \\ x_{n+2N} \eear\right).
$$
whose solution gives an alternative formula for $J_n$.
Noting that the coefficients in this linear system are the same as those
in the first two rows of the equation $\tilde{\Psi}_{n+p}{\bf k}_{n+p} = 0$, it follows that $J_{n+p}=J_n$ for all $n$.
If we consider the left kernel of $\tilde{\Psi}_{n}$ (the kernel of
 $\tilde{\Psi}_{n}^T$ ),
then by symmetry we obtain the recurrence (\ref{Krec}), whose coefficient $K_n$ has period $q$.
\end{prf}
\br  {\em
In the particular case $q=1$, corresponding to $\tilde{A}_{1,N-1}$ Dynkin quivers, the coefficient $K_n$ has period 1, so $K_{n+1}=K_n=\mathcal{K}$ (constant) for all $n$. In that case the recurrence (\ref{Krec}) becomes
\beq\label{k-pn1}  %
x_{n+2N-2}+x_n=\mathcal{K} \, x_{n+N-1},
\eeq
which is a {\em constant coefficient, linear difference equation} for $x_n$.  This was first shown in \cite{fordy_marsh} and
is precisely of the form derived for affine $A$-type quivers in \cite{assem, keller_scher}.}
\er
The quantity $\mathcal{K}$ in (\ref{k-pn1}) can be
considered as a Laurent polynomial of
the initial data $x_1,x_2, \ldots, x_{N}$ for the map $\varphi$ associated with  (\ref{jkrec});
from (\ref{k-pn1})
this is obtained explicitly by repeatedly using the nonlinear recurrence in order to write $\mathcal{K}$ in terms of lower
order iterates until it is given in terms of the $N$ initial data. Moreover, the fact that
it is independent of $n$ means that $\mathcal{K}$ is a first integral 
of $\varphi$.

In the case of general coprime $p,q$, it is straightforward to obtain first integrals of the map $\varphi$
corresponding to (\ref{jkrec}). Indeed, 
the orbit of $J_1$ under
the action of $\varphi$ generates $p$ functions $J_1, J_2, \ldots, J_p$, which can be written
as rational functions of the $N$ initial data via back-substitution in the relation
\beq\label{jdef}
J_{n}=\frac{x_n+x_{n+2q}}{x_{n+q}}.
\eeq
Clearly any cyclically symmetric function of $J_1,J_2, \ldots, J_p$ is a first integral of $\varphi$ and
the first $p$ elementary symmetric functions of these variables are independent functions of the $J_n$.
Given that $J_1,J_2, \ldots, J_p$
are functionally independent as functions of $x_1,x_2,\ldots, x_N$,
any $p$ independent cyclically symmetric functions of the $J_n$ can be picked as first integrals.
Similarly, the action of $\varphi$ generates the functions $K_1, K_2, \ldots, K_q$, and
any $q$ independent cyclically symmetric functions of them
provide another set of independent
first integrals of $\varphi$. This gives a total of $p+q=N$ first integrals.
These cannot all be functionally independent, because that would imply that the map is periodic. However,
the generic orbit generated by (\ref{jkrec}) is not periodic.
Below we describe a single functional  relation between these first integrals
(equation (\ref{k2}) below), by considering the monodromy properties of the linear relations
(\ref{Jrec}) and (\ref{Krec}).

\br {\em
The existence of $N-1$ independent first integrals ${\cal I}_j$,
together with the volume form $\Omega$ in (\ref{vol}),
means that the map $\varphi$ is Liouville integrable
in a rather elementary sense. By applying a method
in \cite{byrnes}, one can pick, say, the first $N-2$
integrals and obtain a Poisson bivector field
$\hat{\cal J}$ which is invariant (or anti-invariant, for odd $N$), namely
$$
\hat{\cal J} = \hat{\Omega}\lrcorner \dd {\cal I}_1\lrcorner
\ldots  \lrcorner \dd {\cal I}_{N-2},
$$
where the $N$-vector field
$\hat{\Omega}$ is 
defined by $\hat{\Omega}\lrcorner \Omega =1$.
By construction, this Poisson structure has Casimirs
${\cal I}_j$ for $j=1,\ldots , N-2$, and
restricting to the two-dimensional symplectic leaves
one has an integrable system with one degree of freedom,
with ${\cal I}_{N-1}$ being the independent first integral.
However, we shall see that the Poisson structure coming from
the two-form (\ref{omega})  leads to more
interesting integrable systems.
}
\er
\subsection{Monodromy matrices and linear relations with constant coefficients}\label{2x2monodromy}

The relation (\ref{Jrec}) implies that the matrix $\Psi_{n}$ satisfies
\beq\label{psirel1}
\Psi_{n+q}=\Psi_n\, {\bf L}_n, \qquad {\bf L}_n=\left(\bear{cc} 0 &  -1 \\ 1  & J_n  \eear\right).
\eeq
Upon taking the ordered product of the ${\bf L}_n$ over $p$ steps, shifting by $q$ each time, we have
the monodromy matrix
\beq\label{mdy1}
{\bf M}_n:=  {\bf L}_{n}{\bf L}_{n+q}\ldots{\bf L}_{n+(p-1)q}=\Psi_n^{-1}\, \Psi_{n+pq}.
\eeq
On the other hand, the recurrence (\ref{Krec}) yields
\beq\label{psirel2}
\Psi_{n+p}= \hat{\bf L}_n \,\Psi_n, \qquad \hat{\bf L}_n=\left(\bear{cc} 0 &  1 \\ -1  & K_n  \eear\right),
\eeq
which gives another monodromy matrix
\beq\label{mdy2}
\hat{\bf M}_n:=  \hat{\bf L}_{n+(q-1)p}\ldots\hat{\bf L}_{n+p}\hat{\bf L}_n=\Psi_{n+pq}\, \Psi_n^{-1}.
\eeq
The cyclic property of the trace implies that
\beq\label{k2}
\mathcal{K}_n:= \mathrm{tr}\, {\bf M}_n = \mathrm{tr}\, \hat{\bf M}_n.
\eeq
Also, since ${\bf L}_n$ has period $p$, shifting $n\to n+p$ in (\ref{mdy1}) and taking the trace implies that
$\mathcal{K}_{n+p}=\mathcal{K}_n$. Similarly, from (\ref{mdy2}) we have $\mathcal{K}_{n+q}=\mathcal{K}_n$.
Now the periods $p$ and $q$ are coprime, and since $\mathcal{K}_n$ has both these periods it follows that
$\mathcal{K}_n=\mathcal{K}=$constant, for all $n$, hence $\mathcal{K}$ is an invariant of $\varphi$.

From the expression (\ref{mdy1}) it further follows that $\mathcal{K}$ is a cyclically symmetric function
of the $J_n$, $n=1,\ldots, p$, while from (\ref{mdy2}) it is also a cyclically symmetric function
of the $K_n$, $n=1,\ldots, q$. Thus we see that the equality of the traces in (\ref{k2})
provides the aforementioned functional relation between these two sets of functions.

We are now ready to show that in general the iterates of (\ref{jkrec}) satisfy a linear relation with
constant coefficients.
The existence of such a relation follows immediately from
(\ref{Jrec}) or (\ref{Krec}), by applying Lemma \ref{perlin},
but the monodromy matrices provide more detailed information about the coefficients.

\begin{thm}\label{Klin}
The iterates of the nonlinear recurrence (\ref{jkrec}) satisfy the linear relation
\beq\label{krec}
x_{n+2pq}+x_n = \mathcal{K}\,x_{n+pq},
\eeq
where  $\mathcal{K}$ is the first integral defined by (\ref{k2}),
with $\mathcal{K}_n=\mathcal{K}$ for all $n$.
\end{thm}

\begin{prf}
Using (\ref{mdy1}) we see that $\Psi_{n+pq}=\Psi_n {\bf M}_n$, so $\Psi_{n+2pq}=\Psi_{n}{\bf M}_n{\bf M}_{n+pq}=\Psi_n {\bf M}_n^2$
by periodicity.  Noting that ${\bf M}_n$ is a $2\times 2$ matrix, with
$\det {\bf M}_n=1$ and $\mathrm{tr}\, {\bf M}_n={\cal K}$, yields
(by Cayley-Hamilton)
$$
\Psi_{n+2pq}-{\cal K} \Psi_{n+pq}+\Psi_n = \Psi_n({\bf M}_n^2-{\cal K} {\bf M}_n+{\bf I})=0.
$$
The $(1,1)$ component of this equation is just (\ref{krec}).
\end{prf}

By making use of the Chebyshev polynomials of the first and second kind, defined by
$T_n(\upzeta ) = \cos (n\theta)$
and $U_n (\upzeta)=\sin ((n+1)\theta) /\sin\theta$,
respectively, for $\upzeta=\cos\theta$,
the linear equation (\ref{krec}) yields
an exact  expression for the iterates of the nonlinear recurrence.
\begin{cor}[Chebyshev polynomials]
\label{cheby}
The recurrence (\ref{jkrec}) has the explicit solution
\beq\label{chebform}
x_{i + npq} = x_i\, T_n (\mathcal{K}/2) + \left(x_{i+pq}-\frac{x_i\mathcal{K}}{2}\right)\,
U_{n-1}(\mathcal{K}/2),
\qquad i=1,\ldots ,pq-1, \qquad for\,\, all\,\,\, n.
\eeq
\end{cor}

\subsection{The structure of monodromy matrices} 

Here we give some relations between the elements of ${\bf M}_n$, which provide
properties of a natural Poisson tensor associated with the functions $J_i$.
Analogous results regarding $\hat{\bf M}_n$ and $K_i$ also hold,
since the structure of ${\bf M}_n$ and $\hat{\bf M}_n$ is the same
up to switching $p\leftrightarrow q$, $J_i\leftrightarrow K_i$ and taking the transpose.
To simplify the presentation, we concentrate on the case of $P_{N}^{(1)}$.
(Similar results hold for $P_{2m}^{(q)}$, with $p+q=2m$ and $p,q$ coprime.)
The remarkable fact is that (when $N$ is even) these properties of the Poisson bracket  are
derived directly from the monodromy matrix. In subsection \ref{pn1-poisson} we proceed to
show how the Poisson algebra of the
$J_i$ is derived from the Poisson bracket between the coordinates $x_j$.

For the case $q=1$, we denote the matrix ${\bf M}_n={\bf L}_n {\bf L}_{n+1}\cdots {\bf L}_{n+p-1}$
by ${\bf M}_n^{(2m)}$, when $p=2m-1, m\geq 1$, and ${\bf M}_n^{(2m+1)}$, when $p=2m, m\geq 1$.

It is important in the calculations below that ${\bf M}_n^{(p+1)}$ depends only
upon the variables $J_n,\dots ,J_{n+p-1}$.
For the moment, this is not really a monodromy matrix,
if we do not assume any periodicity.
The calculations below give a recursive procedure
for building the matrices ${\bf M}_n^{(2m)}$ and ${\bf M}_n^{(2m+1)}$.

The recursion
${\bf M}_n^{(p+3)}={\bf M}_n^{(p+1)}{\bf L}_{n+p}{\bf L}_{n+p+1}$,
with the short-hand notation $\RA=\RA_n^{(p+1)}$,
$\tilde \RA=\RA_n^{(p+3)}$, etc., leads to
\be\label{mtilde}   %
\begin{array}{ll}
\tilde \RA = J_{n+p} \RB-\RA,& \tilde \RB= (J_{n+p}J_{n+p+1}-1)\RB-J_{n+p+1}\RA,\\[3mm]
\tilde \RC=J_{n+p}\RD-\RC, & \tilde \RD=(J_{n+p}J_{n+p+1}-1)\RD-J_{n+p+1}\RC,
\end{array}
\ee  %
so that $\tilde {\cal K}=-{\cal K}+J_{n+p}\RB-J_{n+p+1}\RC+J_{n+p}J_{n+p+1}\RD$.

\begin{lem}[Relations for $\tilde {\bf M}$]  \label{brelns}%
The components of ${\bf M}_n^{(p+1)}$ satisfy the relations
$$
\frac{\pa \RA_n^{(p+1)}}{\pa J_n}=\frac{\pa \RB_n^{(p+1)}}{\pa J_n}=0,\quad \RA_n^{(p+1)}= -\frac{\pa \RC_n^{(p+1)}}{\pa J_n}, \quad
\RB_n^{(p+1)}=-\frac{\pa {\cal K}^{(p+1)}}{\pa J_n},\quad \RC_n^{(p+1)}=\frac{\pa {\cal K}^{(p+1)}}{\pa J_{n+p-1}},
$$
where ${\cal K}^{(p+1)}=\RA_n^{(p+1)}+\RD_n^{(p+1)}$.
\end{lem}  %

\begin{prf}
We just prove this for the even case.  For $m=1$, ${\bf M}_n^{(2)}={\bf L}_n$, which clearly satisfies these relations.
The recursion (\ref{mtilde}) provides us with an inductive step. We have
$$
\frac{\pa\tilde \RA}{\pa J_n} = J_{n+2m-1} \frac{\pa \RB}{\pa J_n}-\frac{\pa \RA}{\pa J_n}=0, \quad
\frac{\pa\tilde \RB}{\pa J_n} = (J_{n+2m-1}J_{n+2m}-1) \frac{\pa \RB}{\pa J_n}+J_{n+2m}\frac{\pa \RA}{\pa J_n}=0.
$$
Then
$$
\frac{\pa\tilde \RC}{\pa J_n} = J_{n+2m-1}\frac{\pa {\cal K}}{\pa J_n}-\frac{\pa \RC }{\pa J_n}=\RA-J_{n+2m-1} \RB=-\tilde \RA,
$$
where we have used $\frac{\pa \RD}{\pa J_n}=\frac{\pa (\RA+\RD)}{\pa J_n}=\frac{\pa {\cal K}}{\pa J_n}$, and
$\frac{\pa\tilde {\cal K}}{\pa J_{n+2m}}=-\RC+J_{n+2m-1}\RD=\tilde \RC$.

\noindent Finally
\bea  %
\frac{\pa\tilde {\cal K}}{\pa J_n} &=& -\frac{\pa{\cal K}}{\pa J_n}+J_{n+2m-1}\frac{\pa \RB}{\pa J_n} -J_{n+2m}\frac{\pa \RC}{\pa J_n}+J_{n+2m-1}J_{n+2m}\frac{\pa {\cal K}}{\pa J_n} \nn\\
&=& (J_{n+2m-1}J_{n+2m}-1) \frac{\pa {\cal K}}{\pa J_n}+J_{n+2m} \RA= -(J_{n+2m-1}J_{n+2m}-1) \RB+J_{n+2m} \RA= -\tilde \RB,  \nn
\eea  %
where again we have used $\frac{\pa \RD}{\pa J_n}=\frac{\pa {\cal K}}{\pa J_n}$.
\end{prf}

We can use the above relations to form {\em recursion operators}
which can be used to build the functions ${\cal K}^{(p+1)}$.
Starting with ${\cal K}^{(2)}=J_n$ and ${\cal K}^{(3)}=J_nJ_{n+1}-2$, we can use
${\cal K}^{(p+3)}={\cal R}^{(p)}{\cal K}^{(p+1)}$, where the {\em recursion operator} is
\be\label{krecursion} %
{\cal R}^{(p)} = J_{n+p}J_{n+p+1}\frac{\pa^2}{\pa J_n\pa J_{n+p-1}}-J_{n+p}\frac{\pa}{\pa J_n}-J_{n+p+1}\frac{\pa}{\pa J_{n+p-1}} + (J_{n+p}J_{n+p+1}-1).
\ee  %

\br  {\em  %
An alternative formula for ${\cal K}$ is given by a link with the dressing chain:
$$
\mathcal{K} = \prod_{j=1}^p \left( 1-\frac{\partial^2}{\partial J_j \partial J_{j+1}}\right)
\, \prod_{k=1}^p J_k .
$$
When $p$ is odd, this formula follows from the results in \cite{shabves}, by setting $\beta_i \to 0$ and $g_i \to J_i$.
}\er  %

\subsubsection{Link with the Poisson structure}

In the case $N=2m$, we can use the monodromy matrix to build the Poisson bracket for the functions $J_n,\, K_n$.
Staying within the context of one quiver $P_{2m}^{(1)}$ for fixed $m$,
we now reinstate the periodicity ${\bf L}_{n+p}={\bf L}_n$.

From (\ref{mdy1}), by periodicity, for all $k$ and $n$ we have
$$
{\bf M}_{n+k}{\bf L}_{n+k}={\bf L}_{n+k}{\bf M}_{n+k+1}.
$$
When $k=0$, this implies
$$
\RA_{n+1}=\RD_n-J_n\RC_{n+1},\quad \RB_{n+1}+\RC_n=J_n(\RD_{n}-\RD_{n+1}), \quad \RC_{n+1}=-\RB_n,\quad \RD_{n+1}=\RA_n-J_n\RB_n .
$$
Shifting $n$, we can write $\RC_n=-\RB_{n-1}=-\RB_{n+p-1}$.  Only two of the remaining equations are independent, leading to
$$
J_n\RB_n=\RA_n-\RD_{n+1},\quad \RB_{n+p-1}-\RB_{n+1}=J_n(\RD_{n+1}-\RD_n).
$$

The equations for $k\neq 0$ are obtained by shifting the indices.  The first equation leads to
$$
\sum_{k=1}^{p-1}(-1)^{k+1}J_{n+k}\RB_{n+k}=\sum_{k=1}^{p-1}(-1)^{k+1}(\RA_{n+k}-\RD_{n+k+1})=\RA_{n+1}-\RA_{n+p}=\RA_{n+1}-\RA_n,
$$
since the remaining sum consists of
$$
\sum_{j=1}^{m-1}(\RA_{n+2j+1}+\RD_{n+2j+1})-\sum_{j=1}^{m-1}(\RA_{n+2j}+\RD_{n+2j})=(m-1){\cal K}-(m-1){\cal K}=0.
$$
On the other hand $\RB_{n+p-1}-\RB_{n+1}=J_n(\RD_{n+1}-\RD_n)$, so
$$
J_n\sum_{k=1}^{p-1}(-1)^{k+1}J_{n+k}\RB_{n+k}+\RB_{n+p-1}-\RB_{n+1}=J_n(\RA_{n+1}-\RA_{n+p}+\RD_{n+1}-\RD_n)=0.
$$
If we take cyclic permutations, we obtain a matrix equation of the form
$$
({\bf P}^{(2)}+{\bf P}^{(0)}){\bf b}=0,\quad \mbox{with}\quad {\bf b}
=(\RB_n,\RB_{n+1},\dots ,\RB_{n+p-1})^T= -\nabla {\cal K},
$$
where the components of the matrix
${\bf P}^{(2)}+{\bf P}^{(0)}$ are those of the
Poisson tensor 
in Lemma \ref{jbra} below, and we have used the formula
$\RB_n=-\frac{\pa {\cal K}}{\pa J_n}$ for all $n$, coming from Lemma \ref{brelns}.

We can summarise these results in
\begin{thm} \label{isomonodromy}
The function $\cal K$, defined by (\ref{k2}) is the Casimir 
of the Poisson bracket of Lemma \ref{jbra}:
$$
({\bf P}^{(2)}+{\bf P}^{(0)})\nabla {\cal K}=0.
$$
\end{thm}

\subsection{Poisson brackets and Liouville integrability for $P_{2m}^{(1)}$}\label{pn1-poisson}

It was proved in \cite{fordy_rec} that the linearisable maps coming from the primitives
$P_N^{(1)}$ (the $\tilde{A}_{1,N-1}$ Dynkin quivers) are Liouville integrable when $N$ is even.
We give the proof here, since it is the basis for
understanding the integrability of the maps for the other  $P_N^{(q)}$ quivers.

The proof starts from the Poisson bracket for the cluster variables.
The matrix $B$ in this
case is nondegenerate, 
having the form
\be  \label{btau}  %
B = \tau_N-\tau_N^T, \quad\mbox{with}\quad \tau_N= \sum_{r=1}^{N-1} {\bf E}_{r+1,r} -{\bf E}_{1,N},
\ee  %
where ${\bf E}_{r,s}$ denotes an element of the standard basis for $gl(N)$. The ``skew rotation'' matrix $\tau_N$ plays an important role in the classification presented in \cite{fordy_marsh}.
The matrix (\ref{btau}) for $N=4$ is obtained by setting $c=0$ in (\ref{s4gen}).

By Theorem \ref{torusred}, the map is symplectic, and hence there is a nondegenerate Poisson bracket of the form (\ref{logcan}), with
$C=B^{-1}$, up to scaling. In accordance with \cite{fordy_rec}, we take
$$
C =\tau_N^T+(\tau_N^T)^3+\dots +(\tau_N^T)^{N-1}= \sum_{s=1}^{\frac{N}{2}}\sum_{r=1}^{N-2s+1} \left(
{\bf E}_{r,r+2s-1} - {\bf E}_{r+2s-1,r}\right),
$$
so that $CB=2\,I$, which gives
\beq\label{xbra}
\{ \, x_j , x_k \, \}=\left\{ \bear{ll}
                               \mathrm{sgn}(k-j)\, x_j x_k, \qquad & k-j \quad \mathrm{odd}, \\[2mm]
                               0 & \mathrm{otherwise},
                               \eear  \right.
\eeq
for $j,k = 1,\ldots , N$, with $\mathrm{sgn}$ denoting the sign function.

The key to the Liouville integrability of the $P_N^{(1)}$ maps is the expression of the Poisson bracket
between the periodic functions $J_n$, which appeared in \cite{honelaur} (for $N=4$) and \cite{fordy_rec}
(for general even $N=2m$).

\begin{lem} \label{jbra}
For even $N=p+1$ and $q=1$, the functions $J_n$ given by (\ref{jdef})
define a Poisson subalgebra of
codimension one in the algebra (\ref{xbra}), with brackets
\beq\label{jbrac}
\{ \, J_j, J_k\, \} = 2\, \mathrm{sgn}(k-j)\, (-1)^{j+k+1}J_jJ_k
+2(\delta_{j,k+1}-\delta_{j+1,k}+\delta_{j+p-1,k}-\delta_{j,k+p-1}),
\eeq
for $j,k=1,\ldots, p$.
\end{lem}

The bracket for the $J_n$ is clearly a sum $\{\, ,\,\}=\{\, ,\,\}_2 + \{\, ,\,\}_0$, corresponding
to the splitting of the Poisson tensor into homogeneous parts
${\bf P}^{(2)}+{\bf P}^{(0)}$, as in Theorem \ref{isomonodromy}.
The quadratic bracket
$\{\, ,\,\}_2$ is log-canonical, while the degree zero bracket can be defined simply by specifying
its only non-zero terms as
\beq\label{dzero}
\{ \, J_{j+1},J_j \, \}_0 = 2 = - \{ \, J_j,J_{j+1} \, \}_0
\eeq
for all $j \bmod\,  p=2m-1$. Since (\ref{jbrac}) is still a Poisson bracket
after scaling $J_j\to\mu J_j $ for arbitrary $\mu$, the brackets $\{ ,\}_0$ and  $\{ ,\}_2$ are compatible,
so define a bi-Hamiltonian structure. This
means that one can use the standard bi-Hamiltonian chain \cite{magri},
defining a sequence of functions $\mathcal{I}_j$ which satisfy
$$
\{ \mathcal{I}_j, \mathcal{I}_k \}_0 = 0 =
\{ \mathcal{I}_j, \mathcal{I}_k \}_2, \qquad
\mathrm{for} \,\,\, \mathrm{all}\quad j,k,
$$
where the sequence starts from $\mathcal{I}_0 = \sum_j J_j$,
the Casimir of the bracket $\{ ,\}_0$ given by (\ref{dzero}),
and finishes with $\mathcal{I}_{m-1}=\prod_j J_j$,
the Casimir of $\{ ,\}_2$.
By Theorem  \ref{isomonodromy}, the function $\mathcal{K}$ defined
by (\ref{k2}) is the generating function for these integrals, so that
\beq\label{kgen}
\mathcal{K} = \sum_{j=0}^{m-1}(-1)^{m+j+1}\mathcal{I}_j,
\eeq where $\mathcal{I}_j$ is the term of
degree $2j+1$ in the variables $J_i$. Since these integrals commute with respect to both
brackets, they commute with respect to the sum   $\{\, ,\,\}_2 + \{\, ,\,\}_0$, and hence provide
$m$ commuting integrals for the map $\varphi$
of the variables $x_j$ in even
dimension $N$, which implies
that the map is Liouville integrable.

In summary, we have
\begin{thm}
For $N=2m$ and $q=1$ the map $\varphi$ defined by (\ref{jkrec})
has $m$ functionally independent Poisson
commuting integrals, given by the terms
of each odd homogeneous degree in the quantity
${\cal K}$, as given by equation (\ref{kgen}).
The map is also superintegrable, having a total of $N-1$ independent first integrals.
\end{thm}

As discussed in subsection \ref{2x2monodromy}, extra first integrals
are obtained by choosing $m-1$ additional cyclically symmetric
functions of $J_1,\ldots , J_p$.

\subsection{Primitives of the form $P_{2m+1}^{(1)}$}

The recurrences (\ref{jkrec})
for $p=2m$ and $q=1$ are given by
\beq\label{oddprim}
x_{n+2m+1}\, x_n = x_{n+2m}\, x_{n+1} + 1.
\eeq
The formula (\ref{btau}) still holds, but now the matrix $B$ is
singular. It has a one-dimensional kernel,
spanned by the vector ${\bf u}=(1,-1,\dots ,1,-1,1)^T$,
which generates the scaling symmetry
\beq\label{lascala}
x_j \rightarrow \la^{(-1)^{j+1}}\, x_j, \qquad \la \in \C^*,
\eeq
and $\mathrm{im}\, B$ is spanned by
\beq \label{oddprimvj}
{\bf v}_j = {\bf e}_j +{\bf e}_{j+1}, \qquad j=1,\ldots , 2m,
\eeq
where ${\bf e}_j$ is the $j$th standard basis vector.
Hence, by  Lemma \ref{symp}, the coordinates
\beq\label{oddprimy}
y_j = x_j\, x_{j+1}, \qquad j=1,\ldots , 2m,
\eeq
are invariant under the scaling (\ref{lascala}),
and the degenerate form (\ref{omega}) pushes
forward to a symplectic form (\ref{yform}) in dimension
$2m$, whose coefficients $\hat{b}_{jk}$
are the matrix elements of
$$
\hat{B} = \tau_{2m}-\tau_{2m}^2+\tau_{2m}^3-\dots +\tau_{2m}^{2m-1},
$$
where $\tau_{2m}$ is the $2m\times 2m$ version of $\tau_N$.  The inverse of this
is the skew-symmetric matrix
$$
\hat{B}^{-1}=\tau_{2m}^T+(\tau_{2m}^T)^2+(\tau_{2m}^T)^3+\dots
+(\tau_{2m}^T)^{2m-1},
$$
with all components above the diagonal equal to $1$,
giving a nondegenerate Poisson bracket for the $y_j$, i.e. 
\beq \label{ybra}
\{ \, y_j , y_k\,\} = \mathrm{sgn}(k-j)\, y_j y_k , \qquad 1\leq j,k\leq 2m.
\eeq

Upon applying the rest of Theorem \ref{torusred}, we see that
(\ref{oddprim}) induces a symplectic map  on the variables $y_i$, given (for $m\geq 2$)
by
\be  \label{ymap}  %
\hat\varphi: \, 
(y_1,y_2,\ldots , y_{2m-1},y_{2m}) 
\mapsto 
(y_2,y_3,\ldots , y_{2m},y_{2m+1}), 
\ee
where
$$
y_{2m+1} 
=
y_2y_4\cdots y_{2m}
(y_2y_4\cdots y_{2m}+y_3\cdots y_{2m-1})/(y_1y_3^2\cdots y_{2m-1}^2)  .
$$
The map is simpler for the case $m=1$,  given in Example \ref{ymap-p31} below.

By the general discussion above, the iterates $x_n$  satisfy
the linear relation (\ref{k-pn1}), where ${\cal K}$ is the
trace of the monodromy matrix ${\bf M}_n$. The latter is given in terms of the
quantities
\beq \label{oddjdef}
J_n = \frac{x_n + x_{n+2}}{x_{n+1}}, \qquad n = 1, \ldots, 2m,
\eeq
which cycle with period $2m$ under the action of the recurrence (\ref{oddprim}).
The polynomial ${\cal K}$ can be expanded as
\beq\label{koddgen}
\mathcal{K} = \sum_{j=0}^{m}(-1)^{m+j}\mathcal{I}_j, \qquad \mathrm{where}
\qquad \mathcal{I}_0 = 2,
\eeq
and each polynomial $\mathcal{I}_j$ is homogeneous of degree $2j$ in the variables
$J_n$. The non-trivial homogeneous components $\mathcal{I}_1, \ldots , \mathcal{I}_{m}$
provide $m$ first integrals for  (\ref{oddprim}), and an additional
$m$ independent first integrals can be obtained by choosing cyclically symmetric
functions of $J_n$ for each odd degree $1,3,\ldots ,2m-1$.

However, not all of these first integrals reduce to functions on the
$2m$-dimensional symplectic manifold with coordinates $y_j$.
Note that the scaling symmetry (\ref{lascala}) acts on the variables (\ref{oddjdef})
according to
\beq\label{jscale}
J_n\longrightarrow \lambda^{2(-1)^{n+1}}\, J_n.
\eeq
Applying this to the formula (\ref{krecursion}) shows
that each component of ${\cal K}$ is invariant under scaling,
which means that the $m$  first integrals $\mathcal{I}_1, \ldots , \mathcal{I}_{m}$
can be rewritten as functions of the scale-invariant variables $y_j$.
The Liouville integrability of the map (\ref{ymap}) then follows,
provided that these $m$ functions are in involution with respect
to the bracket (\ref{ybra}). We shall not pursue the  case of general $m$ further here,
but content ourselves with presenting some low-dimensional examples.

\bex[The primitive $P_3^{(1)}$] \label{ymap-p31} {\em   %
For $m=1$ the map $\varphi$ for the $x_j$ variables  is (\ref{affA2}),
which is associated with $P_3^{(1)}$, the affine $A_2^{(1)}$ Dynkin quiver.
The matrix $B$ has rank two, so in terms of the variables $y_j$ given by
(\ref{oddprimy}) the symplectic form has the log-canonical form (\ref{canon}),
and the induced map of the plane is
$$
\hat\varphi: \qquad \left(\bear{c} y_1 \\ y_2 \eear\right) \mapsto
\left(\bear{c} y_2 \\ y_2(y_2+1)/y_1 \eear\right).
$$
Symmetric functions of the period 2 quantities
$$
J_1=\frac{x_1+x_3}{x_2},\;\; J_2=\frac{x_1x_2+x_2x_3+1}{x_1x_3}
$$
give two first integrals for the map  (\ref{affA2}), namely
$$
J_1+J_2 \qquad \mathrm{and} \qquad  {\cal I}_1= J_1J_2 ={\cal K}+2=\frac{(y_1+y_2)(y_1+y_2+1)}{y_1y_2}.
$$
The latter is defined on the $(y_1,y_2)$ plane, and
$\hat\varphi^*{\cal I}_1={\cal I}_1$, so the symplectic map $\hat\varphi$
with one degree of freedom
has an invariant function  and hence is integrable.
}\eex  %

\bex[The primitive $P_5^{(1)}$] \label{ymap-p51} {\em   %
For $m=2$ the recurrence (\ref{oddprim})
has the first integrals ${\cal I}_1$ and ${\cal I}_2$
given by the homogeneous terms of degree 2 and 4, respectively, in
the expression (\ref{koddgen}):
\beq\label{kp51}
{\cal K} = 2-(J_1J_2+J_2J_3+J_3J_4+J_4J_1)+ J_1J_2J_3J_4={\cal I}_0 - {\cal I}_1+{\cal I}_2,
\eeq
where the $J_n$ are defined by (\ref{oddjdef}).
Picking another pair of cyclically symmetric functions
of $J_n$, of degrees 1 and 3, say, adds two more independent first integrals.

Now defining $y_j$ for $j=1,2,3,4$ by (\ref{oddprimy}),
these variables are endowed with the nondegenerate
Poisson bracket (\ref{ybra}), which
is invariant under the map (\ref{ymap}):
$$
\hat\varphi: \quad
(y_1,y_2,y_3,y_4)\mapsto
\Big(y_2,y_3,y_4 , y_2y_4(y_2y_4+y_3)/(y_1y_3^2)\Big).
$$
To show that this map is Liouville integrable, it is necessary to verify
that $\{ {\cal I}_1 , {\cal I}_2\} =0$ with respect to this bracket.

The terms in the formula (\ref{kp51}) can all be
expressed via  the functions
\beq \label{wi}
w_i=J_i\, J_{i+1},
\eeq
so that
$ {\cal I}_1=w_1+w_2+w_3+w_4$, $ {\cal I}_2=w_1w_3$.
From (\ref{jscale}) it is clear that these $w_i$ are invariant under the
action of the scaling symmetry (\ref{lascala}). This means that they
can be written as functions of $y_j$, viz:
$$
w_1=\frac{(y_1+y_2)(y_2+y_3)}{y_2^2},\quad
w_2=\frac{(y_2+y_3)(y_3+y_4)}{y_3^2}, \quad
w_3=\frac{(y_3+y_4)(y_2y_3+y_1y_3^2+y_2^2y_4)}{y_1y_3^2y_4}.
$$
Under the action of $\hat\varphi$, since the $J_n$ cycle with period 4 under $\varphi$,
the $w_i$ transform as
$$
\hat\varphi^* w_1=w_2,\;\; \hat\varphi^*w_2=w_3,\;\; \hat\varphi^*w_3=w_4=\frac{w_1w_3}{w_2},\;\;
\hat\varphi^*w_4=w_1.
$$
Although only the first three are independent, it is convenient to make use of
$w_4$ as well.

The first three $w_i$ form a three-dimensional Poisson subalgebra  of the $y_j$, which
is non-polynomial:
$$
\{w_1,w_2\}=w_1w_2-w_1-w_2,\quad \{w_1,w_3\}=w_2-\frac{w_1w_3}{w_2}, \quad
\{w_2,w_3\}=w_2w_3-w_2-w_3.
$$
The Casimir of this algebra is
$
{\cal I}_1-{\cal I}_2=w_1+w_2+w_3+\frac{w_1w_3}{w_2}-w_1w_3 =2-{\cal K} .
$

Since ${\cal I}_1$ and ${\cal I}_2$ are both functions defined on this subalgebra,
it follows that $\{ {\cal I}_1,{\cal I}_2\} = \{{\cal I}_1,{\cal K}\} =0$, so the two first
integrals are in involution, as required.
}\eex  %
\br {\em
The Poisson bracket of the four functions $w_i$ can be calculated
in polynomial form as
$$
\{w_1,w_2\}=w_1w_2-w_1-w_2,\quad \{w_1,w_3\}=w_2-w_4,\quad
\{w_1,w_4\}=w_1+w_4-w_1w_4,
$$
with the remaining brackets following from the cyclic property. This bracket
has the two Casimirs
$$
{\cal C}_1=w_1w_3-w_2w_4\quad\mbox{and}\quad
    {\cal C}_2=w_1+w_2+w_3+w_4-w_1w_3,
$$
so that the three-dimensional algebra for $w_1,w_2,w_3$ arises from
the constraint ${\cal C}_1=0$.
}
\er

\subsection{Primitives $P_{N}^{(q)}$ with $q>1$}

For general $q$, to obtain the quiver $P_{N}^{(q)}$ we must modify the formula (\ref{btau}) and write
\be  \label{btauk}  %
B = \tau_N^q-(\tau_N^q)^T.
\ee  %
In this case,
we have to take into account both sets of functions
$J_n,\, n=1,\dots ,p$ and $K_n,\, n=1,\dots ,q$, defined through (\ref{Jrec}) and (\ref{Krec}) respectively, with
the function $\cal K$ being given by the two formulae of (\ref{k2}).
It turns out that the  essential properties of the quantities $J_n$ with period $p$ can
be obtained by considering the $J_n$ in the case of $P_{p+1}^{(1)}$ (the $\tilde{A}_{1,p}$ quiver),
and applying a suitable permutation of indices; and similarly the properties of the
$K_n$ are the same as those of the $J_n$ for $P_{q+1}^{(1)}$, up to a permutation of indices.

Concentrating for the moment on the functions $J_n$, we consider the formula (\ref{mdy1}) for ${\bf M}_n$.
Each of the matrices ${\bf L}_{n+\ell q}$ (after using the cyclic property $J_{n+p}=J_n$) is just one of
the matrices ${\bf L}_n,\, n=1,\dots ,p$.  For coprime $p,q$, each of the $J_n$ appears exactly once in this product
and in a specific order, which defines a permutation $\sigma$ of the integers $(1,\dots ,p)$.  For this
discussion, for a general pair $p,q$ (with $q<p$)
let us use ${\cal K}_{p,q}$ to mean tr$\,\,{\bf M}_n$, considered as a function of the $J_n$.  Then we have
\be\label{kpq}  %
{\cal K} = {\cal K}_{p,q}(J_1,\dots ,J_p)= {\cal K}_{p,1}(J_{\sigma(1)},\dots ,J_{\sigma(p)}).
\ee  %
Similarly, in terms of the functions $K_n$,
the formula (\ref{mdy2}) for $\hat{\bf M}_n$
defines a permutation $\hat\sigma$ of  $(1,\dots ,q)$,
and if we write  ${\cal K}_{q,p}$
to denote tr$\,\hat{\bf M}_n$,
then we have
\beq\label{kpqhat}
{\cal K} = {\cal K}_{q,p}(K_1,\dots ,K_q) =  {\cal K}_{q,1}(K_{\hat{\sigma}(1)},\dots ,K_{\hat{\sigma}(q)}).
\eeq
(There is no risk of confusion between ${\cal K}_{q,p}$ and
${\cal K}_{p,q}$ once we have fixed $q<p$.)

For more detailed properties of the maps defined by (\ref{jkrec}), and the associated quantities
$J_n$ and $K_n$, it is necessary to consider even/odd $N$ separately.


\subsubsection{The even case}

In the case that $N$ is even, when $q$ and  $p$ are coprime it can be shown by elementary row/column operations
on the matrix (\ref{btauk})  that
$\det \, B =4$.
The matrix $B$ is of Toeplitz type, and invertible, with an inverse of the same type
that defines a nondegenerate Poisson structure of the form (\ref{logcan}).
With the choice of scale $C=2B^{-1}$, the Poisson bracket for the $x_j$ is given explicitly by
\beq\label{xqbra}
\{ \, x_j , x_k \, \}
=\left\{ \bear{ll}
                               \mathrm{sgn}(k-j)\, (-1)^{r+s}x_j x_k, \qquad & k-j \quad \mathrm{odd}, \\[2mm]
                               0 & \mathrm{otherwise},
                               \eear  \right.
\eeq
where (by the Euclidean algorithm) the integers $r,s$ are uniquely
determined by writing
$$
\frac{1}{2}\Big(p-|k-j|\Big) = sm -r\ell, \qquad \mathrm{for}\quad 0\leq r <m, \quad 0\leq s\leq \ell,
$$
in terms of the coprime integers $\ell = (p-q)/2$, $m=(p+q)/2$. (Note that $p$ and $q$ are both odd.)

Consider the functions $J_n$ once again, and for fixed $n$ define
the following sequences:
$$
X_j = x_{n+(j-1)q}, \quad j=1,2, \ldots ; \qquad
{J}^\dagger_k = J_{n+(k-1)q}=\frac{X_k+X_{k+2}}{X_{k+1}}, \quad k=1,2,\ldots .
$$
Note that the sequence of ${J}^\dagger_k$ is also periodic with the same period:
${J}^\dagger_{k+p}={J}^\dagger_k$; in fact (up to the choice of $n$, which just gives
an overall cyclic permutation) the ordering of this sequence
corresponds to the permutation
$\sigma$ in (\ref{kpq}), i.e. ${J}^\dagger_k=J_{\sigma (k)}$.
Using the Poisson bracket (\ref{xqbra}), we compute $\{X_1,X_2\} = \{x_n,x_{n+q}\} =X_1X_2$,
$\{X_1,X_3\}=  \{x_n,x_{n+2q}\} = 0$, and so on, and then in each case we find that, for a suitable range of indices,
the $X_j$ satisfy the same Poisson algebra (\ref{xbra}) as the $x_j$ in the $P_{p+1}^{(1)}$ case, so that
the bracket is
$$
\{ \, X_j , X_k \, \}=\left\{ \bear{ll}
                               \mathrm{sgn}(k-j)\, X_j X_k, \qquad & k-j \quad \mathrm{odd}, \\[2mm]
                               0 & \mathrm{otherwise}, \qquad \qquad
\mathrm{for} \quad  | j-k|\leq p .
                               \eear  \right.
$$
To verify the analogue of Lemma \ref{jbra}, it
is sufficient to calculate the brackets $\{J^\dagger_1,J^\dagger_j\}$ for
$j=2,\ldots ,(p+1)/2$ and then use the periodicity,
which shows that the permuted quantities ${J}^\dagger_k$ satisfy
$$
\{ \, J_j^\dagger, J_k^\dagger\, \} = 2\, \mathrm{sgn}(k-j)\, (-1)^{j+k+1}
J_j^\dagger J_k^\dagger
+2(\delta_{j,k+1}-\delta_{j+1,k}+\delta_{j+p-1,k}-\delta_{j,k+p-1}),
$$
for $j,k=1,\ldots, p$.  This is the same as the
Poisson algebra (\ref{jbrac}) for the $J_n$ in the case of $P_{p+1}^{(1)}$. An identical argument
implies that, up to the permutation $\hat\sigma$, the $q$ functions $K_n$ satisfy the algebra (\ref{jbrac})
corresponding to $P_{q+1}^{(1)}$.
\begin{rem} {\em
Theorem \ref{isomonodromy}, together with (\ref{k2}), implies that
the same function ${\cal K}$ is simultaneously the Casimir of the
$J_n$ subalgebra and of the $K_n$ subalgebra.
}
\end{rem}
%
By further direct calculation using the bracket (\ref{xqbra})
one can verify that
\beq\label{commjk}
\{ J_i, K_i\} = 0 \qquad \forall\,  i \qquad\implies \qquad \{ J_i, K_j\} = 0 \qquad \forall\,  i,j,
\eeq
where the second statement follows from the first by repeatedly
shifting $i\rightarrow i+1$, and using
periodicity and the coprimality of $p$ and $q$.

Thus, from (\ref{commjk}), we see that these two  subalgebras Poisson commute with each other.
The $J_n$ subalgebra provides $(p+1)/2$ commuting integrals, which are found by applying the permutation
$\sigma$ to each homogeneous component of the sum (\ref{kgen}) for the case of the
primitive   $P_{p+1}^{(1)}$. Similarly, by applying the permutation
$\hat\sigma$ one obtains  $(q+1)/2$ commuting integrals for the $K_n$ subalgebra. These
two sets of integrals also commute with each other, and the relation  (\ref{k2}) provides a single
constraint, which gives a total of $(p+1)/2+(q+1)/2-1=N/2$ independent commuting integrals,
as required for Liouville integrability.

\bex[The primitive $P_8^{(3)}$] \label{map-p83} {\em   %

The matrix $C=2B^{-1}$ is Toeplitz, with top row
$$
(c_{1,j}) = ( 0 ,-1, 0 , 1 , 0 , 1 , 0 ,- 1),
$$
which specifies the Poisson bracket (\ref{xqbra}) in this case.
This Poisson bracket is invariant under the map
\beq\label{p38map}
\varphi: \qquad
(x_1,\ldots ,x_8)\mapsto  (x_2,\ldots ,x_9), \qquad x_9=\frac{x_4x_6+1}{x_1}.
%
\eeq
The functions $J_n$, which cycle with period 5, are given by
$$ %
J_1=\frac{x_1+x_7}{x_4},\quad J_2=\frac{x_2+x_8}{x_5},\quad
J_3=\frac{x_1x_3+x_4x_6+1}{x_1x_6},
\quad %
J_4=\frac{x_2x_4+x_5x_7+1}{x_2x_7},\quad
J_5=\frac{x_3x_5+x_6x_8+1}{x_3x_8}.
$$
These form a Poisson subalgebra with brackets
$
\{J_1,J_2\}=-2J_1J_2,\quad
\{J_1,J_3\}=-2J_1J_3+2,
$
with all other brackets following from the cyclic property and skew-symmetry.
This is the algebra of $P_6^{(1)}$, after the permutation $\sigma:\, (1,2,3,4,5)\mapsto
(1,4,2,5,3)$.  It provides three commuting functions,
$$
{\cal I}_0=J_1+J_2+J_3+J_4+J_5, \quad 
{\cal I}_1=J_1J_4J_2+J_2J_5J_3+J_3J_1J_4+ J_4J_2J_5+J_5J_3J_1, \quad 
{\cal I}_2=J_1J_2J_3J_4J_5, 
$$
which (from Theorem \ref{isomonodromy})
are the homogeneous components of the Casimir of this subalgebra, namely
$$
{\cal K}=\mathrm{tr}\, {\bf M}_1 = {\cal I}_0-{\cal I}_1+{\cal I}_2.
$$

The generators of the period 3 subalgebra are given by
$$
K_1=\frac{x_1x_3+x_6x_8+1}{x_3x_6},\quad
K_2=\frac{x_1+x_7+x_4(x_1x_2+x_6x_7)}{x_1x_4x_7},\quad
   K_3=\frac{x_2+x_8+x_5(x_2x_3+x_7x_8)}{x_2x_5x_8},
$$
whose Poisson bracket relations are
$$
\{K_1,K_2\}=-2K_1K_2+2,\quad \{K_2,K_3\}=-2K_2K_3+2,\quad \{K_1,K_3\}=2K_1K_3-2.
$$
Up to the permutation $\hat{\sigma}:\, (1,2,3)\mapsto (1,3,2)$,
this is the algebra associated with $P_4^{(1)}$, with Casimir
$$
{\cal K}=\mathrm{tr}\, \hat{\bf M}_1 = -\hat{\cal I}_0+\hat{\cal I}_1,
\qquad \mathrm{where} \quad
\hat{\cal I}_0 = K_1+K_2+K_3,
\quad
\hat{\cal I}_1 =  K_1K_2K_3.
$$
The latter two quantities commute with each other, and we have
the relation
$ {\cal I}_0-{\cal I}_1+{\cal I}_2+\hat{\cal I}_0- \hat{\cal I}_1=0$.

Since (\ref{commjk}) holds, 
any four of the five functions
$ {\cal I}_0,{\cal I}_1,{\cal I}_2,\hat{\cal I}_0, \hat{\cal I}_1$
provide the correct number of independent commuting first
integrals to show Liouville integrability of the 8-dimensional map (\ref{p38map}).
}\eex  %

\subsubsection{The odd case}

When $N$ is odd, then $p$ is odd and $q$ is even, or vice versa. With
$N=2m+1$ we find that, as for the primitives $P_{2m+1}^{(1)}$, the kernel
of the matrix (\ref{btauk}) is spanned by the same vector
${\bf u}=(1,-1,\dots ,1,-1,1)^T$,
orthogonal to the vectors (\ref{oddprimvj}) providing
the symplectic coordinates 
$y_j = {\bf x}^{{\bf v}_j}$, as in (\ref{oddprimy}),
which are invariant under the one-parameter
scaling group (\ref{lascala}). 
The symplectic form obtained via Lemma \ref{symp} gives a nondegenerate Poisson bracket for the $y_j$,
of the form 
\beq \label{yqbra}
\{ \, y_j , y_k\,\} = \epsilon_{jk}\, y_j y_k , \qquad 1\leq j,k\leq 2m.
\eeq
In all examples we find that the Toeplitz matrix
$\hat{B}^{-1}=( \epsilon_{jk})$ has only the entries $1,-1,0$ above the
diagonal, but a concise formula for these $\epsilon_{jk}$ in terms of the
coprime integers $p,q$ is presently unavailable.

By Theorem \ref{torusred}, we have an induced
birational map $\hat\varphi$ in $2m$ dimensions, which is a
Poisson map with respect to the nondegenerate bracket (\ref{yqbra}).
We would like to assert that this is an integrable map.

Assume for the sake of argument that $p$ is odd and $q$ is even.
Then the $J_n$ can be written as functions of $y_j$, and from (\ref{yqbra}) they should satisfy the algebra (\ref{jbrac})
corresponding to $P_{p+1}^{(1)}$, hence providing $(p+1)/2$ commuting integrals.
The scaling-invariant quantities $\hat{w}_n =K_n K_{n+q}$, for $1\leq n \leq q-1$, can also be written in terms of
the $y_j$, and (up to the permutation $\hat\sigma$) should give a Poisson subalgebra isomorphic to
that of the functions (\ref{wi}) for $P_{q+1}^{(1)}$, providing another $q/2$ commuting integrals.
With the constraint (\ref{k2}), this would give $m$ independent commuting integrals in dimension $2m$, as required.

For the rest of this section, we present examples of primitives
$P_{N}^{(q)}$ with odd $N$ and $q>1$.

\bex[The primitive $P_5^{(2)}$] \label{ymap-p52} {\em   %
For $N=5$, $q=2$ the matrix (\ref{btauk}) has null vector ${\bf u}=(1,-1,1,-1,1)^T$, and
im$\, B$ is spanned by the vectors
$${\bf v}_1=(1,1,0,0,0)^T,\, {\bf v}_2=(0,1,1,0,0)^T,\,
{\bf v}_3=(0,0,1,1,0)^T,\,  {\bf v}_4=(0,0,0,1,1)^T.$$
Upon applying Lemma \ref{symp}, the Toeplitz matrix $\hat{B}^{-1}$ is specified by its first row, namely
$
(\epsilon_{1,j})=(0,0, 1, 1),
$
which determines the components of the
nondegenerate Poisson bracket (\ref{yqbra}) for
the variables $y_i=x_ix_{i+1},\; i=1,\dots ,4$.
Explicitly (for indices $j<k$) this is just
$$
 \{y_1,y_3\}= y_1y_3,\;\;  \{y_1,y_4\}= y_1y_4,\;\; \{y_2,y_4\}= y_2y_4,
$$
all other brackets being zero. The Poisson bracket is invariant under the
induced map
\beq\label{p25map}
\hat\varphi: \qquad (y_1,y_2,y_3,y_4)\mapsto
\Big(y_2,y_3,y_4, y_2y_4(y_3+1)/(y_1y_3)\Big).
\eeq

The period 3 functions $J_n$ take the form
$$
J_1=\frac{x_1+x_5}{x_3},\quad J_2=\frac{x_1x_2+x_3x_4+1}{x_1x_4},\quad
J_3=\frac{x_2x_3+x_4x_5+1}{x_2x_5}.
$$
Being invariant under the scaling
symmetry (\ref{lascala}), they
can also be written in terms of the variables $y_i$, as
$$
J_1=\frac{y_1y_3+y_2y_4}{y_2y_3},\quad J_2=\frac{y_2(y_1+y_3+1)}{y_1y_3},\quad
J_3=\frac{y_3(y_2+y_4+1)}{y_2y_4}.
$$
The Poisson brackets between these functions follow by the cyclic property
from
$
\{J_1,J_2\}=1-J_1J_2,
$
which, up to rescaling by a factor of 2 and applying
the permutation
$\sigma : (1,2,3)\mapsto (1,3,2)$,
is the bracket (\ref{jbrac})
of the $J_n$ for $P_4^{(1)}$. The $J_n$ subalgebra provides
a pair of first integrals in involution, namely 
$$
{\cal I}_0=J_1+J_2+J_3, \qquad {\cal I}_1=J_1J_2J_3,
$$
which is sufficient for the map (\ref{p25map}) to be Liouville integrable.
(Since these functions are totally symmetric,
not just cyclically symmetric, the permutation $\sigma $ plays no role.)

The period 2 quantities $K_n$, which are not invariant under the scaling
(\ref{lascala}), 
are
$$
K_1=\frac{x_1x_2+x_4x_5+1}{x_2x_4},\quad
K_2=\frac{x_1+x_5+x_3(x_1x_2+x_4x_5)}{x_1x_3x_5}.
$$
As for the case of $P_3^{(1)}$ in Example \ref{ymap-p31},
the product
$K_1K_2$ is invariant under the scaling symmetry, so
can be written in terms of the variables $y_i$. In fact,
from (\ref{k2}) we have
$
{\cal K} =  {\cal I}_1-{\cal I}_0 = K_1K_2-2,
$
where (by Theorem \ref{isomonodromy}) ${\cal K}$
is the Casimir of the bracket
for the $J_n$.  %
}\eex  %

\bex[The Case $P_7^{(2)}$] \label{ymap-p72} {\em   %
In the case $N=7$, $q=2$,
the Toeplitz matrix $\hat{B}^{-1}$ is specified by its first row, namely
$
(\epsilon_{1,j})=(0, 1, 1, 0, 0, 1),
$
which determines the components of the nondegenerate Poisson bracket
(\ref{yqbra}) for the variables $y_i=x_ix_{i+1},\; i=1,\dots ,6$.
This bracket is preserved by the 6-dimensional map
$$
\hat\varphi:\quad
(y_1,\ldots , y_6)\mapsto
(y_2,\ldots ,y_7),
\qquad
y_7=  \frac{y_2y_6(y_3y_5+y_4)}{y_1y_3y_5},
$$
which (by Theorem \ref{torusred}) is induced
from the map $\varphi$ defined by the recurrence (\ref{jkrec})
with $p=5$, $q=2$.

The functions $J_n$, which cycle with period 5 under the action of $\varphi$,
can be written in terms of $y_i$ thus:
\bea  %
&& J_1=\frac{y_1y_3+y_2y_4}{y_2y_3},\quad J_2=\frac{y_2y_4+y_3y_5}{y_3y_4},
\quad J_3=\frac{y_3y_5+y_4y_6}{y_4y_5}, \nn\\
&&  J_4=\frac{y_2y_4+y_1y_3y_4+y_2y_3y_5}{y_1y_3y_5},\quad
J_5=\frac{y_3y_5+y_2y_4y_5+y_3y_4y_6}{y_2y_4y_6}. \nn
\eea  %
The Poisson subalgebra generated by these functions is specified by
$$
\{J_1,J_2\}=J_1J_2,\quad \{J_1,J_3\}=J_1J_3-1,
$$
with all other brackets following from the cyclic property and skew-symmetry.
By rescaling by a factor of 2 and
applying a permutation to order these functions as $(J_1,J_3,J_5,J_2,J_4)$,
this  is seen to isomorphic to the algebra of the $J_n$ for $P_6^{(1)}$, so
we find the commuting functions
$$
{\cal I}_0 =J_1+J_3+J_5+J_2+J_4,\quad
{\cal I}_1=J_1J_3J_5+J_2J_4J_1+J_3J_5J_2+J_4J_1J_3+J_5J_2J_4,\quad
{\cal I}_2=J_1J_3J_5J_2J_4.
$$
Of course, the ordering is unimportant for the totally symmetric functions
${\cal I}_0$ and ${\cal I}_2$.

The period 2 quantities, which scale as
$K_1\rightarrow \la^{2}K_1$, $K_2\rightarrow \la^{-2}K_2$
under (\ref{lascala}), are
$$
K_1=\frac{x_2+x_6+x_4(x_1x_2+x_6x_7)}{x_2x_4x_6},\quad
K_2=\frac{x_1x_3+x_1x_7+x_5x_7+x_3x_5(x_1x_2+x_6x_7)}{x_1x_3x_5x_7}.
$$
From (\ref{k2}), the scaling-invariant combination $K_1K_2$
can be written in terms of $y_i$, via
$$
K_1K_2-2={\cal K} ={\cal I}_0-{\cal I}_1+{\cal I}_2 .
$$
}\eex  %

\bex[The Case $P_7^{(3)}$] \label{ymap-p73} {\em   %
For $N=7$, $q=3$, the  first row of
the Toeplitz matrix $\hat{B}^{-1}$ is
$
(\epsilon_{1,j})=(0, 0, 0,  1, 1,0).
$
This  defines the nondegenerate Poisson bracket
(\ref{yqbra}), which is invariant under the map
$$
\hat\varphi:\quad
(y_1,\ldots , y_6)\mapsto
(y_2,\ldots ,y_7),
\qquad
y_7= \frac{y_2y_4y_6(y_4+1)}{y_1y_3y_5} ,
$$
in terms of the variables $y_i=x_ix_{i+1},\; i=1,\dots ,6$.

The functions $J_n$, with period 4, are not invariant under (\ref{lascala}),
but $w_n=J_nJ_{n+1}, \, n=1,\dots,4$ are:
\bea  %
&& w_1=\frac{(1+y_1+y_4)(y_1y_3y_5+y_2y_4y_6)}{y_1y_3y_4y_5},\quad
w_2=\frac{(1+y_1+y_4)(1+y_2+y_5)}{y_1y_5}, \nn\\
&&  w_3=\frac{(1+y_2+y_5)(1+y_3+y_6)}{y_2y_6},\quad w_4=\frac{w_1w_3}{w_2}, \nn
\eea  %
of which only three are independent (since $w_1w_3=w_2w_4$). By periodicity and skew-symmetry, all
of their Poisson brackets follow from
$$
\{w_1,w_2\}=w_1+w_2-w_1w_2,\quad \{w_1,w_3\}=w_4-w_2.
$$
By applying the permutation $\sigma: \, (1,2,3,4)\mapsto (1,4,3,2)$,
this is seen to be isomorphic to the algebra for $P_5^{(1)}$,
as in Example \ref{ymap-p51}.
Hence we have two functions in involution, namely
$$
{\cal I}_1=w_1+w_4+w_3+w_2,\qquad {\cal I}_2=w_1w_3,
$$
where the ordering is unimportant since both functions
${\cal I}_1$ and ${\cal I}_2$ are invariant under $\sigma$.

The necessary third function in involution is derived from the quantities $K_n$, which cycle
with period 3. Being invariant under the scaling (\ref{lascala}), they
can be written in terms of the variables $y_i$:
$$
K_1=\frac{y_3(y_1+y_5+1)}{y_2y_4},\quad K_2=\frac{y_4(y_2+y_6+1)}{y_3y_5},\quad
   K_3=\frac{y_1y_3y_5(y_3+1)+y_2y_4y_6(y_4+1)}{y_1y_3y_4y_6}.
$$
They generate a three-dimensional
Poisson subalgebra with the same relations as for the subalgebra in $P_4^{(1)}$
(up to a factor of 2), i.e.
$$
\{K_1,K_2\}=K_1K_2-1,\quad \{K_2,K_3\}=K_2K_3-1,\quad \{K_1,K_3\}=-K_1K_3+1.
$$
There are two first integrals that are commuting functions defined on this subalgebra,
which we denote by
$$
\hat{\cal I}_0=K_1+K_2+K_3,
\qquad
\hat{\cal I}_1=K_1K_2K_3, $$
and the joint Casimir of the two subalgebras is given by
$
{\cal K} = 2-{\cal I}_1+{\cal I}_2=\hat{\cal I}_1-\hat{\cal I}_0.
$
Since $\{K_i,w_j\}=0$ for all $i,j$, we may
use any three of the functions ${\cal I}_1,{\cal I}_2,\hat{\cal I}_0,\hat{\cal I}_1$ to show Liouville integrability.
}\eex  %

\section{Linearisable recurrences from
$P_{2m}^{(q)}-P_{2m}^{(m)}+P_{2(m-q)}^{(m-q)}$ quivers}
\setcounter{equation}{0}
\label{pert}

In this section we consider the family of recurrences (\ref{comprec}),
as in case (iii) of Theorem \ref{zeroe}, which come from the quivers  of the form
$P_{2m}^{(q)}-P_{2m}^{(m)}+P_{2(m-q)}^{(m-q)}$.
It is convenient to rewrite each recurrence as
\beq
x_{n+N}\, x_n = x_{n+p}\, x_{n+q} + x_{n+m}, \qquad p+q=N=2m,
\label{pqrec}
\eeq
which (for fixed $m$) gives a different recurrence for each $q=1,\ldots,m-1$.
The associated matrix $B$ is given by
\be  \label{btau3}  %
B = \tau_{2m}^q-(\tau_{2m}^q)^T-\tau_{2m}^m+\hat\tau_{2(m-q)}^{m-q},
\ee  %
where $\tau_N$ is defined in (\ref{btau}), and $\hat\tau_{2(m-q)}$ denotes
the $N\times N$ matrix obtained
by adding $q$ left and right columns and upper and lower rows of zeros
to the $2(m-q)\times 2(m-q)$ matrix $\tau_{2(m-q)}$. The simplest
examples of the quivers corresponding to such $B$ are shown in
Figure 1.

Observe that if $\mbox{gcd}(m,q)=r>1$ then the quiver consists of
$r$ disjoint copies of the same type of quiver, but with
the parameters $q$ and $m$ replaced by the coprime
integers $q/r$ and $m/r$, respectively,
and similarly
(\ref{pqrec}) decouples into $r$ copies of the corresponding
recurrence. Therefore we shall assume that $\mbox{gcd}(m,q)=1$
from now on. With this assumption it follows from $p+q=2m$ that 
$\mbox{gcd}(p,q)=1$ or $2$ only.
The case $\mbox{gcd}(p,q)=1$ has a very similar structure to that of the
primitives  $P_{N}^{(q)}$ for even $N$, but the case $\mbox{gcd}(p,q)=2$ has
several new features, so we will need to distinguish between these
two cases in due course.

The family (\ref{pqrec})  has some basic properties
that are analogous to those of the family in
case (ii) of Theorem \ref{zeroe},
as described in the previous section.
In particular, all of the recurrences
(\ref{pqrec}) are linearisable, and they have two sets of
periodic functions, with periods
$p,q$ respectively, which lead to the construction of
first integrals.

\subsection{More linear relations with periodic coefficients}

By analogy with (\ref{frieze}), the recurrence (\ref{pqrec})  can be written in the form
\beq \label{2frieze}
\det \,\Psi_n = \left| \bear{cc} x_n & x_{n+q} \\
                                 x_{n+p} & x_{n+N}
\eear\right| = x_{n+m}.
\eeq
The above identity  is the relation for a 2-frieze  \cite{morier}, and  it implies that the iterates
of (\ref{pqrec}) can be placed in the form an infinite 2-frieze. Using Dodgson condensation once again to condense a $3\times 3$
determinant, we have
$$
\det\tilde{\Psi}_n =(x_{n+m}x_{n+N+m}-x_{n+m+q}x_{n+m+p})/x_{n+N}=1,
$$
with $\tilde{\Psi}_n$ as in (\ref{3by3}). Then condensing the appropriate  $4\times 4$ matrix $\Delta_n$ in terms of
$3\times 3$ minors yields
$$
\det \Delta_n =  \left| \bear{cccc}x_n & x_{n+q} & x_{n+2q} &  x_{n+3q}\\
                               x_{n+p} & x_{n+N} & x_{n+N+q} & x_{n+N+2q} \\
 x_{n+2p} & x_{n+N+p} & x_{n+2N} & x_{n+2N+q} \\
 x_{n+3p} & x_{n+N+2p} & x_{n+2N+p} & x_{n+3N}

 \eear\right|=0.
$$
As in the case of the primitives considered before, the left and right kernels of the singular matrix  $\Delta_n$
yield linear relations between the $x_n$.

\begin{lem} \label{PQ}
The iterates of the recurrence (\ref{jkrec}) satisfy the linear relations
\beq\label{Prec}
x_{n+3q}-J_{n+m}\,x_{n+2q}+ J_{n}\, x_{n+q} - x_n = 0,
\eeq
\beq\label{Qrec}
x_{n+3p}-K_{n+m}\,x_{n+2p}+ K_{n}\, x_{n+p} - x_n = 0,
\eeq
whose coefficients are periodic functions of period $p,q$ respectively,
that is
$$
J_{n+p}=J_n, \qquad K_{n+q}=K_n, \qquad  for\,\,  all \, \, n.
$$ 
\end{lem}
\begin{prf}
Upon solving $\Delta_n{\bf k}_n =0$, using the first three rows,
it is convenient to normalise the first entry of ${\bf k}_n$ to be $-1$, and solve
a $3\times 3$ system to find the other three entries.
Then from  Cramer's rule and $\det\tilde{\Psi}_n =1$ the fourth entry must be $+1$, so that this
vector 
has the form
${\bf k}_n=(-1,J_n,\hat{J}_n,1)^T$.
The $3\times 3$  linear system coming from the first three rows of the equation
$\Delta_{n+p}{\bf k}_{n+p} =0$ is the same as that coming from the last three rows
of $\Delta_n{\bf k}_n =0$, which implies that $J_{n+p}=J_n$ and $\hat{J}_{n+p}=\hat{J}_n$.
Applying Cramer's rule in the first three rows of $\Delta_n{\bf k}_n =0$ together with Dodgson condensation
also implies that
\beq\label{cform}
J_n =\left|\bear{ccc} x_n & x_{n+2q} & x_{n+3q}\\
                        x_{n+p} & x_{n+N+q} & x_{n+N+2q}\\
x_{n+2p} & x_{n+2N} & x_{n+2N+q} \eear \right|
=\frac{1}{x_{n+4m-p}}\,  \left|\bear{cc}E_n & x_{n+5m-2p} \\
     E_{n+p} & x_{n+5m-p} \eear\right|,
\eeq
and similarly
\beq\label{dform}
\hat{J}_n = \frac{1}{x_{n+2m}}\,  \left|\bear{cc}E_{n+2m-p} & x_{n+m} \\
                               E_{n+2M} & x_{n+m+p} \eear\right|,
\qquad \mathrm{where} \qquad E_n =  \left|\bear{cc}x_{n} & x_{n+2q} \\
                             x_{n+p} & x_{n+N+q} \eear\right|.
\eeq
Permuting the first and second columns of the determinant in the identity $\det\tilde{\Psi}_n =1$
and expanding in terms of $2\times 2$ minors, and then doing the same thing after permuting the second and third
columns instead, leads to the formulae
\beq\label{eids}
\left|\bear{cc} E_{n} & x_{n+m}   \\
      E_{n+p} & x_{n+m+p} \eear\right| = -x_{n+p},
\qquad
\left|\bear{cc}E_{n} & x_{n+3m-p} \\
                    E_{n+p} & x_{n+3m} \eear\right| =x_{n+4m-p},
\eeq
respectively. Then the combination
$x_{n+2m}x_{n+5m-p}(J_{n+m}+\hat{J}_n)$ can be rewritten as a sum of determinants,
whose entries can be expanded  using each of the identities (\ref{eids}), to yield
$$
\bear{l}
\quad
\left|\bear{cc} x_{n+2m}E_{n+m} & x_{n+6m-2p} \\
       E_{n+m+p} x_{n+2m}  & x_{n+6m-p} \eear\right|
 + \left|\bear{cc} x_{n+5m-p}E_{n+2m-p} & x_{n+m} \\
       x_{n+5m-p}E_{n+2m} & x_{n+m+p} \eear\right|
       \\ \\
= \left|\bear{cc} x_{n+2m}E_{n+m} & x_{n+6m-2p} \\
        x_{n+2m+p}E_{n+m}+ x_{n+m+p}   & x_{n+6m-p} \eear\right|
+\left|\bear{cc} x_{n+5m-2p}E_{n+2m} +x_{n+6m-2p} & x_{n+m} \\
        x_{n+5m-p}E_{n+2m} & x_{n+m+p} \eear\right|
\\ \\
= E_{n+m}   \left|\bear{cc} x_{n+2m} & x_{n+2m+2q} \\
                 x_{n+2m+p} & x_{n+2m+N+q} \eear\right|
+E_{n+2m}\left|\bear{cc} x_{n+m+2q}& x_{n+m} \\
              x_{n+m+N+q}& x_{n+m+p} \eear\right|=0,
\eear
$$
as required.
This proves (\ref{Prec}), and the relation (\ref{Qrec}) follows by symmetry, from considering the left kernel
of $\Delta_n$.
\end{prf}
\begin{rem}\label{penta}
{\em
The four-term linear relations (\ref{Prec}) and (\ref{Qrec}),
together with $\det\tilde{\Psi}_n =1$, should be compared with those of the
pentagram map \cite{pentagram}, but there
the coefficients of the second and third terms are independent.
}
\end{rem}
\begin{rem}
\label{gcd1}
{\em
When $q=1$ the coefficient $K_n$ has period 1, so $K_{n+1}=K_n={\cal K}$
for all $n$, and the recurrence (\ref{Qrec}) is just the
{\it constant coefficient, linear difference equation}
\beq\label{q1krec}
x_{n+3N-3} -{\cal K}\, x_{n+2N-2}+{\cal K}\, x_{n+N-1}-x_n =0.
\eeq
An immediate consequence of the latter relation is an inhomogeneous
version of (\ref{k-pn1}), namely
$$
x_{n+2N-2} -({\cal K}-1)\, x_{n+N-1} + x_n = F_n, \qquad
\mathrm{where}\qquad F_{n+N-1}= F_n,
$$
for some quantity $F_n$ (which has period $p=N-1$ in this case).
Thus, by a minor modification of Corollary \ref{cheby}, one can
find explicit formulae for $x_n$ in terms of Chebyshev polynomials.
}
\end{rem}
\begin{rem}
\label{gcd2}
{\em
When $q=2$,
the coefficients in the recurrence (\ref{Qrec})
have period 2, so $K_{n+2}=K_n$ for all $n$, and (since
$\mbox{gcd}(m,q)=1$ implies that $m$ is odd) we have $K_{n+m}=K_{n+1}$, whence
\beq\label{q2krec}
x_{n+3p} -K_{n+1}\, x_{n+2p}+K_n\, x_{n+p}-x_n =0.
\eeq
This is a four-term linear relation, whose coefficients
alternate with the parity of the index $n$.
}
\end{rem}

The existence of the periodic quantities
$J_n$ and $K_n$ means that, as for the case of primitives
considered in the last section, one can construct first integrals by taking
cyclically symmetric functions of each of these sets of quantities.
When $\mbox{gcd}(p,q)=1$ we find one relation between these two
sets of quantities, and when $\mbox{gcd}(p,q)=2$ we find two relations,
which will be discussed below.

\subsection{Monodromy and linear relations with constant coefficients}
\label{3x3monodromy}

This subsection follows closely the discussion of subsection \ref{2x2monodromy} for the case
of primitives. However, it will subsequently be necessary to refine the discussion further,
depending on whether   $\mbox{gcd}(p,q)=1$ or $2$.

The relation (\ref{Prec}) implies that the matrix $\tilde\Psi_{n}$ satisfies
\beq\label{psi3rel1}
\tilde\Psi_{n+q}=\tilde\Psi_n\, {\bf L}_n, \qquad {\bf L}_n=\left(\bear{ccc} 0 & 0 & 1 \\ 1 & 0 & -J_n
                \\ 0  & 1 & J_{n+m}  \eear\right).
\eeq
On the other hand, the recurrence (\ref{Qrec}) yields
\beq\label{psi3rel2}
\tilde\Psi_{n+p}= \hat{\bf L}_n \,\tilde\Psi_n, \qquad \hat{\bf L}_n=\left(\bear{ccc} 0 &  1 & 0 \\
                                                                          0 & 0 & 1 \\
                                                                          1  & -K_n & K_{n+m}  \eear\right).
\eeq

As before, upon taking the ordered product of the ${\bf L}_n$ over $p$ steps, shifting by $q$ each time, we have the monodromy matrix
\beq\label{mdy31}
{\bf M}_n:=  {\bf L}_{n}{\bf L}_{n+q}\ldots{\bf L}_{n+(p-1)q}=\tilde\Psi_n^{-1}\,\tilde \Psi_{n+pq}.
\eeq
Taking the ordered product of the $\hat{\bf L}_n$ over $q$ steps, shifting by $p$ each time, gives another monodromy matrix
\beq\label{mdy32}
\hat{{\bf M}}_n:=  \hat{\bf L}_{n+(q-1)p}\ldots\hat{\bf L}_{n+p}\hat{\bf L}_n=\tilde\Psi_{n+pq}\, \tilde\Psi_n^{-1}.
\eeq
From the cyclic property of the trace it follows that
\beq\label{k32}
\mathcal{K}_n:= \mathrm{tr}\, {\bf M}_n = \mathrm{tr}\, \hat{\bf M}_n.
\eeq

The periodicity of ${\bf L}_n$, together with  
(\ref{mdy31}), 
implies that $\mathcal{K}_{n+p}=\mathcal{K}_n$, and similarly, from (\ref{mdy32}),
we have $\mathcal{K}_{n+q}=\mathcal{K}_n$.
If the periods $p$ and $q$ are coprime, then 
$\mathcal{K}_n=\mathcal{K}=$constant, for all $n$,
hence $\mathcal{K}$ is a first integral for the map $\varphi$
corresponding to (\ref{pqrec}). However, if $\mbox{gcd}(p,q)=2$ holds instead then
we have $\mathcal{K}_{n+2}=\mathcal{K}_n$, so this quantity has period 2.

Once again the general result of Lemma \ref{perlin} can be applied here, to show
that the iterates of (\ref{pqrec}) satisfy a linear relation with constant coefficients.

\begin{propn}\label{K3pq}
The iterates of the nonlinear recurrence (\ref{pqrec}) 
satisfy a linear relation of order $3pq$ with constant coefficients.
It is a four-term relation
if $\mbox{gcd}(p,q)=1$, and  
a seven-term relation if $\mbox{gcd}(p,q)=2$.
\end{propn}
\begin{prf}
From (\ref{Prec}), Lemma \ref{perlin} implies that $x_n$ satisfies a linear recurrence of order
$3pq$ with constant coefficients, for which the gaps between indices of adjacent terms with
nonzero coefficients are of size $p$. On
the other hand, (\ref{Qrec}) implies a linear recurrence of the same order with gaps of
size $q$.  Thus the actual size of the gaps must be the lowest common multiple of $p$ and $q$.
Hence, when  $p$ and $q$ are coprime, the gaps are of size $pq$, giving a four-term relation,
while $\mbox{gcd}(p,q)=2$ gives a seven-term relation with gaps of size $pq/2$.
\end{prf}

Below we provide refined versions of the preceding result, with more precise details of the coefficients, by
considering the cases $\mbox{gcd}(p,q)=1$ and $\mbox{gcd}(p,q)=2$ separately.

\subsection{The case $\mbox{gcd}(p,q)=1$}
%

The  discussion of the case where $p$ and $q$ are coprime is almost identical to that for the
primitives $P_N^{(q)}$ with $N$ even, as in subsection
4.7 above. The integers $p$ and $q$ are both odd, and for the matrix (\ref{btau3})
in each case we find that  det$\,B=4$ whenever $p$ or $q$ is divisible by 3, and
det$\,B=1$ otherwise. From such a nondegenerate matrix $B$ we get an
invariant Poisson bracket of the form (\ref{logcan}),
which is specified uniquely (up to scale) by the Toeplitz matrix $C=B^{-1}$.
However,  in this case
a general closed-form expression for the entries of $C$, analogous
to the formula (\ref{xqbra}) for
the Poisson bracket of the even primitives, is not available to us at present.

We now consider the associated functions $J_n$ and $K_n$, of periods $p$ and $q$ respectively.
From the expression (\ref{mdy31}), the first integral $\mathcal{K}$
defined by (\ref{k32}) is a cyclically symmetric polynomial
in the $J_n$, $n=1,\ldots, p$, and from (\ref{mdy32}) it is also a cyclically symmetric polynomial
in the $K_n$, $n=1,\ldots, q$. Thus the equality of the traces in (\ref{k32})
provides a single functional relation between these two sets of functions.
Note that we also have the same phenomenon as in (\ref{kpq}) with regard to the
different expressions for ${\cal K}$ as a function of the quantities $J_n$, when we compare the cases with
$q=1$ and $q>1$, for the same value of $p$: up to the action of a suitable permutation $\sigma$ of
$(1,\ldots, p)$, the two expressions are identical; and the analogous
statement applies
to ${\cal K}$ considered as a function of the $K_n$.

Now observe that all of the preceding comments concerning ${\cal K}$ apply equally well to the quantity 
$$
\tilde{\cal K} :=
\mathrm{tr}\, {\bf M}^{-1}_n = \mathrm{tr}\, \hat{\bf M}^{-1}_n,
$$
which is also a first integral (so this definition holds for any $n$).
We would like to assert that in fact
$\tilde{\cal K}={\cal K}$.
To see this, note that tr$\, \hat{\bf L}_n = K_{n+m}$, and tr$\, \hat{\bf L}^{-1}_n = K_{n}$.
Thus, in the case $q=1$, when $K_n={\cal K}=\,$constant, we have ${\cal K}=\mathrm{tr}\,\hat{\bf M}_n
=\mathrm{tr}\,\hat{\bf M}^{-1}_n=\tilde{\cal K}$ by (\ref{mdy32}). This then implies that, in terms of functions
of $J_1,\ldots ,J_p$, we have
 \beq\label{kt32}
\mathrm{tr}\,{\bf M}_n=\mathrm{tr}\,{\bf M}^{-1}_n
\eeq
for $q=1$, and clearly this identity remains true when $q>1$, since (for fixed $p$)
the functions ${\cal K}$ and $\tilde{\cal K}$ are obtained from the case $q=1$ by applying the same
permutation $\sigma$ to both sides. This allows us to make a more precise statement than Proposition \ref{K3pq}.

\begin{thm}\label{Klin3}
When $\mbox{gcd}(p,q)=1$,
the iterates of the nonlinear recurrence (\ref{pqrec})  satisfy the linear relation
\beq\label{krec3}
x_{n+3pq}- \mathcal{K}\,x_{n+2pq}+\mathcal{K}\,x_{n+pq}-x_n=0,
\eeq
where  $\mathcal{K}$ is the first integral defined by (\ref{k32}).
\end{thm}
\begin{prf}
Using (\ref{mdy31}) we see that $\tilde\Psi_{n+pq}=\tilde\Psi_n {\bf M}_n$, so $\tilde\Psi_{n+2pq}=\tilde\Psi_{n+pq}{\bf M}_n{\bf M}_{n+pq}=\tilde\Psi_n {\bf M}_n^2$, by periodicity of ${\bf M}_n$,
and similarly $\tilde\Psi_{n+3pq}=\tilde\Psi_n {\bf M}_n^3$.  Applying Cayley-Hamilton to both ${\bf M}_n$
and ${\bf M}_n^{-1}$, and noting
that det$\,{\bf M}_n=1$ and tr$\,{\bf M}_n={\cal K}$, as well as (\ref{kt32}), yields
$$
\tilde\Psi_{n+3pq}-{\cal K} \tilde\Psi_{n+2pq}+{\cal K} \tilde\Psi_{n+pq}-\tilde\Psi_n = \tilde\Psi_n({\bf M}_n^3-{\cal K} {\bf M}_n^2+{\cal K} {\bf M}_n-{\bf I})=0.
$$
The $(1,1)$ component of this equation is just (\ref{krec3}).
\end{prf}

\br[Another construction of the integral for $P_3^{(1)}$]  %
{\em
We may think of the map $\hat\varphi$ in Example \ref{ymap-p31}
as coming from the recurrence
$y_{n+2}y_n=y_{n+1}^2+y_{n+1}$, which is of the same form as (\ref{pqrec}),
with $p=q=m=1$, although not directly obtained from a cluster mutation.
Replacing $x_n\to y_n$ and ${\cal K}\to \hat{\cal K}$
in (\ref{krec3}), and solving,  leads to the first integral
$$
\hat{\cal K} =\frac{y_{n+3}-y_n}{y_{n+2}-y_{n+1}}.
$$
When this is written purely in terms of $y_n, y_{n+1}$, using the map, we find the  previously obtained
quantity
$$
{\cal I}_1=\hat{\cal K}+1=\frac{(y_n+y_{n+1})(y_n+y_{n+1}+1)}{y_ny_{n+1}}.
$$
}
\er  %

For the Liouville integrability of the map $\varphi$ corresponding to
(\ref{pqrec}), the counting of first integrals
appears to be the same as for the even primitives $P_{2m}^{(q)}$.
The two sets of quantities $J_n$ and $K_n$ generate Poisson subalgebras of dimensions $p$ and $q$,
which should contain $(p+1)/2$ and $(q+1)/2$ commuting integrals, respectively, including the Casimir
$\cal K$, which is common to both subalgebras. Taking the constraint (\ref{k32}) into account,
this produces $m=(p+q)/2$ independent commuting integrals in terms of the $x_j$, as required.

However, while the above  counting argument is plausible,
it rests on some unproven assumptions.
Everything relies on the structure of the Poisson bracket for the $J_n$  in the case $q=1$, since all
other subalgebras of $J_n$ or $K_n$ should be isomorphic to one with $q=1$.
Yet in general (excluding $m=2$),
this Poisson bracket consists of 
three homogeneous parts (of degrees $0, 1$ and $2$), which
(unlike the bracket (\ref{jbrac}) for the primitive $P_{2m}^{(1)}$)
does not have an obvious splitting into a bi-Hamiltonian pair.
Moreover, while every cyclically symmetric polynomial function of the $J_n$ is a first integral 
of the map $\varphi$, we do not yet have an algorithm for selecting an involutive set of them.

\begin{rem}{\em
In \cite{pentagram} a quadratic Poisson structure is presented
for the coefficients of the four-term
linear relations for twisted polygons, together with
corresponding monodromy matrices and commuting first integrals
for the pentagram map.
However, a Dirac reduction of this bracket to the case of (\ref{Prec})
or (\ref{Qrec}) gives only the trivial bracket.
A general approach to Poisson structures related to twisted polygons
is described in \cite{marshall}, which should shed some light
on the situation here.
}
\end{rem}

For want of more general statements, we illustrate the foregoing
discussion with several examples of the integrable systems that
arise in this case.


\bex[The quiver $P_{4}^{(1)}-P_{4}^{(2)}+P_{2}^{(1)}$] \label{ex-p4142} {\em   %
This example was first studied in an ad hoc
way in \cite{honelaur}.
The appropriate matrix (\ref{btau3}),
which corresponds to the first quiver in Figure 1, is obtained by
setting $c=1$ in (\ref{s4gen}).
The explicit form of the map is
\be  \label{p4142map}  %
\varphi : \qquad (x_1,x_2,x_3,x_4) \mapsto  (x_2,x_3,x_4,x_5), \qquad
 x_5=  \frac{x_2x_4+x_3}{x_1}.
\ee  %
The invariant Poisson bracket (\ref{logcan}) for this map is given by the
Toeplitz matrix $C=2B^{-1}$, with top row
$
     (c_{1,j})=(0,1,1,2).
$
The period 3 functions $J_i$ take the form
$$
J_1=\frac{x_3(x_2+x_1x_3)+x_4(x_1^2+x_2^2)}{x_1x_2x_4},\quad
J_2=\frac{x_1x_4+x_2^2+x_3^2}{x_2x_3},\quad
J_3=\frac{x_3(x_2+x_1x_3)+x_4(x_1x_4+x_2^2)}{x_1x_3x_4}.
$$
The Poisson brackets between these three functions follow by the cyclic property
from
\beq\label{N4jbr}
\{J_1,J_2\}=J_1J_2-2J_3.
\eeq
This example is exceptional, in that the bracket for the $J_i$
is the sum of only {\em two} homogeneous terms:
$$
{\bf P}={\bf P}^{(2)}+{\bf P}^{(1)},\quad\mbox{where}\;\;\;
     {\bf P}^{(2)}=\left(\begin{array}{ccc}
                   0 & J_1J_2 & -J_1J_3 \\
                  -J_1J_2 & 0 & J_2J_3 \\
                   J_1J_3 & -J_2J_3 & 0
                      \end{array} \right), \quad
                       {\bf P}^{(1)}=\left(\begin{array}{ccc}
                   0 & -2J_3 & 2J_2 \\
                   2J_3 & 0 & -2J_1 \\
                   -2J_2 & 2J_1 & 0
                      \end{array} \right).
$$
Each of the tensors specified by ${\bf P}^{(1)}$ and ${\bf P}^{(2)}$
satisfies the Jacobi identity, so (since their sum is a Poisson tensor)
they define a compatible pair of Poisson brackets.

The first integrals 
$$
\CH_1=J_1^2+J_2^2+J_3^2, \qquad \CH_2=J_1J_2J_3
$$
satisfy the bi-Hamiltonian ladder 
$$
 {\bf P}_1\nabla \CH_1=0,\qquad  {\bf P}_2\nabla \CH_1={\bf P}_1\nabla \CH_2,
\qquad {\bf P}_2\nabla \CH_2=0,
$$
so they commute with respect to the bracket defined by (\ref{N4jbr}).
The quantity
\beq\label{k31eq}
{\cal K}=3-\CH_1+\CH_2
\eeq
provides the Casimir of this bracket;
following (\ref{kpq}),  we will find it useful to denote this quantity by
${\cal K}_{3,1}$.
}\eex  %

\bex[The quiver $P_{6}^{(1)}-P_{6}^{(3)}+P_{4}^{(2)}$] \label{ex-p6163} {\em   %

Mutation of the second quiver in Figure 1 gives the map
$$
\varphi : \qquad (x_1,x_2,x_3,x_4,x_5,x_6) \mapsto  (x_2,x_3,x_4,x_5,x_6,x_7), \qquad
  x_7=  \frac{x_2x_6+x_4}{x_1}.
$$
The invariant Poisson bracket for this map is given by (\ref{logcan}),
with the coefficients specified by the Toeplitz matrix $C=B^{-1}$ with top row
$
     (c_{1,j})=(0,0,1,0,1,1). 
$
The functions $J_i$, which cycle with period 5 under the action of $\varphi$, take the form
\bea  %
&& J_1=\frac{x_1x_5+x_2x_6+x_3x_4}{x_2x_5},\quad
J_2=\frac{x_4(x_3+x_1x_5)+(x_1+x_3)x_2x_6}{x_1x_3x_6},\quad
J_3=\frac{x_3(x_2+x_4)+x_1x_5)}{x_2x_4},  \nn\\
&& J_4=\frac{x_4(x_3+x_5)+x_2x_6)}{x_3x_5},\quad
J_5=\frac{x_4(x_3+x_1x_5)+(x_1x_5+x_2x_3)x_6}{x_1x_4x_6}.  \nn
\eea  %
The Poisson bracket between these five functions follow by the cyclic property
from
$$
\{J_1,J_2\}=-J_1J_2-J_4+1,\quad \{J_1,J_3\}=2J_1J_3.
$$
This Poisson bracket is the sum of three homogeneous terms,
$$
{\bf P}={\bf P}^{(2)}+{\bf P}^{(1)}+{\bf P}^{(0)}, \quad\mbox{with}\;\;\; {\bf P}^{(2)}_{ik}=c^{(2)}_{ik}J_iJ_k,\;\;\; {\bf P}^{(1)}_{ik}=c^{(1)}_{ik}J_{k+2},\;\;\;
   {\bf P}^{(0)}_{ik}=c^{(0)}_{ik},
$$
where $c^{(\ell)}_{ik}$ are the Toeplitz matrices with top rows given by
$$
(c^{(2)}_{1,k})=(0,-1,2,-2,1),\quad (c^{(1)}_{1,k}) =(0,-1,0,0,1),
\quad (c^{(0)}_{1,k})=(0,1,0,0,-1).
$$
Both ${\bf P}^{(0)}$ and ${\bf P}^{(2)}$ satisfy the Jacobi identity,
but ${\bf P}^{(1)}$ does not, so we cannot think of this sum as some sort of Poisson compatibility.

The Casimir for the 5-dimensional Poisson algebra generated by
the $J_i$ is the trace of the monodromy matrix,
as in (\ref{k32}).  It can be written as the sum
\beq\label{k51}
{\cal K}=\CH_1-\CH_2+\CH_3,
\eeq
where each of the components is a first integral:
$$
\CH_1=\sum_{i=1}^5(J_i-J_iJ_{i+1}),\quad
\CH_2=\sum_{i=1}^5(J_iJ_{i+1}J_{i+2}-J_iJ_{i+1}^2J_{i+2}),\quad
\CH_3=\prod_{i=1}^5J_i.
$$
We find that $\{ \CH_i ,\CH_j \} =0$ for all $i,j$, so the 6-dimensional
Poisson map $\varphi$ has the correct number of
first integrals in involution.
For later comparison, we denote ${\cal K}$ in (\ref{k51}) by ${\cal K}_{5,1}$.
}\eex  %

\bex[The quiver $P_{8}^{(1)}-P_{8}^{(4)}+P_{6}^{(3)}$] \label{ex-p8184} {\em   %

The map obtained from mutation of this quiver is
$$
\varphi : \qquad (x_1,x_2,x_3,x_4,x_5,x_6,x_7,x_8)
\mapsto  (x_2,x_3,x_4,x_5,x_6,x_7,x_8,x_9), \qquad
x_9=  \frac{x_2x_8+x_5}{x_1}.
$$
The corresponding non-singular matrix $B$
defines an invariant Poisson bracket (\ref{logcan})
with matrix $C=B^{-1}$.
The top row of the Toeplitz matrix $C$ is
$
     (c_{1,j})=(0,1,0,0,1,1,0,1).
$
The functions $J_i$ with period 7 can be determined from $J_1$,
which takes the form
$$
 J_1=\frac{x_1x_6+x_2x_7+x_3x_5}{x_2x_6},
$$
The remaining six functions are obtained by applying
$\varphi^*J_i=J_{i+1}$,
with $(\varphi^*)^7J_i = J_i$.
The Poisson bracket relations
between these functions follow by the cyclic property
from
$$
\{J_1,J_2\}=2J_1J_2-J_5,\quad \{J_1,J_3\}=-J_1J_3+1,\quad
\{J_1,J_4\}=-J_1J_4.
$$
Again, this Poisson bracket is the sum of three homogeneous terms,
${\bf P}={\bf P}^{(2)}+{\bf P}^{(1)}+{\bf P}^{(0)}$,
where ${\bf P}^{(0)}$ and ${\bf P}^{(2)}$ satisfy the Jacobi identity,
but ${\bf P}^{(1)}$ does not.

The Casimir of the Poisson subalgebra generated
by the $J_i$
is again $\cal K$ (the trace of the monodromy matrix),
but in this case
it is not clear how to split
the Casimir into four pieces that Poisson commute,
as required for Liouville integrability.
Of course, there are still
seven functionally independent invariant functions
(built from cyclically symmetric combinations of the $J_i$)
and we expect that four commuting functions exist.
}\eex  %

\begin{figure}[htb]
\centering
\psfrag{1}{$1$}\psfrag{2}{$2$}\psfrag{3}{$3$}\psfrag{4}{$4$}\psfrag{5}{$5$}\psfrag{6}{$6$}
\subfigure[
Example \ref{ex-p4142}.]{
\includegraphics[width=3.5cm]{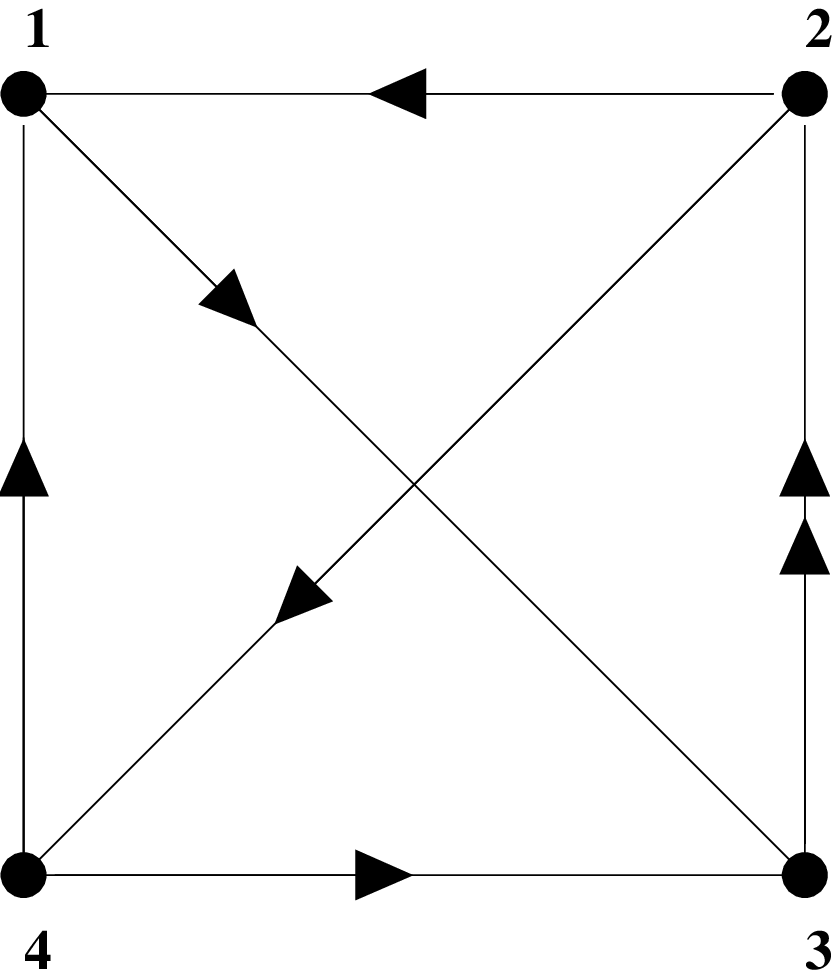}\label{subfig:p4142fig}
} \qquad\qquad
\subfigure[
Example \ref{ex-p6163}.]{
\includegraphics[width=4cm]{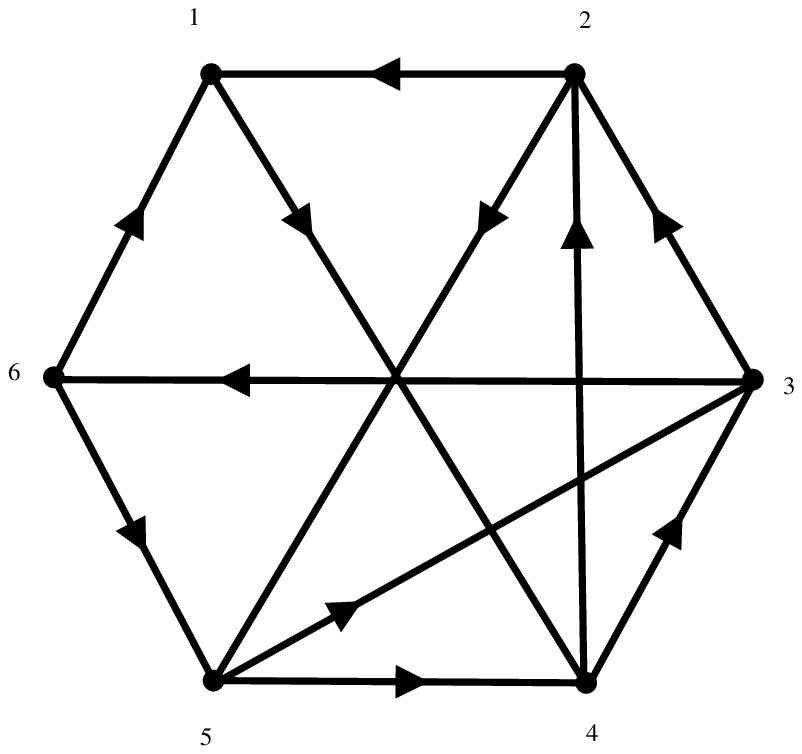}\label{subfig:p6163fig}
} \qquad\qquad
\subfigure[
Example \ref{ex-p6263}.]{
\includegraphics[width=4cm]{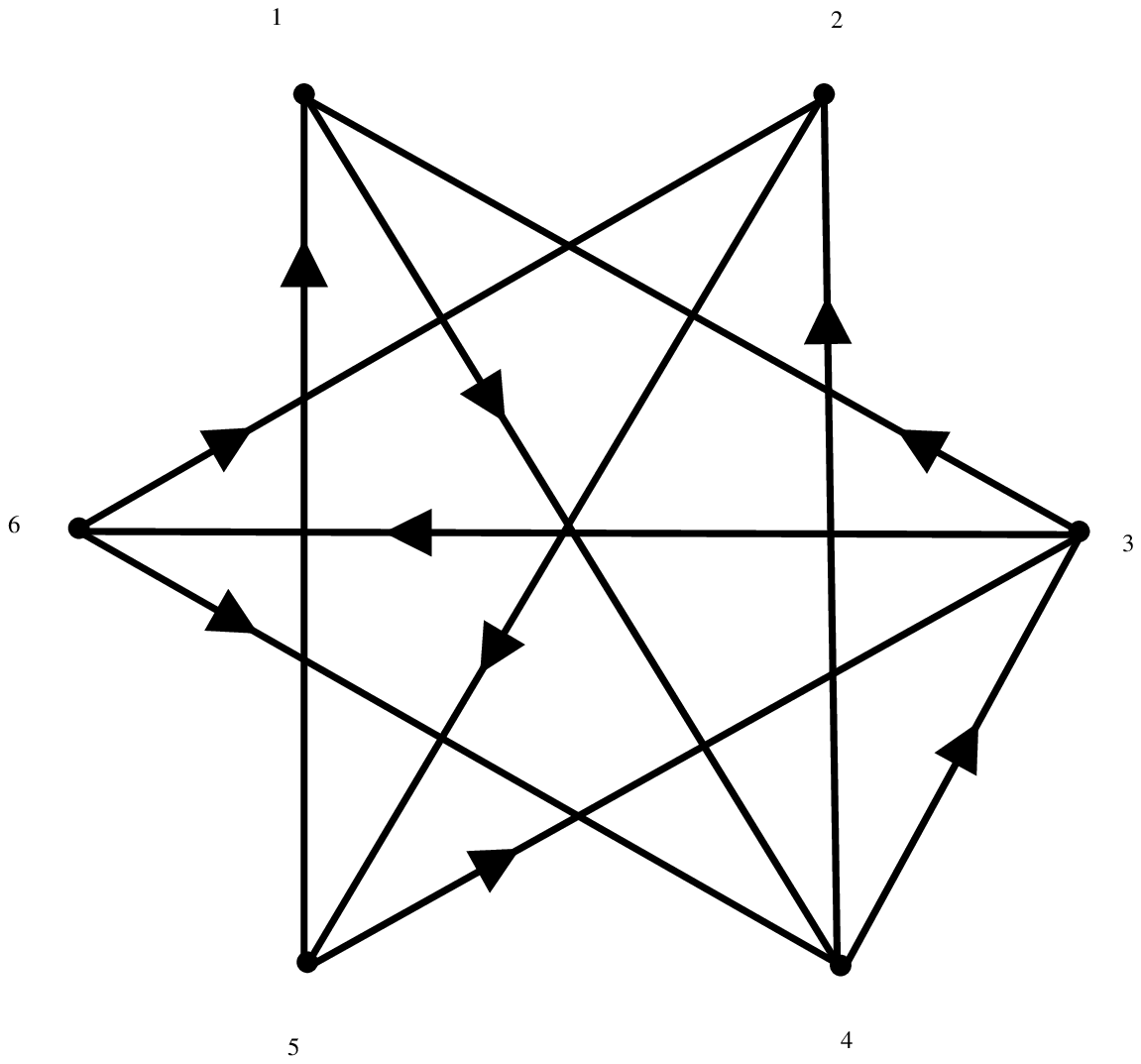}\label{subfig:p6263fig}
} \caption{The first three quivers in this class.}\label{quivers}
\end{figure}


\bex[The quiver $P_{8}^{(3)}-P_{8}^{(4)}+P_{2}^{(1)}$]
\label{ex-p8384}  {\em   %
For this quiver, with $p=5$ and $q=3$, the recurrence is
$$
x_nx_{n+8}=x_{n+3}x_{n+5}+x_{n+4}.
$$
The corresponding birational map $\varphi$
is Poisson with respect to a log-canonical bracket (\ref{logcan})
defined by the inverse of the associated matrix $B$.
Upon setting $C=2B^{-1}$,
the bracket is given by the first row of this Toeplitz matrix:
$
(c_{1,j})= (0,1,1,0,-1,1,2,1).
$

Since $\mbox{gcd}(p,q)=1$,  we have the same phenomenon as in the discussion of
(\ref{kpq}). From (\ref{mdy31}), we have
$$
{\bf M}_n:=  {\bf L}_{n}{\bf L}_{n+3}{\bf L}_{n+6}{\bf L}_{n+9}{\bf L}_{n+12}= {\bf L}_{n}{\bf L}_{n+3}{\bf L}_{n+1}{\bf L}_{n+4}{\bf L}_{n+2},
$$
since ${\bf L}_n$ has period 5, so the permutation
$\sigma :\, (1,2,3,4,5)\mapsto (1,4,2,5,3)$ gives the trace as
$$
{\cal K}={\cal K}_{5,3}(J_1,J_2,J_3,J_4,J_5)=
{\cal K}_{5,1}(J_1,J_4,J_2,J_5,J_3),
$$
where ${\cal K}_{5,1}$ is given by the formula (\ref{k51})
in Example \ref{ex-p6163}. Similarly, from (\ref{mdy32}) we find
$$
\hat{{\bf M}}_n:=  \hat{\bf L}_{n+10}\hat{\bf L}_{n+5}\hat{\bf L}_n=\hat{\bf L}_{n+1}\hat{\bf L}_{n+2}\hat{\bf L}_n,
$$
where $\hat{\bf L}_n$ has period 3, so with the same notation
as in (\ref{kpqhat}) we also have
$$
{\cal K} = {\cal K}_{3,5}(K_1,K_2,K_3)= {\cal K}_{3,1}(K_1,K_3,K_2),
$$
where ${\cal K}_{3,1}$ is given by (\ref{k31eq})
in Example \ref{ex-p4142}.
Since ${\cal K}_{3,1}$ is totally symmetric, the permutation
$\hat\sigma : \, (1,2,3)\mapsto (1,3,2)$
makes no difference to the result.

The explicit forms of the  period 5 functions $J_i$,
appearing as entries in the matrices ${\bf L}_n$,
all derive from
$$
 J_1=\frac{x_1x_3x_8+x_3x_5x_7+x_4x_6x_8+x_4x_7}{x_3x_4x_8}
$$
by acting with the map $\varphi$.
From the above bracket for the $x_j$, determined by the matrix $C$,
the Poisson brackets between the $J_i$ can be calculated as
$$
\{J_1,J_2\}=4J_1J_2,\quad \{J_1,J_3\}=2J_1J_3+2J_2-2,
$$
with all other brackets being deduced through the cyclic property.
The resulting 5-dimensional
Poisson subalgebra is isomorphic to that for the $J_i$ in
Example \ref{ex-p6163}, as can be seen by applying the permutation
$\sigma$ and rescaling by an overall factor of $2$
(which depends on the choice of scale for $c_{ij}$).
Therefore, subject to this permutation,
it follows that the same three functions $\CH_i$
are in involution:
$$
\CH_1=\sum_{i=1}^5(J_i-J_iJ_{i+3}),\quad
\CH_2=\sum_{i=1}^5(J_iJ_{i+1}J_{i+3}-J_iJ_{i+1}J_{i+3}^2),\quad
\CH_3=\prod_{i=1}^5J_i .
$$

The required fourth function that commutes
with these $\CH_i$ is derived from the algebra of the $K_i$,
which appear as entries in the matrices  $\hat{\bf L}_n$.
In terms of $x_j$, we have
$$
K_1 = \frac{x_1}{x_6}+\frac{x_6}{x_1}+ \frac{x_2}{x_3x_6}+
\frac{x_5}{x_1x_4}+ \frac{x_8}{x_4x_7}+ \frac{x_2x_8}{x_3x_7},
$$
with $K_2=\varphi^* K_1$ and $K_3=(\varphi^*)^2 K_1$ providing
the other two functions which cycle with period 3 under the
action of the map. The $K_i$ generate a 3-dimensional
Poisson subalgebra, whose brackets are all determined by acting with
$\varphi$ on the single relation
$$
\{K_1,K_2\}=-K_1K_2+2K_3.
$$
This algebra is isomorphic to that of the $J_i$ in
Example \ref{ex-p4142}
(as is seen immediately by applying the permutation
$\hat\sigma$), so it contains the two commuting
quantities
$$
\hat\CH_1=K_1^2+K_2^2+K_3^2,\qquad \hat\CH_2=K_1K_2K_3.
$$

The subalgebras of $J_i$ and $K_i$ share the joint Casimir
$
{\cal K} = \CH_1-\CH_2+\CH_3 = 3-\hat\CH_1+\hat\CH_2,
$
which gives a single relation between these two sets of functions.
Since $\{K_i,J_j\}=0$, for all $i,j$, we can take any four
of the first integrals $\CH_1,\CH_2,\CH_3,\hat\CH_1,\hat\CH_2$
as a commuting set, which demonstrates that this 8-dimensional
map $\varphi$ is integrable in the Liouville sense.
}\eex  %

\subsection{The case $\mbox{gcd}(p,q)=2$}

The  discussion of the case where $p$ and $q$ are both even
involves some new features. Upon setting $p=2\hat p$, $q=2\hat q$ with
$\mbox{gcd}(\hat{p},\hat{q})=1$, the fact that $\mbox{gcd}(m,q)=1$
implies that $m=\hat{p}+\hat{q}$ is odd,
hence either $\hat q$ is odd and $\hat p$ is even, or vice versa.
For the matrix (\ref{btau3})
in each case we find that  det$\,B=0$
whenever $\hat p$ or $\hat q$ is divisible by 3, and
det$\,B=9$ otherwise. These two possibilities lead to
quite different behaviour, so eventually we shall have to distinguish
between them. 
For the time being we concentrate on the associated linear relations,
which do not depend on whether $B$ is degenerate or not.

Since $q$ is even, observe that the ordered product of matrices
${\bf L}_n$ in (\ref{mdy31}) now cycles only through indices with the same
parity, so the product cycles twice through $\hat{p}=p/2$ terms. Thus
we see that ${\bf M}_n$ is a perfect square, and it is convenient
to take the square root
$$
{\bf M}_n^{1/2}= {\bf L}_n  {\bf L}_{n+q}\ldots {\bf L}_{n+(\hat{p}-1)q},
$$
and similarly for $\hat{\bf M}_n^{1/2}$.
We know that in this situation
the quantity ${\cal K}_n$ in (\ref{k32}) has period 2,
but for our purposes it is
more useful to consider the quantity
\beq\label{kstar}
\mathcal{K}^*_n:= \mathrm{tr}\, {\bf M}_n^{1/2} =
\mathrm{tr}\, \hat{\bf M}_n^{1/2},
\eeq
which cycles with period 2 for the same reasons.

In this setting, with $\mbox{gcd}(p,q)=2$, the algebraic structure
in terms of the functions
$J_n$ and $K_n$ is based on that for
the case $q=2$ (up to suitable permutations),
similarly to the way that for $\mbox{gcd}(p,q)=1$, and for the
primitives, this structure is based on the case $q=1$.

Now when $q=2$ we have the quantities $K_n$ with period 2,
giving
$$\mathrm{tr}\, \hat{\bf M}_n ^{1/2}=\mathrm{tr}\, \hat{\bf L}_n=K_{n+1},
\qquad
\mathrm{tr}\, \hat{\bf M}_n ^{-1/2}=\mathrm{tr}\, \hat{\bf L}_n^{-1}=K_{n},
$$
and hence the identity
\beq\label{per2tr}
\mathrm{tr}\, {\bf M}_n^{-1/2}=\mathrm{tr}\, {\bf M}_{n+1}^{1/2}=
\mathcal{K}^*_{n+1}
\eeq
holds. For even $q>2$, with $p$ fixed,
the formula for $\mathrm{tr}\, {\bf M}_n^{1/2}$ as a function
of $J_1,\ldots ,J_p$ is identical to that for the case $q=2$,
up to a permutation, which means that the formula (\ref{per2tr})
holds in general.

\begin{thm}\label{Klin3k}
When $\mbox{gcd}(p,q)=2$,
the iterates of the nonlinear recurrence (\ref{pqrec})
satisfy the linear relation
\beq\label{krec3k}
x_{n+3pq/2}- \mathcal{K}^*_n\,x_{n+pq}+\mathcal{K}_{n+1}^*\,x_{n+pq/2}-x_n=0,
\eeq
where  $\mathcal{K}_n^*$ is the period 2 quantity defined by (\ref{kstar}).
\end{thm}
\begin{prf}
This follows by essentially the same argument as in
the proof of Theorem \ref{Klin3}, applying Cayley-Hamilton to
${\bf M}_n^{1/2}$ and ${\bf M}_n^{-1/2}$, and making use of
(\ref{per2tr}).
\end{prf}
\begin{rem}\label{twocas}
{\em
With respect to the functions $J_n$ and $K_n$,
the main new feature in this case, compared with the
case $\mbox{gcd}(p,q)=1$, is that here we have
two quantities $\mathcal{K}_1^*$ and $\mathcal{K}_2^*$,
so the identity (\ref{kstar}) for $n=1,2$
provides two independent relations between these two sets of functions.
Given that these are the only relations, the cyclically symmetric functions
of the $J_n$,  and those of the $K_n$, together provide $N-2$
independent first integrals.
}
\end{rem}
We are now ready to present a further refinement of Proposition \ref{K3pq}.

\begin{thm} \label{Klin3cas}
When $\mbox{gcd}(p,q)=2$,
the iterates of the nonlinear recurrence (\ref{pqrec})
satisfy the constant coefficient, linear relation
\beq\label{krec3c}
x_{n+3pq}-{\cal B}\, x_{n+5pq/2} +\mathcal{C}\,x_{n+2pq}
-{\cal D}\, x_{n+3pq/2} + {\cal C}\, x_{n+pq}-\mathcal{B}\,x_{n+pq/2}+x_n=0,
\eeq
where  the non-trivial coefficients are first integrals, given by
$$
{\cal B} =\mathcal{K}_1^*+\mathcal{K}_2^* , \qquad
{\cal C} = \mathcal{K}_1^* \mathcal{K}_2^*+\mathcal{K}_1^*+\mathcal{K}_2^* ,
\qquad
{\cal D} =(\mathcal{K}_1^*)^2+(\mathcal{K}_2^*)^2+2.
$$
\end{thm}
\begin{prf} For the sake of argument, suppose that
$\hat q$ is odd and $\hat p$ is even.
To get the seven-term relation (\ref{krec3c}), 
we consider the $6\times 6$ matrix that is specified in terms of its
$(j,k)$ entry by $\Phi_n = (x_{n+p\hat{q}(j-1)+\hat{q}(k-1)}) $.
The linear recurrence (\ref{Prec}) implies that
$$
\Phi_{n+\hat q} = \Phi_n \, {\bf L}_n^*, \qquad \mathrm{where} \qquad
{\bf L}_n^*=\left(\bear{cccccc}
0 & 0 & 0 & 0 & 0 & 1 \\
1 & 0 & 0 & 0 & 0 & 0 \\
0 & 1 & 0 & 0 & 0 & -J_n \\
0 & 0 & 1 & 0 & 0 & 0 \\
0 & 0 & 0 & 1 & 0 & J_{n+m} \\
0 & 0 & 0 & 0 & 1 & 0
\eear \right).
$$
The monodromy matrix corresponding to $p$ iterations
of the latter equation is
$$
{\bf M}_n^* =  {\bf L}_n^* \, {\bf L}_{n+\hat q}^*\ldots
{\bf L}_{n+(p-1)\hat q}^*,
$$
so that
\beq\label{mstar}
\Phi_{n+p\hat q} = \Phi_n \, {\bf M}_n^*.
\eeq
We need to show that the characteristic polynomial of
the $6\times 6$ matrix $ {\bf M}_n^*$ is given by
\beq\label{chi}
\chi (\zeta )
= \zeta^6 -{\cal B}\, \zeta^5 +{\cal C}\, \zeta^4 - {\cal D}\, \zeta^3
+ {\cal C}\, \zeta^2 - {\cal B}\, \zeta + 1,
\eeq
as once we have determined this the recurrence  (\ref{krec3c})
follows immediately, by applying
the Cayley-Hamilton theorem to (\ref{mstar}).

It is convenient to conjugate all of the $6\times 6$ matrices
by the permutation matrix corresponding to
$(1,2,3,4,5,6)\rightarrow (1,3,5,2,4,6)$, which reduces
everything to calculations with $3\times 3$ blocks.
This gives
$$
{\bf L}_n^* \sim
\left(\bear{cc}  \mathbf{0} & {\bf L}_n \\  \mathbf{1} &  \mathbf{0}
\eear \right), \qquad
{\bf L}_n^* \, {\bf L}_{n+\hat q}^*\sim
\left(\bear{cc}  {\bf L}_n & \mathbf{0}   \\  \mathbf{0} &  {\bf L}_{n+\hat{q}}
\eear \right),
$$
so that that for the monodromy matrix we have 
$$
{\bf M}_n^* \sim
\left(\bear{cc}
{\bf L}_n  {\bf L}_{n+q}\ldots {\bf L}_{n+(\hat{p}-1)q} & \mathbf{0}
\\  \mathbf{0} &
{\bf L}_{n+\hat{q}} {\bf L}_{n+\hat{q}+q}\ldots {\bf L}_{n+\hat{q}+(\hat{p}-1)q}
\eear \right)
\sim
\left(\bear{cc}    {\bf M}_n^{1/2} & \mathbf{0}   \\  \mathbf{0} &
{\bf M}_{n+1}^{1/2}
\eear \right).
$$
Thus the characteristic polynomial of ${\bf M}_n^*$
factors as the product of the  characteristic polynomials
of ${\bf M}_n^{1/2}$ and ${\bf M}_{n+1}^{1/2}$,
$$
\chi (\zeta ) = (\zeta^3 - {\cal K}_1^*\, \zeta^2 + {\cal K}_2^*\, \zeta -1)\,
(\zeta^3 - {\cal K}_2^*\, \zeta^2 + {\cal K}_1^*\, \zeta -1),
$$
which multiplies out to give (\ref{chi}) with
the correct coefficients ${\cal B}, {\cal C},{\cal D}$.

An analogous argument holds when $\hat q$ is even and $\hat p$ is odd.
\end{prf}

To discuss the Liouville integrability of the systems
that appear when $\mbox{gcd}(p,q)=2$, it is necessary to
give a separate treatment according to whether the matrix
$B$ is invertible or not.

\subsubsection{Nondegenerate $B$ matrix}

When matrix $B$ is  nondegenerate, it appears
that 
the Liouville integrability of the symplectic map $\varphi$
should follow by very similar arguments to those for the case
$\mbox{gcd}(p,q)=1$ (or for the primitives with even $N$).
The map  preserves
a Poisson bracket of the form (\ref{logcan}),
which is specified uniquely (up to scale) by the Toeplitz matrix $C=B^{-1}$.
There are the two sets of quantities, $J_n$ and $K_n$, with periods
$p$ and $q$ respectively, and by Remark \ref{twocas} these
provide $N-2$ independent first integrals,
but it is necessary to find $m=N/2$ integrals in involution.

Given that the $J_n$ generate a Poisson subalgebra of dimension $p$,
and that both of the quantities ${\cal K}_1^*$ and
${\cal K}_2^*$ defined by (\ref{kstar}) are Casimirs, with the symplectic
leaves  being of dimension $p-2=2(\hat{p}-1)$, a further $\hat{p}-1$
independent commuting functions of the $J_n$ are required in order
to define an integrable
system on this subalgebra. Similarly, given that
$\{J_i, K_j\}=0$ for all $i,j$, and
the $K_n$ produce a $q$-dimensional subalgebra, with the same functions
${\cal K}_1^*$ and ${\cal K}_2^*$ as Casimirs,
it is necessary to have an additional $\hat{q}-1$ independent
functions that define an integrable system in terms of $K_n$ alone.
The quantities ${\cal B}$  and ${\cal C}$ which appear as
the coefficients in the linear relation (\ref{krec3c}) are first integrals,
as well as being joint Casimirs for the two subalgebras, so combining
these with the two sets of first integrals gives a total
of $2+( \hat{p}-1) +(\hat{q}-1) =m$ independent functions
in involution, as required.

Since we do not have a general proof of all the foregoing
assertions, here we will merely illustrate the discussion
with the simplest example of this kind, which arises
for $p=4$, $q=2$.

\bex[The quiver $P_{6}^{(2)}-P_{6}^{(3)}+P_{2}^{(1)}$] \label{ex-p6263} {\em
Mutation of the third quiver in Figure 1 gives the map
\beq\label{p4q2}
\varphi : \qquad (x_1,x_2,x_3,x_4,x_5,x_6)
\mapsto  (x_2,x_3,x_4,x_5,x_6,x_7), \qquad
x_7=  \frac{x_3x_5+x_4}{x_1}.
\eeq
The inverse of the  non-singular matrix $B$ defines a
Poisson bracket (\ref{logcan}), invariant with respect to $\varphi$.
Taking $C=3B^{-1}$, the top row of this Toeplitz matrix is
$
     (c_{1,j})=(0,1,1,0,2,2) .
$

The period 4 functions $J_i$, generated from $J_1$ by the action
of the map, are
\bea  %
&& J_1=\frac{x_1x_2x_6+x_2x_4x_5+x_3x_4x_6+x_3x_5}{x_2x_3x_6},\quad
J_2=\frac{x_1x_6+x_2x_3+x_4x_5}{x_3x_4}, \nn\\
&& J_3=\frac{x_1x_3x_4+x_2x_3x_5+x_1x_5x_6}{x_1x_4x_5},\quad
J_4=\frac{x_1x_3x_5+x_2x_4x_6+x_1x_2x_4x_5+x_1x_3x_4x_6+
    x_2x_3x_5x_6}{x_1x_2x_5x_6}.  \nn
\eea  %
The Poisson brackets between these functions all
follow by the cyclic property
from
\beq\label{jp4q2}
\{J_1,J_2\}=3J_1J_2-3,\quad \{J_1,J_3\}=3(J_2-J_4).
\eeq
From the first equality in (\ref{kstar})  
we have the two functions
$$
{\cal K}_1^*=J_2J_4-J_1-J_3, \qquad {\cal K}_2^*=J_1J_3-J_2-J_4,
$$
which are both Casimir functions of the 4-dimensional
algebra generated by the $J_i$, but cycle with period 2 under
the map. Symmetric functions of ${\cal K}_1^*$
and ${\cal K}_2^*$ are first integrals as well as being Casimirs,
so we may take  the quantities
$$
{\cal B}={\cal K}_1^*+{\cal K}_2^*
, \qquad
%
{\cal C}={\cal K}_1^*{\cal K}_2^* + {\cal B}
,
$$
as in Theorem \ref{Klin3cas}. The homogeneous
components of ${\cal B}$ are automatically first integrals,
and they Poisson commute:
$$
{\cal B} =-\CH_1 +\CH_2
$$
where
$$
\CH_1=J_1+J_2+J_3+J_4,\quad \CH_2=J_1J_3+J_2J_4 ,
\quad\mbox{with}\quad
\{\CH_1,\CH_2\}=0.
$$
Hence, we may take $\CH_1, {\cal B},{\cal C}$ as
three independent first integrals in involution,
which proves Liouville integrability of the 6-dimensional
map (\ref{p4q2}).
Clearly there are other choices, for instance  $\CH_1, \CH_2,{\cal C}$,
which would do just as well.
}
\eex  %
\br {\em
Note that in the preceding example, since $q=2$ we have
$K_1={\cal K}_2^*$ and $K_2={\cal K}_1^*$, which
Poisson commute with each other, so the subalgebra of the
$K_i$ is trivial.
}
\er

\subsubsection{Degenerate $B$ matrix}

When $B$ is degenerate, which only happens when either
$p$ or $q$ is a multiple of 6, the
maps that arise have features that
make them much more like the
odd primitives $P_{2m+1}^{(q)}$ than
the other cases with even $N$.

For these particular cases, the matrix $B$ has a two-dimensional
kernel, which is spanned by two integer vectors of the form
$$
{\bf u}_1 =(1,1,0,-1,-1,0, \ldots )^T,
\qquad
{\bf u}_2 =(0,1,1,0,-1,-1, \ldots )^T,
$$
where in each vector the components continue to repeat the same blocks of six
numbers,
until the final block which is truncated (of length 2 or 4, since
$N=2m$ is even and not a multiple of 6).
Hence there is a two-parameter scaling group, which acts by
\beq\label{lascala2}
(x_1,x_2,x_3,x_4,x_5,x_6, \ldots )
 \rightarrow
(\la \,x_1,\la\, \mu \, x_2,\mu\, x_3,
\la^{-1}\, x_4,\la^{-1}\mu^{-1}\,x_5,\mu^{-1}\,x_6,
\ldots ), \qquad (\la ,\mu ) \in (\C^*)^2.
\eeq
This action extends to all the iterates $x_n$ of (\ref{pqrec});
the pattern repeats itself
on each successive block of six adjacent iterates.

Now $\mathrm{im}\, B$ is spanned by
\beq\label{degvec}
{\bf v}_j = {\bf e}_j -{\bf e}_{j+1}+{\bf e}_{j+2}, \qquad j=1,\ldots , 2m-2,
\eeq
where ${\bf e}_j$ is the $j$th standard basis vector.
Hence, by  Lemma \ref{symp}, the coordinates
\beq\label{degy}
y_j = \frac{x_j\,x_{j+2}} {x_{j+1}}, \qquad j=1,\ldots , 2m-2,
\eeq
are invariant under the scaling (\ref{lascala2}),
and the degenerate form (\ref{omega}) pushes
forward to a symplectic form (\ref{yform}) in dimension $2m-2$.
The coefficients of the latter are obtained from a
skew-symmetric  matrix
$\hat{B}=(\hat{b}_{jk})$, whose
inverse provides a nondegenerate Poisson bracket for the $y_j$, i.e. 
\beq\label{toep}
\{ \, y_j , y_k\,\} = \epsilon_{jk}\, y_j y_k , \qquad 1\leq j,k\leq 2m-2,
\eeq
where $\hat{B}^{-1} = (\epsilon_{jk})$ must be
a Toeplitz matrix,  because the coordinates
(\ref{degy}) transform as $\varphi^*y_j = y_{j+1}$ under the
map corresponding to (\ref{pqrec}).
Upon applying the rest of Theorem \ref{torusred}, we see that
this induces a symplectic map $\hat\varphi$
on the variables $y_j$.

To prove the Liouville integrability of all of the maps $\hat\varphi$
that arise in this way, we require 
a general expression for the coefficients $\epsilon_{jk}$
in (\ref{toep}), which is  presently lacking.
However, it is possible to give a plausible argument for the
counting of first integrals, which agrees with
all examples we have checked so far.

Suppose, for the sake of argument, that $6|q$, hence 6$\not | p$.
The scaling action  of $(\C^*)^2$ on
the $x_j$, as in (\ref{lascala2}), extends to an action
on the coefficients $J_n$ that appear in the linear
equation (\ref{Prec}). However, the indices of the terms
$x_{n+jq}$ for $j=0,1,2,3$ differ by multiples of 6,
which means that they all scale the same way, and hence
the period $p$ quantities $J_n$ are invariant under this
scaling, and can be expressed in terms of the $y_j$
given by (\ref{degy}). Given that the $J_n$ generate
a $p$-dimensional Poisson subalgebra with respect to the bracket
(\ref{toep}), with two Casimirs given by the quantities ${\cal K}_1^*$ and
${\cal K}_2^*$ as in (\ref{kstar}), a further $\hat{p}-1$
independent commuting functions of the $J_n$ are needed
to have an integrable
system defined on this subalgebra. Similarly, the scaling action
(\ref{lascala2}) extends to an action
on the coefficients $K_n$ in (\ref{Qrec}), but now the terms
$x_{n+jp}$ for $j=0,1,2,3$ do {\it not} all
differ by multiples of 6, so they scale differently.
This means that there is a non-trivial scaling
action of the two-parameter group $(\C^*)^2$
on the quantities $K_n$, $n=1,\ldots , q$, for which
there should be $q-2$ invariant monomials. The
invariant monomial functions of $K_n$,
which we denote by $w_i$ for $i=1,\ldots, q-2$, can
also be written in terms of the original variables $x_j$.
The fact that they are invariant under (\ref{lascala2}) means
that these $w_i$ can  be written as functions of the
symplectic coordinates $y_j$ as well. Given that the $w_i$
generate a $(q-2)$-dimensional Poisson subalgebra
in the $(2m-2)$-dimensional space with the bracket (\ref{toep}),
and that the quantities ${\cal K}_1^*$ and
${\cal K}_2^*$ are Casimirs for this subalgebra too,
an integrable system is defined on the $(q-4)$-dimensional
symplectic leaves by an additional set of $\hat{q} -2$
commuting functions of the $w_i$.
Supposing further that $\{J_i, w_j\}=0$ for all $i,j$,
we take the quantities ${\cal B}$  and ${\cal C}$
from (\ref{krec3c}), which are both first integrals
and joint Casimirs for the two subalgebras, and combining
these with the two sets of first integrals gives a total
of $2+( \hat{p}-1) +(\hat{q}-2) =m-1$ independent functions
in involution, as required
for the Liouville integrability of the symplectic map $\hat\varphi$
in dimension $2m-2$.

Since we do not have a general proof of all the preceding
assertions, here we will only present
the simplest example of this kind, which arises for $p=6$, $q=4$.

\bex[The quiver $P_{10}^{(4)}-P_{10}^{(5)}+P_{2}^{(1)}$] \label{ex-p6q4} {\em
The recurrence arising from mutation of this quiver is
\beq\label{p6q4}
x_{n+10}\, x_n = x_{n+6}\,x_{n+4} +x_{n+5}.
\eeq
The matrix $B$ has a two-dimensional kernel,  spanned by
the integer vectors
$$
{\bf u}_1 =
(1,1,0,-1,-1,0,1,1,0,-1)^T, \qquad
{\bf u}_2 = (0,1,1,0,-1,-1,0,1,1,0)^T ,
$$
which generate the action of a two-parameter scaling group
on the iterates of (\ref{p6q4}), as given in
(\ref{lascala2}). Taking the scaling-invariant
variables $y_j$, given by (\ref{degy}) for
$j=1,\ldots ,8$,  we apply Lemma \ref{symp}
to find the symplectic form $\hat\om$ expressed
in these coordinates, which leads to the
nondegenerate Poisson bracket specified by
\beq\label{yp6q4bra}
\{ y_1,y_5\} =y_1y_5, \qquad
\{ y_1,y_6\} =-y_1y_6, \qquad
\{ y_1,y_7\} =y_1y_7,
\eeq
where all other brackets are either zero or follow
from the Toeplitz property/skew-symmetry.
From Theorem \ref{torusred}, this Poisson bracket is preserved
by the induced map
\beq\label{yp6q4}
\hat\varphi : \qquad (y_1,\ldots ,y_{8})
\mapsto  (y_2,\ldots,y_{9}), \qquad
y_{9}=  \frac{y_4y_5y_6(y_5+1)}{y_1y_2y_8}.
\eeq

Note that (in contrast to the foregoing discussion) in this example
$q$ is {\it not} a multiple of 6, but rather $p$ is.
The period 6 functions $J_n$ are given 
by the formula
$$
J_n =\frac{x_n}{x_{n+4}} +\frac{x_{n+7}}{x_{n+3}}
+\frac{x_{n-1}x_{n+8}}{x_{n+3}x_{n+4}}, \quad n=1,\ldots ,6,
$$
where each of the above quantities can be written
as a function of $x_1,\ldots ,x_{10}$ by iterating the recurrence
(\ref{p6q4}) either forwards or backwards.
Since the gaps between the indices of the terms $x_j$
in the linear relation (\ref{Prec}) are not all multiples of 6,
these  $J_n$ are not invariant under the scaling action (\ref{lascala2}),
but instead they transform as follows:
$$
J_1\to \la^2\mu J_1, \quad 
J_2\to \la\mu^2 J_2,\quad 
J_3\to \la^{-1}\mu J_3, \quad
J_4\to \la^{-2}\mu^{-1} J_4, \quad
J_5\to \la^{-1}\mu^{-2} J_5, \quad
J_6\to \la\mu^{-1} J_6 .
$$
There are four independent scaling-invariant monomial functions
of the $J_n$, but it is convenient to consider the five
functions given by
$
w_1=J_1J_4, \;\; w_2=J_2J_5, \;\; w_3=J_3J_6,
\;\; w_4=J_1J_3J_5, \;\; w_5 = J_2J_4J_6,
$
which satisfy the single relation
\beq\label{wreln}
w_1w_2w_3=w_4w_5.
\eeq
Since these $w_i$ can also be expressed in terms of
$x_1,\ldots ,x_{10}$, the scaling-invariance implies that
they can be written as functions of the symplectic
coordinates $y_j$ as well. For instance, we have the formula
$$
w_1 =\frac{(y_1y_2y_8+y_4y_5y_6+y_5y_6y_8+y_5y_6)
(y_1y_2+y_1y_6+y_5y_6+y_1+y_6)}{y_1y_2y_5y_6y_8},
$$
and analogous formulae can be obtained for
$w_2=\hat{\varphi}^*w_1$, $w_3=(\hat{\varphi}^*)^2w_1$
using the map (\ref{yp6q4}); there are different
expressions for $w_4$ and  $w_5=\hat{\varphi}^*w_4$,
but these are more unwieldy so they are omitted.
Calculating the Poisson brackets between these functions,
we have a subalgebra with four generators $w_1,w_2,w_3,w_4$,
but this is more conveniently expressed with the
extra function $w_5$ included.  The brackets
can be determined from  the  relations
\beq\label{5dim}
\{w_1,w_2\} =w_1w_2 -w_4-w_5,
\qquad \{w_1,w_4\} = w_1(w_2-w_3),
\eeq
with all other brackets following by applying the map and
noting that $w_1$ cycles with period 3,
and $w_4$ has period 2,
so $(\hat{\varphi}^*)^3w_1=w_1$ and so $(\hat{\varphi}^*)^2w_4=w_4$.
The 5-dimensional algebra defined by
(\ref{5dim}) has three Casimirs, given by
$$
{\cal K}_1^*=3-w_1-w_2-w_3+w_5, \qquad
{\cal K}_2^*=3-w_1-w_2-w_3+w_4, \qquad
\hat{\cal C} = w_1w_2w_3-w_4w_5,
$$
where (as functions of $J_n$)
the quantities ${\cal K}_1^*$ and ${\cal K}_2^*$  come from (\ref{kstar}),
by taking the trace of the monodromy matrix ${\bf M}_n^{1/2}$
for $n=1,2$, while fixing $\hat{\cal C}=0$ corresponds to the
constraint (\ref{wreln}).
The Casimir function
$
{\cal B} =  {\cal K}_1^* + {\cal K}_2^* = 3-2\hat\CH_1+\hat\CH_2,
$
is also a first integral, with the components
$$
\hat\CH_1 = J_1J_4+J_2J_5+J_3J_6= w_1+w_2 +w_3, \qquad \hat\CH_2 =
J_1J_3J_5 +J_3J_4J_6 =w_4 +w_5,
$$
which are homogeneous functions
of the $J_n$, are also first integrals, and
Poisson commute: $\{ \hat\CH_1 ,\hat\CH_2 \}=0$.
Either of the latter two functions
defines an integrable system on the
two-dimensional symplectic leaves of the
algebra generated by the $w_i$.
The function ${\cal C} = {\cal K}_1^* {\cal K}_2^* +{\cal B}$
is another first integral that is also a Casimir.

The functions $K_n$, which cycle with period 4, are specified by the formula
$$
K_n= \frac{x_n}{x_{n+6}}+ \frac{x_{n+9}}{x_{n+3}}+ \frac{x_{n+1}}{x_{n+2}x_{n+6}}+ \frac{x_{n+8}}{x_{n+3}x_{n+7}}+ \frac{x_{n+1}x_{n+8}}{x_{n+2}x_{n+7}},
\qquad n=1,\ldots ,4.
$$
These quantities are  invariant under the scaling (\ref{lascala2}), so they can also be
written in terms of the iterates of (\ref{yp6q4}):
$$
K_n = \frac{y_{n+1}+y_{n+6}+y_ny_{n+1}+y_{n+1}y_{n+6}+y_{n+6}y_{n+7}}{y_{n+3}y_{n+4}},
\qquad n=1,\ldots ,4,
$$
where $\hat\varphi^*y_n=y_{n+1}$ defines the sequence of $y_n$ for all $n\in\Z$.
Upon using (\ref{yp6q4bra}) we find the relations
$$
\{K_1,K_2\} = -K_1K_2+1, \qquad \{K_1,K_3\}=-K_2+K_4,
$$
which provide all the Poisson brackets between the $K_n$ by applying the cyclic property.
Comparing with (\ref{jp4q2}) and scaling by a factor of $-3$, this four-dimensional Poisson
subalgebra is seen to be isomorphic to the algebra of the $J_i$ in Example \ref{ex-p6263}.
Hence the Casimirs of this subalgebra are
$$
{\cal K}_1^*=K_2K_4 -K_1-K_3,  \qquad
{\cal K}_2^*=K_1K_3 -K_2-K_4,
$$
and, with ${\cal B}=-\CH_1+\CH_2$,  two first integrals in involution are
$$
\CH_1=K_1+K_2+K_3+K_4, \qquad \CH_2=K_1K_3+K_2K_4.
$$
From (\ref{kstar}), these Casimirs are shared with the subalgebra generated by the $J_n$,
and the two different formulae for ${\cal B}$ imply that the first integrals are related according to
$$
3-2\hat\CH_1+\hat\CH_2  +\CH_1-\CH_2=0.
$$

Since $\{w_i,K_j\}=0$ for all $i,j$, the functions ${\cal B},{\cal C},\hat\CH_1,\CH_1$
provide four first integrals that Poisson commute, as required for Liouville integrability
of the 8-dimensional symplectic map (\ref{yp6q4}).
}
\eex

\section{Integrable maps from Somos sequences} \label{somosmaps}

\setcounter{equation}{0}

The quadratic recurrences (\ref{somosN}) (case (iv) of Theorem \ref{zeroe}), are referred to as
three-term Gale-Robinson recurrences \cite{gale, rob}.
We mow consider the slightly more
general case where these recurrences have coefficients:
\beq \label{GR}
x_{n+N}\, x_n = \alpha \, x_{n+N-p}\, x_{n+p} +\beta \,  x_{n+N-q}\, x_{n+q}.
\eeq

These can be included by adding extra nodes to the quiver for the coefficient-free
recurrence (see Section 10 in \cite{fordy_marsh}). The coefficients are
frozen variables, attached to these new nodes, which  do not change under mutations.
For three-term Gale-Robinson recurrences, one can add two extra nodes,
corresponding to the parameters $\al$, $\beta$, so a quiver with $N+2$ nodes is obtained
(Proposition 10.4. in \cite{fordy_marsh}).  It is then straightforward to check that the
presymplectic form $\om$  in (\ref{omega}), defined by
the same $N\times N$ skew-symmetric matrix $(b_{jk})$ as for the coefficient-free case,
is preserved by (\ref{GR}).  Hence Theorem \ref{torusred} can be applied directly to
the latter, to obtain a reduced symplectic map for suitable variables $y_j$.

Below we outline two different approaches to showing that the map for the variables $y_j$ is integrable.
The first way is to use the fact that all of the recurrences (\ref{GR}) are  ordinary difference equations
that arise as reductions of the Hirota-Miwa equation, which is an integrable partial difference
equation with three independent variables. The Lax pair of the Hirota-Miwa equation
allows one to obtain Lax pairs for its reductions, and the associated spectral curves
provide first integrals in terms of the $y_j$. The second way is to find Somos-type recurrences
of higher order that are satisfied by the iterates of (\ref{GR}). The coefficients of these Somos-$k$
relations, for certain $k>N$, can also provide first integrals (analogous to the first integrals which appear
as coefficients in {\it linear} recurrence relations for the iterates of the families (ii) and (iii)).

Recently, Goncharov and Kenyon have found Somos recurrences arising as discrete symmetries
of classical integrable systems associated with dimer models on a torus \cite{gk}.
A further connection with relativistic analogues of the Toda lattice appeared in \cite{eager}.

\subsection{Reductions of the Hirota-Miwa equation}

The  relations (\ref{GR})
all arise by reduction of the Hirota-Miwa (discrete KP) equation, which is
the bilinear partial difference equation
\beq\label{dkp}
T_{1}\, T_{-1}=T_{2}\, T_{-2}+T_{3}\, T_{-3}.
\eeq
In the above, $T=T(n_1,n_2,n_3)$ is a function of three independent variables,
and to denote shifts we have used
$T_{\pm j} = T|_{n_j\to n_j\pm 1}$. If we set
\beq\label{redntau}
T(n_1,n_2,n_3)=\exp \left(\sum_{i,j}S_{ij}n_in_j\right)\, \uptau (n),
\eeq
where
$S=(S_{ij})$ is a symmetric matrix and $n=n_0+\updelta_1n_1+\updelta_2n_2+\updelta_3n_3$, then  $\uptau(n)$
satisfies the ordinary difference equation
$$
\uptau (n+\updelta_1)\uptau(n-\updelta_1)=\alpha \, \uptau(n+\updelta_2)\uptau(n-\updelta_2) +\beta\,  \uptau(n+\updelta_3)\uptau(n-\updelta_3)
$$
where $\alpha = \exp(2(S_{22}-S_{11}))$, $\beta =\exp(2(S_{33}-S_{11}))$. Upon taking
$x_n=\uptau (n-\updelta_1)$ with $\updelta_1=\frac{1}{2}N$, $\updelta_2= \frac{1}{2}(N-2p)$,
$\updelta_3= \frac{1}{2}(N-2q)$, this becomes (\ref{GR}).

In the combinatorics literature, equation (\ref{dkp}) is referred to as the octahedron recurrence,
which has the Laurent property (shown in \cite{fz}).  The Laurent property for three-term Somos
(or Gale-Robinson) recurrences of the form (\ref{GR}) then follows by the reduction (\ref{redntau}).

The Hirota-Miwa equation (\ref{dkp})  has a scalar Lax pair (see equation (3.8) in \cite{zabrodin},
for instance): it is the compatibility condition for the linear system given by
\beq \label{kplax}
\bear{rcrcr}
 T_{-1,3}\, \psi_{1,2} & + & T\, \psi_{2,3} & = & T_{2,3}\, \psi , \\
T\, \psi_{-1,2} & + & T_{-1,3}\, \psi_{2,-3} & = & T_{-1,2}\, \psi ,
\eear
\eeq
in terms of the scalar function $\psi = \psi (n_1,n_2,n_3)$,
with the same notation for shifts as before.
Using the latter, one can use the reduction (\ref{redntau})
to obtain Lax pairs for all of the Somos recurrences
(\ref{GR}), which leads directly
to spectral curves whose coefficients are
conserved quantities. Here we briefly illustrate how this works for the cases $N=4$ and $N=5$ only;
reducing the Lax pair becomes more involved as $N$ increases.

The general Somos-4 recurrence with coefficients is
\beq\label{s4ab}
x_{n+4}\, x_n = \al \, x_{n+3}\, x_{n+1} + \beta x_{n+2}^2.
\eeq
By taking the same monomials as in (\ref{s4yj}), this reduces to the map
\beq\label{s4mapab}
(y_1\, , \, y_2) \mapsto \Big( y_2\, ,\, (\alpha y_2 +\beta)/(y_1y_2^2) \Big) ,
\eeq
which becomes (\ref{s4map}) in the case
$\alpha=\beta =1$, and
preserves the same symplectic from $\hat\om$.  This means that
(\ref{s4mapab}) has the invariant
Poisson bracket
\beq\label{s4bra}
\{ y_1\, ,\, y_2\} =y_1\, y_2.
\eeq

The recurrence (\ref{s4ab}) arises from the reduction (\ref{redntau}) with $\updelta_1=2$,
$\updelta_2=1$, $\updelta_3=0$.
Upon taking $S_{jk}=0$ for $j\neq k$, without loss of generality, and setting
$$
\psi (n_1,n_2,n_3) = \exp \Big(\sum_{j=1}^3 n_j\log\lambda_j+S_{jj}n_j^2\Big)\,  \uptau (n)\, \phi (n),
\qquad \mathrm{with }\qquad \lambda_1 = \frac{e^{-S_{11}}}{\sqrt{\zeta \xi} }, \quad \la_2 = \la_1\zeta,
\quad \la_3 = e^{2S_{22}}\, \la_2,
$$
the scalar linear equations (\ref{kplax}) reduce to a $2\times 2$  linear system for
the vector ${\bf w} = (\phi (n),\phi (n+1))^T$. Up to shifts 
of indices,  the coefficients of the latter can
all be written in terms of the $y_j$  given in (\ref{s4yj}), as well as the spectral parameters
$\zeta$ and $\xi$, as follows.
\bex[Somos-4 Lax pair]
\label{s4lax}
{\em
The
Lax pair for the map (\ref{s4mapab}) takes the form 
\beq \label{laxs4}
{\bf L} \, {\mathbf w} = \xi {\mathbf w}, \qquad
\tilde{{\mathbf w}} ={\bf M} \, {\mathbf w} ,
\eeq
where the tilde denotes the shift $n\to n+1$. The matrices ${\bf L}={\bf L}(\zeta )$,
${\bf M}={\bf M}(\zeta )$
are functions of $y_j$ and the spectral parameter $\zeta$,  given by
$$
{\bf L} = \left(\bear{cc}
-\frac{(\al y_1+\beta)}{y_1y_2} \, \zeta & -\al y_1 \zeta + \frac{(\al y_1+\beta)}{y_1y_2}  \\
\frac{\al}{y_1}\, \zeta^2 -\zeta & \left(-y_1y_2-\frac{\al}{y_1}\right)\, \zeta +1
\eear \right),
\qquad
{\bf M} = \left(\bear{cc} 0 & 1 \\ -\frac{1}{y_1y_2}\, \zeta & \frac{1}{y_1y_2} \eear\right).
$$
The discrete Lax equation $\tilde{\bf L} {\bf M}={\bf M} {\bf L}$ holds if and only if
the map (\ref{s4mapab}) does.  The spectral curve is
$$
\mathrm{det}\, ({\bf L}(\zeta ) - \xi \, {\mathbf 1}) = \xi^2+(H_1\, \zeta -1)\xi+ \al^2\zeta^3+\beta\zeta^2 =0,
$$
in which the coefficient of $\zeta \xi$ is the first integral
\beq\label{s4ham}
H_1=y_1y_2+\frac{\alpha}{y_1}+\frac{\alpha}{y_2}+\frac{\beta}{y_1y_2}.
\eeq
The level sets of $H_1$ are biquadratic curves of genus one in the $(y_1,y_2)$ plane, and the map is
a particular instance of the QRT family \cite{qrt1}.
}
\eex

Applying the reduction (\ref{redntau}) to the Hirota-Miwa equation (\ref{dkp}) with $\updelta_1=5/2$,
$\updelta_2=3/2$, $\updelta_3=1/2$ leads to the general form of the Somos-5 recurrence
with coefficients, which we denote by $\tilde{\alpha}$, $\tilde{\beta}$:
\beq
\label{s5ab}
x_{n+5}\, x_n =\tilde{\alpha}, x_{n+4}\, x_{n+1}+ \tilde{\beta}\, x_{n+3}\, x_{n+2}.
\eeq
From the appropriate $B$ matrix, one obtains $y_1=x_1x_4/(x_2x_3)$,
$y_2=x_2x_5/(x_3x_4)$ as coordinates in the plane (see \cite{sigma}), satisfying the Poisson bracket (\ref{s4bra}),
and the corresponding map $\hat\varphi$ 
is also of QRT type, namely
\beq\label{s5mapab}
\hat\varphi : \qquad
(y_1\, , \, y_2) \mapsto \Big( y_2\, ,\, (\tilde{\alpha} y_2 +\tilde{\beta})/(y_1y_2) \Big) .
\eeq
Similar calculations to those in the Somos-4 case yield the appropriate reduction of (\ref{kplax}).
\bex[Somos-5 Lax pair]
\label{s5lax}
{\em
The map (\ref{s5mapab}) arises as the compatibility condition $\tilde{\bf L} {\bf M}={\bf M} {\bf L}$
for a linear system of the form (\ref{laxs4}), where
\beq\label{s5lm}
{\bf L} ={\bf C}_0 +{\bf C}_1\, \zeta + {\bf C}_2 \, \zeta^2, \qquad
 {\bf M} ={\bf C}_0 + \left(\begin{array}{cc} 0 & 0 \\
-y_1 & 0 \end{array} \right)\, \zeta,
\eeq
$$ \mathrm{with }\qquad
{\bf C}_0=\left(\begin{array}{cc} 0 & 1 \\
0 & 1 \end{array} \right) , \quad {\bf C}_1=\left(\begin{array}{cc} -y_1 & -\left(y_2+\frac{\tilde{\alpha}}{y_1}\right) \\
-y_1 &  -\left(y_2+\frac{\tilde{\alpha}}{y_1}+\frac{\tilde{\alpha}}{y_2}+\frac{\tilde{\beta}}{y_1y_2}\right)\end{array} \right) , \quad {\bf C}_2= \left(\begin{array}{cc} \tilde{\alpha} & 0 \\
\tilde{\alpha}+\frac{(\tilde{\alpha}y_1+\tilde{\beta})}{y_2} & \tilde{\alpha} \end{array} \right).
$$
The coefficient of $\zeta \xi$ in the equation for the spectral curve, that  is
$$
\mathrm{det}\, ({\bf L}(\zeta ) - \xi \, {\mathbf 1}) = \xi^2-(2\tilde{\alpha}\zeta^2-\tilde{J}\, \zeta +1)\xi+ \tilde{\alpha}^2\zeta^4+\tilde{\beta}\zeta^3 =0,
$$
gives a first integral whose level sets are cubic (also biquadratic) curves of genus one, that is
\beq\label{s5j}
\tilde{J}=y_1+y_2+\tilde{\alpha}\left(\frac{1}{y_1}+ \frac{1}{y_2}\right) + \frac{\tilde{\beta}}{y_1y_2}.
\eeq
}
\eex
\begin{rem}\label{hgsint}{\em
The first integral  (\ref{s4ham}) for Somos-4 can be rewritten in terms of the cluster variables, so
that it becomes a ratio  of homogeneous polynomials of total degree 4 in $x_1,x_2,x_3,x_4$,
with the denominator just being $x_1x_2x_3x_4$.
Similarly, in the case of Somos-5 the first integral (\ref{s5j}) can be rewritten as a ratio of  homogeneous polynomials of
degree 5, with the denominator $x_1x_2x_3x_4x_5$. It turns out that  (\ref{s5ab}) has another rational
first integral, also of degree 5 in terms of the cluster variables, which can be written 
as
$$
\tilde{I} =f_1f_2f_3+  \tilde{\alpha}\left(\frac{1}{f_1}+ \frac{1}{f_2}+ \frac{1}{f_3}\right)
+\frac{\tilde{\beta}}{f_1f_2f_3} \qquad \mathrm{with}  \quad f_j=\frac{x_jx_{j+2}}{x_{j+1}^2}
\qquad  \mathrm{for} \quad  j=1,2,3
$$
(see Proposition 2.3 in  \cite{hones5}). However, the quantity $\tilde{I}$ is not defined
on the $(y_1,y_2)$ plane, where there is only one first integral (as required
for the Liouville-Arnold theorem).
}
\end{rem}

\subsection{Bilinear relations of higher order}

In   \cite{swartvdp}, Swart and van der Poorten proved that sequences generated by Somos-4 recurrences also
satisfy quadratic  (Somos-type) relations of order $k$, for all $k\geq 4$. They also
noted that for Somos-5 sequences, there are Somos-$k$ relations of all odd orders
$k=5,7,9,\ldots$. Moreover, the coefficients of the higher order relations are constant along orbits,
which means that, as long as they are not trivially constant, they provide first integrals.

In this subsection we explain how to obtain first integrals for Somos-7 recurrences using associated quadratic
(bilinear) relations of higher order. The analogous results for Somos-6 recurrences are in \cite{hones6}.

To present the results concisely, it is convenient to consider the most general form of a Somos-7 recurrence,
which is the four-term Gale-Robinson relation
\beq\label{s7gen}
x_{n+7}\, x_n = \upalpha \, x_{n+6}\, x_{n+1} +  \upbeta \, x_{n+5}\, x_{n+2} +\upgamma \, x_{n+4}\, x_{n+3}.
\eeq
With all three terms on the right hand side, this
does not arise from a cluster algebra. Nevertheless, the general   Somos-7  recurrence can be
obtained as a reduction of the cube recurrence (Miwa's equation), and in \cite{fz} this was
used to prove the Laurent property for all four-term Gale-Robinson recurrences, including (\ref{s7gen}).
In each of the cases where one of the parameters $\upalpha,\upbeta,\upgamma$
vanishes, (\ref{s7gen}) reduces to a bilinear relation with two terms on the right hand side,
corresponding to a different cluster algebra in each case.

Our main result on the family of recurrences (\ref{s7gen}) is stated as follows and the remainder of this
subsection is devoted to its proof.
\begin{thm} \label{s7ints}
The Somos-7 recurrence (\ref{s7gen}) has three independent first integrals,
denoted $\mathcal{H}_1,\mathcal{H}_2,\hat{I}$,
which are  rational  functions (in fact, Laurent polynomials) of degree 7 in $x_1,\ldots, x_7$.
$\hat{I}$ can be written
as
\beq\label{I}
\begin{array}{rcl}
\hat{I} & =  & \upbeta \, f_1f_2f_3f_4f_5 +\upgamma (f_1f_2f_3 + f_2f_3f_4+f_3f_4f_5)
+\upalpha \upbeta \left( \frac{1}{f_1}+ \frac{1}{f_2}+\frac{1}{f_3}+\frac{1}{f_4}+\frac{1}{f_5}\right)
\\
&&
+ \upalpha \upgamma \left(\frac{1}{(f_1f_2f_3} + \frac{1}{ f_2f_3f_4}+\frac{1}{f_3f_4f_5} \right)
+\frac{\upbeta^2} {  f_1f_2f_3f_4f_5 }
+\upbeta \upgamma\left( \frac{1}{f_1f_2f_3^2f_4^2f_5} +\frac{1}{f_1f_2^2f_3^2f_4f_5} \right)
+\frac{\upgamma^2}{f_1f_2^2f_3^3f_4^2f_5} ,
\end{array}
\eeq
in terms of the variables $f_j=x_jx_{j+2}/x_{j+1}^2$, $j=1,2,3,4,5$,
while $\mathcal{H}_1$ and $\mathcal{H}_2$ are given in terms of $y_j = x_{j}x_{j+3}/(x_{j+1}x_{j+2})$, as
in Proposition \ref{higherbil} below. For $\upalpha =0$ or $\upgamma=0$, Somos-7
reduces to a Liouville integrable map $\hat\varphi$ in four dimensions, while for $\upbeta =0$ it reduces
to an integrable map of the plane. 
\end{thm}

The equation (\ref{s7gen}) admits the action of the two-parameter scaling group
$(\C^*)^2$, via
\beq\label{2torus}
x_n \rightarrow \la \, \mu^n\, x_n
\eeq
for non-zero $\la ,\mu$. The variables $f_j$ are invariants
under this scaling symmetry. In terms of these variables, the Somos-7 recurrence
 (\ref{s7gen}) is transformed
to a recurrence of fifth order, namely
\beq\label{s7frec}
f_{n+5}f_{n+4}^2f_{n+3}^3f_{n+2}^3f_{n+1}^2f_{n}=\upalpha \, f_{n+4}f_{n+3}^2f_{n+2}^2f_{n+1}
+\upbeta \, f_{n+3}f_{n+2}+\upgamma.
\eeq
Being of odd order, (\ref{s7gen}) has a further scaling symmetry depending on the parity of $n$:
\beq \label{extrasym}
x_n\rightarrow \nu^{(-1)^n}x_n, \qquad \nu \in\C^*.
\eeq
The variables $f_n$ have the symmetry $f_n\rightarrow \nu^{\pm 2}f_n$ for
even/odd $n$ respectively. The variables $y_j=f_jf_{j+1}$ are the invariants
under the combined scaling group and lead to
\beq\label{s7yrec}
y_{n+4}y_{n+3}y_{n+2}^2 y_{n+1}y_{n}=\upalpha\, y_{n+3}y_{n+2} y_{n+1}
+\upbeta \, y_{n+2}+\upgamma.
\eeq
Of the cluster algebra subcases, we shall see that the case
$\upalpha =0$ and the case $\upgamma =0$, the 4D map defined by (\ref{s7yrec}) is symplectic,
while for $\upbeta =0$, additional scaling symmetries allow reduction to the plane.

The quantities $\mathcal{H}_1,\mathcal{H}_2,\hat{I}$ in Theorem \ref{s7ints} can all
be written in terms of $f_j$, so the recurrence (\ref{s7frec}) has
three independent first integrals. The rational function $\hat{I}$ is not invariant under (\ref{extrasym}),
which means that it does not provide a first integral for (\ref{s7yrec}),
but both  $\mathcal{H}_1$ and $\mathcal{H}_2$ do.

The quantities $\mathcal{H}_1$ and $\mathcal{H}_2$
appear in coefficients of bilinear relations of higher order. Since the Somos-7 recurrence
is invariant under the three-parameter family of scalings defined by (\ref{2torus}) and
(\ref{extrasym}), one expects there to be relations of odd order having the same symmetry.
The first non-trivial relation is the Somos-11 recurrence (\ref{s11}) below.  It can be seen that
a combination of $\mathcal{H}_1$ and $\mathcal{H}_2$
appears in one of the coefficients.  Since this coefficient remains constant along each orbit of
(\ref{s7gen}), it provides a non-trivial first integral.  A second independent integral is provided by
the Somos-13 recurrence (\ref{s13}), given in the following.
\begin{propn}\label{higherbil}
The iterates of the Somos-7 recurrence (\ref{s7gen}) also satisfy the
Somos-11 recurrence
\beq\label{s11}
\bear{rcl}
x_{n+11}\, x_n & = & -\upbeta \, x_{n+10} \, x_{n+1}+\upgamma(\upgamma-\upalpha^2)x_{n+8}\,x_{n+3}
-\upalpha \upbeta \upgamma\,  x_{n+7}\, x_{n+4}
\\
&&
+(\upalpha^5+2\upalpha \upgamma^2 +2\upbeta^3+\upbeta \mathcal{H}_1+\upalpha^2 \mathcal{H}_2)\,
x_{n+6}\, x_{n+5} ,
\eear
\eeq
as well as the Somos-13 recurrence
\beq\label{s13}
x_{n+13}\, x_n=-\upbeta \upgamma \, x_{n+11}\,x_{n+2}
+\upGamma\, x_{n+9}\, x_{n+4}+ \upDelta\, x_{n+8}\, x_{n+5}
+\upTheta \, x_{n+7}\,x_{n+6},
\eeq
where
$$
 \upGamma = \upalpha^4\upgamma+\upalpha^2\upgamma^2-\upalpha \upbeta^3+\upgamma^3
+\upalpha\upgamma \mathcal{H}_2,
\quad
\upDelta = \upalpha^5 \upbeta+\upalpha^3\upbeta\upgamma+4\upalpha\upbeta\upgamma^2+\upbeta^4
+\upbeta^2\mathcal{H}_1
+\upbeta(\upalpha^2+\upgamma)\mathcal{H}_2,
$$
$$
\upTheta = \upalpha^7+3\upalpha^3\upgamma^2+2\upalpha^2\upbeta^3-\upalpha\upgamma^3
+\upalpha^2\upbeta \mathcal{H}_1+\upalpha^4 \mathcal{H}_2
$$
and, in terms of $y_j = x_{j}x_{j+3}/(x_{j+1}x_{j+2})$ for $j=1,2,3,4$,
the first integrals $\mathcal{H}_1$ and $\mathcal{H}_2$ are given  by
\beq\label{s7h12} \bear{rcl}
\mathcal{H}_1 & = &
\upgamma y_1y_2y_3y_4 +\upalpha\upbeta (y_1y_3+y_1y_4+y_2y_4)
+\upalpha\upgamma\left(y_1 +y_2+y_3+y_4 +\frac{y_1y_3}{y_4}+\frac{y_2y_4}{y_1}\right) \\
&&
+\upalpha^2\upbeta\left(\frac{1}{y_1}+\frac{1}{y_2}+\frac{1}{y_3}+\frac{1}{y_4}+\frac{y_2}{y_1y_3}+\frac{y_3}{y_2y_4}\right)
+ \upbeta\upgamma \left( \frac{1}{y_1}+\frac{1}{y_2}+\frac{1}{y_3}+\frac{1}{y_4}\right) \\
&&
+\upalpha^2\upgamma
\left(\frac{1}{y_1y_2}+\frac{1}{y_2y_3}+\frac{1}{y_3y_4}+\frac{1}{y_1y_3}+\frac{1}{y_2y_4} \right)
+\upgamma^2\left( \frac{1}{y_1y_3}+\frac{1}{y_2y_4} \right)
+\upalpha\upbeta^2\left( \frac{1}{y_1y_2y_4}+\frac{1}{y_1y_3y_4} \right)
\\
&&
+\upalpha\upbeta\upgamma\left( \frac{2}{y_1y_2y_3y_4}+\frac{1}{y_1y_2^2y_4} +\frac{1}{y_1y_3^2y_4}\right)
+ \upalpha\upgamma^2 \left( \frac{1}{y_1y_2^2y_3y_4}+\frac{1}{y_1y_2y_3^2y_4} \right) ,
\\
\mathcal{H}_2& = &
\upgamma (y_1y_2y_3 +y_2y_3y_4)+\upalpha\upbeta (y_1+y_2+y_3+y_4)
+ \upalpha\upgamma \left( \frac{y_1}{y_4}+\frac{y_4}{y_1}\right)
+   \upalpha^2 \upbeta \left( \frac{1}{y_1y_3}+\frac{1}{y_2y_4}\right)
\\
&&
+    \upbeta\upgamma \left( \frac{1}{y_1y_2}+\frac{1}{y_2y_3}+\frac{1}{y_3y_4}\right)
+ \upgamma (\upalpha^2+\upgamma ) \left( \frac{1}{y_1y_2y_3}+\frac{1}{y_2y_3y_4}\right)
+\frac{ \upalpha \upbeta^2}{y_1y_2y_3y_4}
\\
&&
+\upalpha\upbeta\upgamma\left( \frac{1}{y_1y_2^2y_3y_4}+\frac{1}{y_1y_2y_3^2y_4} \right)
+\frac{ \upalpha \upgamma^2}{y_1y_2^2y_3^2y_4} .
\eear
\eeq
\end{propn}

Having obtained the first integrals, we now consider each of the three cases corresponding to cluster algebras
separately, and explain how the reduction result of Theorem \ref{torusred} applies in each case.

\vspace{.1in}
\noindent
{\bf The case $\mathbf{\upalpha =0}$:} When $\upalpha =0$, the recurrence
(\ref{s7gen}) arises from a cluster algebra defined by a 7-node quiver.
The latter comes from a  $7\times 7$ matrix, specified in terms of its columns by
$$
B=\left( -{\bf v}_3,    -{\bf v}_4,{\bf v}_1+{\bf v}_2+2{\bf v}_3+{\bf v}_4, -{\bf v}_1+{\bf v}_4,
-{\bf v}_1-2{\bf v}_2-{\bf v}_3-{\bf v}_4,{\bf v}_1,{\bf v}_2 \right),
$$
where im$\, B$ is spanned by the vectors
\beq\label{vj}
{\bf v}_j={\bf e}_j-{\bf e}_{j+1}-{\bf e}_{j+2}+{\bf e}_{j+3} \qquad
\mathrm{for}\qquad  j=1,2,3,4,
\eeq
where ${\bf e}_j$ is the $j$th standard basis vector.
In this case, ker$\, B$ is spanned by the three integer vectors
\beq\label{kerb}
{\bf u}_1=(1,1,1,1,1,1,1)^T, \qquad
{\bf u}_2=(1,2,3,4,5,6,7)^T, \qquad {\bf u}_3=(1,-1,1,-1,1,-1,1)^T.
\eeq
These three integer vectors produce
a three-dimensional group of scaling transformations,
$${\bf x} \rightarrow \la^{{\bf u}_1} \cdot \mu^{{\bf u}_2}\cdot \nu^{{\bf u}_3}\cdot  {\bf x},$$
which coincides  with the scalings  defined by (\ref{2torus}) and
(\ref{extrasym}).  By Lemma \ref{symp}
there  is a symplectic form, given in terms of the scale-invariant monomials
$y_j = {\bf x}^{{\bf v}_j}$ for $j=1,2,3,4$, as
$$
\hat\om = \frac{ \dd y_1 \wedge \dd y_3}{y_1y_3}+
\frac{ \dd y_2 \wedge \dd y_3}{y_2y_3}+
\frac{ \dd y_2 \wedge \dd y_4}{y_2y_4}.
$$
This yields the Poisson bracket
\beq\label{PBs7azero}
\{y_j,y_{j+1}\}=0, \qquad \{ y_j, y_{j+2} \} = -y_j y_{j+2}, \qquad
\{ y_j, y_{j+3} \} = y_j y_{j+3}.
\eeq
In terms of $y_j$ one finds the map
$$
\hat\varphi : \qquad (y_1,y_2,y_3,y_4) \mapsto \Big(y_2,y_3,y_4,(\upbeta y_3+\upgamma)/(y_1y_2y_3^2y_4)\Big),
$$
which is equivalent to iteration of (\ref{s7yrec}) with $\upalpha =0$, and
preserves the nondegenerate Poisson bracket (\ref{PBs7azero}).
Setting $\upalpha =0$ in the two first integrals in (\ref{s7h12})
and computing their bracket gives
$
\{ \mathcal{H}_1,\mathcal{H}_2\} |_{\upalpha =0} =0,
$
so this is a Liouville integrable system in 4D.

\vspace{.1in}
\noindent
{\bf The case $\mathbf{\upbeta =0}$:} In this case the matrix $B$ is specified by
$
B= (-\hat{{\bf v}}_2 , \hat{{\bf v}}_1,\hat{{\bf v}}_2,-\hat{{\bf v}}_1+\hat{{\bf v}}_2, -\hat{{\bf v}}_1,
-\hat{{\bf v}}_2,\hat{{\bf v}}_1),
$
where $\hat{{\bf v}}_j={\bf v}_j+{\bf v}_{j+1}+{\bf v}_{j+2}$ for $j=1,2$.
The kernel of $B$ is 5-dimensional, being spanned by the integer vectors
${\bf u}_2$ and  ${\bf u}_3$ together with
$
{\bf u}_4 = (1,0,0,1,0,0,1)^T, \;\; {\bf u}_5 =  (0,1,0,0,1,0,0)^T, \;\; {\bf u}_6 =  (0,0,1,0,0,1,0)^T.
$
These five independent integer vectors give an action of the algebraic torus $(\C^*)^5$ on ${\bf x}$
by scaling transformations, and the scalings  (\ref{2torus}) and
(\ref{extrasym}) form a three-parameter subgroup,
since ${\bf u}_1={\bf u}_4+{\bf u}_5+{\bf u}_6$.
The invariants under the full 5-parameter scaling group are
$$\hat{y}_j = {\bf x}^{\hat{{\bf v}}_j}  = x_jx_{j+5}/(x_{j+2}x_{j+3})
= y_jy_{j+1}y_{j+2},$$
and in terms of these
one obtains the general Lyness-2 recurrence with coefficients, that is
\beq\label{lyness2}
\hat{y}_{n+2}\,  \hat{y}_{n}=\upalpha \, \hat{y}_{n+1}+\upgamma,
\eeq
which is equivalent to iteration of a map in the $(\hat{y}_1,\hat{y}_2)$ plane and of  QRT type \cite{qrt1},
with invariant symplectic form
$(\hat{y}_1\hat{y}_2)^{-1} \dd \hat{y}_1 \wedge  \dd \hat{y}_2$. The first integral $\mathcal{H}_1$ does not reduce to the plane,
as it is not invariant under the full 5-dimensional scaling group, but
$\mathcal{H}_2$ is fully invariant, and reduces to a first integral of (\ref{lyness2}):
$$
\mathcal{H}_2|_{\upbeta =0} = \upgamma \, (\hat{y}_1+  \hat{y}_2) +\upalpha\upgamma \, \left(
\frac{\hat{y}_1}{\hat{y}_2}+\frac{\hat{y}_2}{\hat{y}_1}\right) +\upgamma (\upalpha^2+\upgamma ) \left(
\frac{1}{\hat{y}_1}+\frac{1}{\hat{y}_2}\right)+\frac{\upalpha\upgamma^2}{\hat{y}_1\hat{y}_2}.
$$
The level sets of the latter are cubic (and biquadratic) plane curves of genus one.

\vspace{.1in}
\noindent
{\bf The case $\mathbf{\upgamma =0}$:} This is very similar to the case    $\upalpha =0$,
since the recurrence
comes from the rank 4 matrix 
$$
B=\left( -{\bf v}_2-{\bf v}_4,  {\bf v}_1+{\bf v}_2+{\bf v}_4,-{\bf v}_1+{\bf v}_2, -{\bf v}_2+{\bf v}_3,
-{\bf v}_3+{\bf v}_4,-{\bf v}_1-{\bf v}_3-{\bf v}_4,{\bf v}_1+{\bf v}_3\right),
$$
with the same vectors
${\bf v}_j$ as in (\ref{vj}). Since
both ker$\, B$ and im$\, B$ are the same as
for $\upalpha =0$, there is the same scaling group
with invariants $y_j={\bf x}^{{\bf v}_j}$.
The symplectic form in this case is
$$
\hat\om = \frac{ \dd y_1 \wedge \dd y_2}{y_1y_2}+
\frac{ \dd y_1 \wedge \dd y_4}{y_1y_4}+
\frac{ \dd y_3 \wedge \dd y_4}{y_3y_4},
$$
and (up to an overall constant) this gives the unique
log-canonical Poisson bracket
\beq\label{PBs7czero}
\{y_j,y_{j+1}\}=y_jy_{j+1}, \qquad \{ y_j, y_{j+2} \} = 0 =
\{ y_j, y_{j+3} \} 
\eeq
that is preserved by the map
$$
\hat\varphi : \qquad (y_1,y_2,y_3,y_4) \mapsto \Big(y_2,y_3,y_4,(\upalpha y_2y_4+ \upbeta )/(y_1y_2y_3y_4)\Big).
$$
The latter map in 4D corresponds to iteration of (\ref{s7yrec}) with $\upgamma =0$. 
Setting $\upgamma =0$ in (\ref{s7h12}) and
computing the bracket 
using  (\ref{PBs7czero}) gives
$
\{ \mathcal{H}_1,\mathcal{H}_2\} |_{\upgamma =0} =0,
$
so the two first integrals are involution, as required.
\begin{rem} \label{s7integ}{\em
We have shown that for the two parameter subcases of (\ref{s7gen}), the reduced map
is integrable in the Liouville sense, either in 4D or 2D.  However, there is a different Poisson structure
in each case. When $\upalpha\upbeta\upgamma \neq 0$, it is easy to check that there is no
log-canonical Poisson bracket in the variables $y_j$ that is compatible with (\ref{s7yrec}).
Nevertheless, we expect that
there is a compatible Poisson structure for which
the two first integrals in (\ref{s7h12}) are in involution, so that
(\ref{s7yrec}) defines a 4D map that is Liouville integrable.
}
\end{rem}

\subsubsection*{Acknowledgments:} The authors would like to thank the Isaac
Newton Institute, Cambridge for hospitality in 2009 during the Programme on Discrete
Integrable Systems, where this collaboration began.
We are also grateful to A. Veselov and A. Zelevinsky for helpful comments.

\end{document}